%% file: MNRAS_main.tex
\documentclass[fleqn,usenatbib]{mnras}
\usepackage[T1]{fontenc}
\usepackage{graphicx}	
\usepackage{amsmath}	
\usepackage{amssymb}	

\usepackage{newtxtext,newtxmath}
\usepackage{subfigure}
\newcommand{\wazp}{\textsc{W}a\textsc{ZP}}
\newcommand{\RM}{redMaPPer}
\newcommand{\SZ}{\textsc{SZ}}
\newcommand{\redMaPPer}{\RM}
\newcommand{\rich}{N_{gals}}

\newcommand{\phz}{\textsc{photo-z}}

\newcommand{\Munit}{M_\odot h^{-1}}

\newcommand{\zwazp}{z_{\wazp}}
\newcommand{\zsz}{z_{SZ}}
\newcommand{\zrm}{z_{RM}}

\newcommand{\grizy}{$g, r, i, z, Y$ }
\newcommand{\griz}{$g, r, i, z$ }
\newcommand{\magG}{m_{g}}
\newcommand{\magR}{m_{r}}
\newcommand{\magI}{m_{i}}
\newcommand{\magZ}{m_{z}}

\newcommand{\wazpweblink}{\footnote{\href{https://www.linea.gov.br/catalogs/wazp/}{https://www.linea.gov.br/catalogs/wazp/}}}

\usepackage{eso-pic}
  \newcommand{\DESnumbers}{
  \AddToShipoutPictureBG*{%
    \AtPageUpperLeft{%
      \hspace{0.75\paperwidth}%
      \raisebox{-5.5\baselineskip}{%
        \makebox[0pt][l]{\textnormal{DES-2017-0241}}
  }}}%

  \AddToShipoutPictureBG*{%
    \AtPageUpperLeft{%
      \hspace{0.75\paperwidth}%
      \raisebox{-6.5\baselineskip}{%
        \makebox[0pt][l]{\textnormal{FERMILAB-PUB-20-419-AE}}
  }}}%
}

\title[The WaZP cluster sample of the DES Y1]{The WaZP galaxy cluster sample of the Dark Energy Survey Year 1}

\author[M. Aguena et al.]{
    M.~Aguena,$^{1,2}$\thanks{E-mail: aguena@linea.gov.br}
    C.~Benoist,$^{1,3}$
    L.~N.~da Costa,$^{1,4}$
    R.~L.~C.~Ogando,$^{1,4}$
    J.~Gschwend,$^{1,4}$
    \newauthor
    H.~B.~Sampaio-Santos,$^{1,4}$
    M.~Lima,$^{1,5}$
    M.~A.~G.~Maia,$^{1,4}$
    S.~Allam,$^{6}$
    S.~Avila,$^{7}$
    D.~Bacon,$^{8}$
    E.~Bertin,$^{10,9}$
    \newauthor
    S.~Bhargava,$^{11}$
    D.~Brooks,$^{12}$
    A.~Carnero~Rosell,$^{13,14}$
    M.~Carrasco~Kind,$^{15,16}$
    J.~Carretero,$^{17}$
    M.~Costanzi,$^{18,19}$
    \newauthor
    J.~De~Vicente,$^{20}$
    S.~Desai,$^{21}$
    H.~T.~Diehl,$^{6}$
    P.~Doel,$^{12}$
    S.~Everett,$^{22}$
    A.~E.~Evrard,$^{23,24}$
    I.~Ferrero,$^{25}$
    A.~Fert\'e,$^{26}$
    \newauthor
    B.~Flaugher,$^{6}$
    P.~Fosalba,$^{27,28}$
    J.~Frieman,$^{29,6}$
    J.~Garc\'ia-Bellido,$^{7}$
    P.~Giles,$^{11}$
    R.~A.~Gruendl,$^{15,16}$
    G.~Gutierrez,$^{6}$
    \newauthor
    S.~R.~Hinton,$^{30}$
    D.~L.~Hollowood,$^{22}$
    K.~Honscheid,$^{31,32}$
    D.~J.~James,$^{33}$
    T.~Jeltema,$^{22}$
    K.~Kuehn,$^{34,35}$
    \newauthor
    N.~Kuropatkin,$^{6}$
    O.~Lahav,$^{12}$
    P.~Melchior,$^{36}$
    R.~Miquel,$^{17,37}$
    R.~Morgan,$^{38}$
    A.~Palmese,$^{29,6}$
    \newauthor
    F.~Paz-Chinch\'{o}n,$^{16,39}$
    A.~A.~Plazas,$^{36}$
    A.~K.~Romer,$^{11}$
    E.~Sanchez,$^{20}$
    B.~Santiago,$^{1,40}$
    M.~Schubnell,$^{24}$
    \newauthor
    S.~Serrano,$^{27,28}$
    I.~Sevilla-Noarbe,$^{20}$
    M.~Smith,$^{41}$
    M.~Soares-Santos,$^{24}$
    E.~Suchyta,$^{42}$
    G.~Tarle,$^{24}$
    C.~To,$^{43}$
    \newauthor
    D.~L.~Tucker,$^{6}$
    and R.D.~Wilkinson,$^{11}$
    \\
    (Affiliations can be found after the references)
}

\date{Accepted 16 Febuary}
\pubyear{2021}

\begin{document}
\label{firstpage}
\pagerange{\pageref{firstpage}--\pageref{lastpage}}
\maketitle

\input{files/info.tex}
\input{files/wazp_info.tex}
\begin{abstract}
{We present a new (2+1)D galaxy cluster finder based on photometric redshifts called Wavelet Z Photometric (\wazp) applied to DES first year (Y1A1) data. The results are compared to clusters detected by the South Pole Telescope (SPT) survey and the \RM\ cluster finder, the latter based on the same photometric data. 
}
   {\wazp\ searches for clusters in wavelet-based density maps of galaxies selected in photometric redshift space without any assumption on the cluster galaxy populations.
   The comparison to other cluster samples was performed with a matching algorithm based on angular proximity and redshift difference of the clusters.
It led to the development of a new approach to match two optical cluster samples, following an iterative approach to minimize incorrect associations.
   }
   {The \wazp\ cluster finder applied to DES Y1A1 galaxy survey (\VacArea\ deg$^2$ up to $\magI = 23$ mag) led to the detection of \WazpClusters\ galaxy
   clusters with redshifts $0.05 < z < 0.9$ and richness $N_{gals}\geq5$.
   Considering the overlapping regions and redshift ranges between the DES Y1A1 and SPT cluster surveys, all \SZ\ based SPT clusters are recovered by the \wazp\ sample.
   The comparison between \wazp\ and \RM\ cluster samples showed an excellent overall agreement for clusters with richness $\rich$ ($\lambda$ for \RM) greater than 25 (20), with 95\% recovery on both directions.
  Based on the cluster cross-match we explore the relative fragmentation of the two cluster samples and investigate the possible signatures of unmatched clusters.  
}
\end{abstract}




\DESnumbers
\maketitle
   
	\section{Introduction}

The abundance and clustering properties of galaxy clusters have been shown to be powerful probes to constrain cosmological models, provided that their astrophysical properties are well characterized and linked to theoretical predictions \citep[e.g.,][]{Lima05,Vik09,Man10,Ben13,Wei13,Ade15,Cos20}.

Galaxy clusters can be detected from X-ray observations  (\citealt{Kim07,Lloyd11,Adami18})
and from the Sunyaev-Zel’dovich (SZ) effect \citep{Ble15}, but on-going and future large photometric surveys constitute a very promising approach to build large controlled galaxy cluster samples for both cosmological and astrophysical studies. These include the Kilo Degree Survey (KIDS,  \citealt{kids2013}), the Dark Energy Survey (DES, The Dark Energy Survey Collaboration \citealt{Fla05}), Pan-STARRS \citep{kaiser2002}, the Legacy Survey of Space and Time (LSST,  \citealt{LSST2009}) and the European Space Agency Cosmic Vision mission (Euclid, \citealt{euclid2011}).

However, detecting and characterizing clusters through their galaxy component remains a nontrivial task, especially when considering lower mass or higher redshift clusters. One has to distinguish between gravitationally bound groups of galaxies and projection effects due to the underlying large scale distribution of galaxies. Projection effects not only impact detection, but also several fundamental properties of detected clusters, such as centering, redshift, and mass proxy (e.g., cluster richness).

Many automated algorithms were developed in the last three decades to overcome these difficulties. Automatic optical cluster finders can generally be described as algorithms searching for cluster scale galaxy overdensities. Galaxies are first filtered (or weighted) following prescriptions to increase the detection contrast relative to background galaxies. The main techniques used for searching galaxy overdensities include kernel smoothing \citep[e.g.,][]{shectman1985,lumsden1992,adami2010,gladders2000}, Friends-of-Friends \citep[e.g.,][]{botzler2004,trevese2007,Wen12} or Voronoi tesselation \citep[e.g.,][]{ramella1999,Soa11}. These 
techniques have been applied to galaxy catalogs that are
usually previously filtered in one or several dimensions (e.g., magnitudes, colors, or photometric redshifts). More sophisticated approaches assume an underlying cluster model (e.g., density profile, luminosity function, color content) and identify clusters in likelihood maps based on matched filter techniques \citep[e.g.,][]{postman1996,olsen1999,Koe07,olsen2008,Ryk14,Ryk16,bellagamba2018}.

A typical assumption of optical cluster finders is to consider the presence of a red sequence of galaxies \citep{gladders2000,Koe07,Hao10,Mur12}. In low redshift clusters, the most luminous galaxies define a tight sequence in the color-magnitude diagram, the so-called "E/S0 ridge line", or "red sequence". Red sequence galaxies have very uniform colors and are among the reddest galaxies at a given redshift. Because of the strong 4000 \AA break in their rest-frame spectra, their color is tightly correlated with redshift and can be used to estimate cluster redshifts. This feature has been observed in rich clusters up to $z\sim 1.7$ \citep[e.g.,][]{mei2009,george2011,wetzel2013,Stra19}.
However, some galaxy clusters observed at high redshifts can display appreciable star formation, even in cluster cores \citep[e.g.,][]{brodwin2013}, weakening the red sequence.

Cluster finders such as maxBCG \citep{Koe07}, \RM\ \citep{Ryk14,Ryk16} or RedGOLD \citep{licitra2016} rely on the red sequence for cluster detection and redshift estimate. In the context of recent surveys, cluster finders not based on the red sequence usually rely on photometric redshifts. An alternative based on the knee of the cluster luminosity function was also used in the context of surveys with a limited number of passbands \citep[e.g.,][]{postman1996,olsen1999}.

Even if current automated optical cluster finders are all able to identify rich clusters, evaluating their performances over broad ranges of masses and redshifts and deriving the selection function of the resulting cluster samples remain highly complex tasks. On the theoretical side, this requires the development of ever more realistic simulated galaxy catalogs. On the observational side, we need multiple surveys covering the same area at different frequency domains to detect clusters through a variety of signatures.

There is not a unique methodological framework to evaluate {and compare} the performances of optical cluster finders. A variety of approaches have been proposed, based either on mock galaxy catalogs (e.g., \cite{Adam2019} and references therein), or on real data, or even on a mix of the two approaches \citep[e.g.,][]{goto2002,kim2002,Ryk14,Cos19}.

Within simulations assumptions, simulation-driven methods provide a truth table useful for comparison, with clusters embedded in realistic large scale structures.
These methods also offer a direct link between galaxy clusters and dark matter halos. However, they rely on sophisticated modeling that so far does not fully reproduce all observed galaxy properties, specially at high redshift \citep[e.g.,][]{derose19}. In addition to this fundamental problem, mock catalogs do not usually reproduce the variety and complexity of defects occurring in observed images and introduced at the stage of source extraction and classification. DES has recently started to deal with this using the Balrog algorithm \citep{suchyta16}, which embeds simulations into real data and should accompany future releases.

Addressing the cluster selection function based on real data is necessarily limited by the absence of an absolute reference to confront the results of any cluster finder. Nonetheless, useful information can be extracted from the cross-match of a given optical cluster sample with detections based on different tracers that do not suffer from the same projection effects \citep[e.g.,][]{Sar15}
A better understanding of the galaxy cluster selection function can also be improved from cross-matching samples from different optical cluster finders. The resulting samples may differ not only due the different adopted physical assumptions but also due to the details of cluster finder implementation \citep{Asc16, AguLim18}, or even the way the algorithms deal with specific features of real data (e.g. noise, missing data, star/galaxy separation, etc.).

DES has produced galaxy cluster samples with the \RM\ algorithm which were published in  \citet{Ryk16} and \citet{McC19}, based on DES Science Verification and DES-Y1 data releases, respectively. These samples led to several studies focusing on the mass-richness relation
 (\citealt{Mel15, Sar15, Zha16, Pal16, Sar17, Mel17,Per18,McC19,Ble20, Per20}).
Complementary analyses were also performed on cluster luminosity function \citep{Zha17}, baryon content \citep{Chi18}, and cluster miscentering relative to X-ray detections \citep{Zha19}. These clusters were also used for the detection of voids \citep{Pol19}.
Most of these studies contribute to the work on cosmological constraints using DES first year release \RM\ clusters \citep{Cos20}.

In this paper, we present the Wavelet Z Photometric (\wazp) cluster finder and apply it to DES-Y1 data. \wazp\ is an optical cluster finder designed to detect clusters based mainly on the spatial clustering of galaxies using photometric redshift information. The primary motivation for developing \wazp\ is to limit assumptions on the properties of cluster galaxies such as the presence of a red sequence, the shape of their luminosity function or radial profile, assumptions that may impact cluster detection, in particular at high redshift or at lower mass regime.

Here, the \wazp\ DES-Y1 sample is compared to cluster samples obtained from the SPT survey based on the \SZ\ effect and those obtained by the \RM\ cluster finder on the same DES-Y1 data set. The first comparison allows to test how well the \wazp\ algorithm recovers the massive clusters detected by the SPT. The second comparison, for which the two samples have similar cluster densities, gives insights on the relative completnesses of the two optical cluster samples, and on the derived properties of the common detections.
Variations may occur in the samples due to the different assumptions made in terms of cluster modelling. They may also occur due to different uses of the underlying galaxy dataset as the
\wazp\ algorithm uses magnitude information from all bands through photometric redshifts and $i$-band as a reference band, whereas \RM\ uses combinations of band pairs (colors) to select likely red sequence galaxies and $z$-band as a reference.
Considered survey coverage can therefore be slightly different with one approach or the other. Depth variability in all bands will also impact differently cluster detection with each algorithm.
While the comparisons performed here provide a valuable heuristic approach to partly qualify the cluster samples, a complete evaluation of the \wazp\ sample requires to address a quantitative assessment of its purity, a work that will be presented in a companion paper based on mock galaxy catalogs.

The present paper is organized as follows.
In Section~\ref{sec:DESdata}, we describe the DES Y1 data used in our analysis.
In Section~\ref{sec:WAZP}, we describe the main properties of the \wazp\ cluster finder.
In Section~\ref{sec:results}, we present our main results on DES-Y1 data.
In Section~\ref{sec:matches} we matched the derived \wazp\ cluster samples to the \SZ\ sample and to the \RM\ sample obtained from DES-Y1 data.
Finally, in Section~\ref{sec:discussion} we analyze the properties of the catalog,
and discuss the differences between our catalog and the others compared in Section~\ref{sec:matches}.

Throughout this work, we fix cosmological parameters from the Planck results \citep{Ade15} for a flat $\Lambda$CDM model with $\Omega_m=0.308$ and $H_0=67.8$ km s$^{-1}$~Mpc$^{-1}$.

	\section{Data}
    \label{sec:DESdata}

The DES is an imaging survey covering 5,000 deg$^2$ in 5 bands (\grizy) \citep[e.g.,][]{Fla05,Die16,Abb16}.
In this paper, we use the DES Year 1 data release, which has been extensively studied by the DES collaboration \cite[e.g.,][]{Trox18, Shipp18}.
The DES built the Dark Energy Camera \citep[DECam,][]{Fla15} with a field of view diameter of 2.2 deg covered by 520 Megapixels distributed on a mosaic of 62 CCDs that are extra sensitive on the red part of the electromagnetic spectrum, enhancing its capability of observing high redshift galaxies.
DECam is installed on CTIO 4-meter Blanco telescope prime focus, and its observations follow a strategy that optimizes pointings based on properties like weather and moon phase \citep{Nei19}. The images are reduced and calibrated by the DES Data Management (DESDM) team at the National Center for Supercomputing Applications (NCSA).
The DESDM pipeline includes the reduction of single-exposure images, their co-addition into deeper images, source extraction and calibration, all resulting in the creation of the main scientific catalog \citep{Mor18}.

The DES Year 1 Annual Release \citep[Y1A1,][]{Abb18} co-added catalog used in this analysis covers a total area of $\sim$1,520~deg$^2$, split into two main wide regions. One of them has an area of $\sim$140~deg$^2$ overlapping the SDSS Stripe 82 area \citep{Aih11}. The other part has an area of $\sim$1,380~deg$^2$ overlapping the South Pole Telescope footprint \citep[SPT,][]{Car11}. In the following, we will refer to these two regions as Y1-S82 and Y1-SPT, respectively. They were observed with three to four exposures in each filter \citep{Drl18Gold}.

In addition to catalogs, we also use ancillary maps to track defects and foreground objects (e.g., bright stars, very bright galaxies, and globular clusters) all over DES footprint. The resulting coverage map is represented by a detection fraction map, where pixels have values of area fraction from 0 to 1. We also use systematic maps to track observing conditions across the footprint, such as number of exposures, seeing, and airmass \citep{Lei16SysSV}. These maps are combined to produce depth maps based on galaxy magnitude limits, as described in \citet{Ryk15depth}. All maps are recorded in \textsc{Healpix} format (\textsc{nside} = 4096) \citep{healpix}.

The Y1A1 coadd catalog and maps produced by DES DM were transferred to the Laborat\'orio Interinstitucional de e-Astronomia (LIneA)\footnote{\href{http://www.linea.gov.br}{http://www.linea.gov.br}} and ingested into the database associated to the DES Science Portal (henceforth, the Portal) as described in \citet{Fau18}.
We used the Portal infrastructure to create a galaxy Value Added Catalog (VAC) tailored for galaxy cluster search based on photometric redshift. The creation of the galaxy VAC includes: computation of photometric redshifts, star-galaxy classification, and pruning regions and objects to produce a
clean galaxy catalog with well controlled levels of completeness and homogeneity. Along with the VAC, a final footprint map in \textsc{Healpix} format is created, reflecting the selection and pruning applied to this VAC.

The computation of photometric redshifts relies on the machine-learning algorithm \textsc{DNF} (Directional Neightbourhood Fitting, \citealt{deV16}), operated in Euclidean Neighborhood Fitting (ENF) mode since tests using DES Y1 data have shown that ENF mode is considerably faster, while providing similar results as in DNF mode. \textsc{DNF} uses as input observables SExtractor MAG\_AUTO magnitudes \citep{Ber96}.
\textsc{DNF} was trained with a large sample of spectroscopic redshifts extracted from a compilation of 29 public surveys intercepting the DES footprint. 
Quality flags of all surveys are brought to a common standard following OzDES approach \citep{Yuan15ozdes}. As described in \citet{GSC18}, sources with flags 0 and 1 have unknown redshift, flag 2 redshifts are not reliable, flag 3 redshift reliability is above 90\% confidence, and flag 4 is attributed to a trusted redshift (over 99\% confidence).
The DES photometric catalog is matched to this spectroscopic redshift sample with a 1.0~arcsec search radius and down to $\magI=23$ mag, producing a catalog of 101,971 galaxies with a mean redshift of 0.63, and covering the redshift range $z=0-1.1$.
Although color and magnitude distributions of this spectroscopic sample differ from the global photometric set under study, we stress that it does cover the same color-magnitude ranges with the exception of faint low redshift galaxies (typically magnitudes fainter than 19 and redshifts below 0.15). The spectro-photometric catalog is then randomly split into a training and a validation sets. Details about all the steps carried out to compute photometric redshifts in the DES Science Portal for Y1A1 data are described in \citet{GSC18}.

Star-galaxy classification follows a morphological prescription developed within the DES consortium called \texttt{MODEST}, described by equations~(3) and (4) of \citet{Sev18}. It mainly depends on the SExtractor \texttt{SPREAD\_MODEL} \citep{Des12,Buoy13} and its error, assessing how extended is the source to the local PSF. Note that this classification is based on the DES $i$-band and extends to the faintest sources.

Based on bad regions maps \citep{Drl18Gold}, areas around bright stars, bleed trails, bright foreground galaxies, or globular clusters \citep[from][]{Har10} are removed from the footprint. 
Pixels with an effective coverage $\le$ 0.1 are discarded from our analysis. We also exclude regions
not covered simultaneously by \griz bands with a minimum of 90s total exposure time.
Besides region-based filtering, we also discard individual sources based on SExtractor \texttt{FLAGS} (only sources with \texttt{FLAGS} $\le$ 3 are kept), apply a magnitude cut ($\magI \ge 15$), and color cuts
($-2.0 \le \magG-\magR \le 10.0$;
 $-2.0 \le \magR-\magI \le 4.0$;
 $-2.0 \le \magI-\magZ \le 4.0$).

\begin{figure}
	\includegraphics[scale=1]{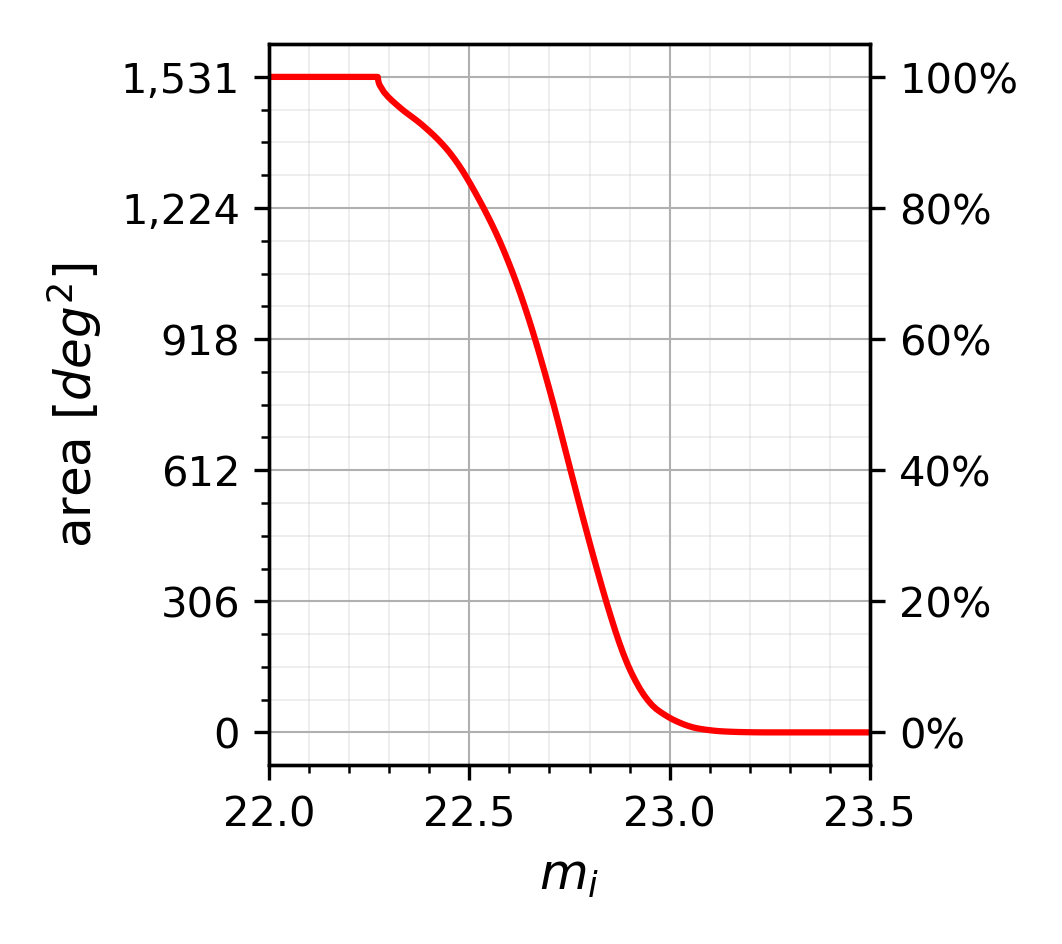}
	\caption{The effective area coverage of DES-Y1 as a function of the limiting magnitude in the $i$-band.
	}
	\label{fig:depth_10sig}
\end{figure}

\begin{figure}
	\includegraphics[scale=1]{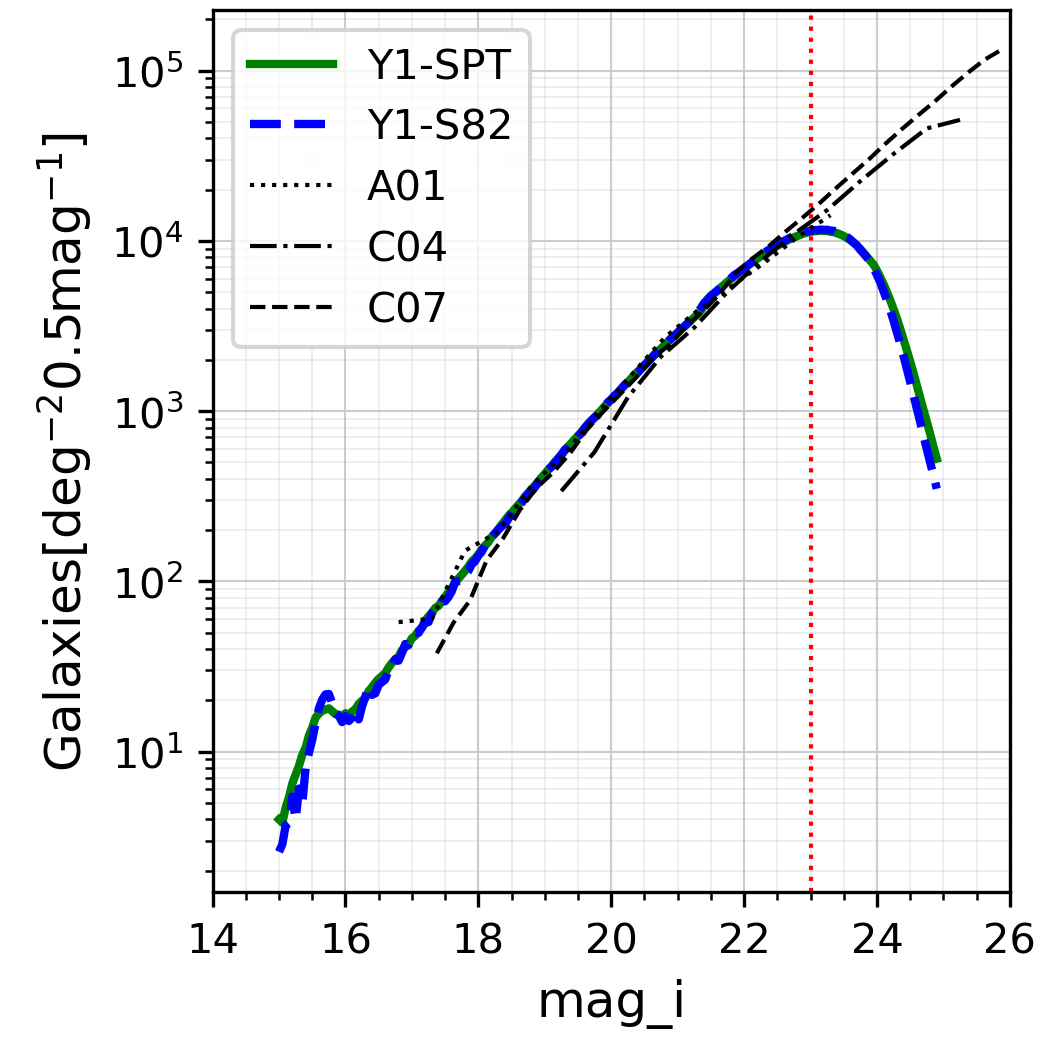}
	\caption{Galaxy density for Y1-S82 (thick blue solid line) and Y1-SPT (thick green dashed line) compared to \citealt{Arn01} (A01), \citealt{Cap04} (C04) and \citealt{Cap07} (C07).	Magnitude limits of 22.5 mag and 23 mag are also noted as red vertical dotted and dashed line, respectively.
	}
	\label{fig:maglim_nc}
\end{figure}

Depth maps, defined here as 10$\sigma$ limiting magnitude maps, were built following the method described in \citealt{Ryk15depth}. 
These maps correspond to the limit where the flux is at least 10 times its variance $\sigma$, computed from the magnitude errors. 
This definition assures a galaxy completeness larger than 90\%. Figure~\ref{fig:depth_10sig} shows what survey area fraction is covered at a given 10$\sigma$ $i$-band limiting magnitude (the $i$-band being the reference band in this work). The whole survey is at least as deep as $\magI=22.27$ and shallower than $\magI=23.25$ with half of the survey area reaching $\magI=22.7$. 
In Figure~\ref{fig:maglim_nc} we compare galaxy number counts for Y1-S82 and Y1-SPT with number counts from \cite{Arn01} and \cite{Cap07} as compiled by Nigel Metcalfe\footnote{http://astro.dur.ac.uk/~nm/pubhtml/counts/counts.html}. The surveys used for comparison are deeper than our own, and number counts are comparable through a wide range of magnitudes up to $\magI\sim 22.5$ mag, beyond which we observe an increasing deficit of galaxies, consistent with the median depth of the survey.

Based on the survey depth shown above, the present analysis considers galaxies down to a limiting magnitude $\magI=23$ (98\% of the survey area). The resulting galaxy VACs for Y1-S82 and Y1-SPT contain, respectively, \VacSetGals\ and \VacSptGals\ galaxies, in a total of \VacGals\ galaxies
covering \VacArea~deg$^2$. Both regions have similar galaxy number densities and mean photometric redshift (Table~\ref{tab:vacproperties}).

\begin{figure}
	\includegraphics[scale=1]{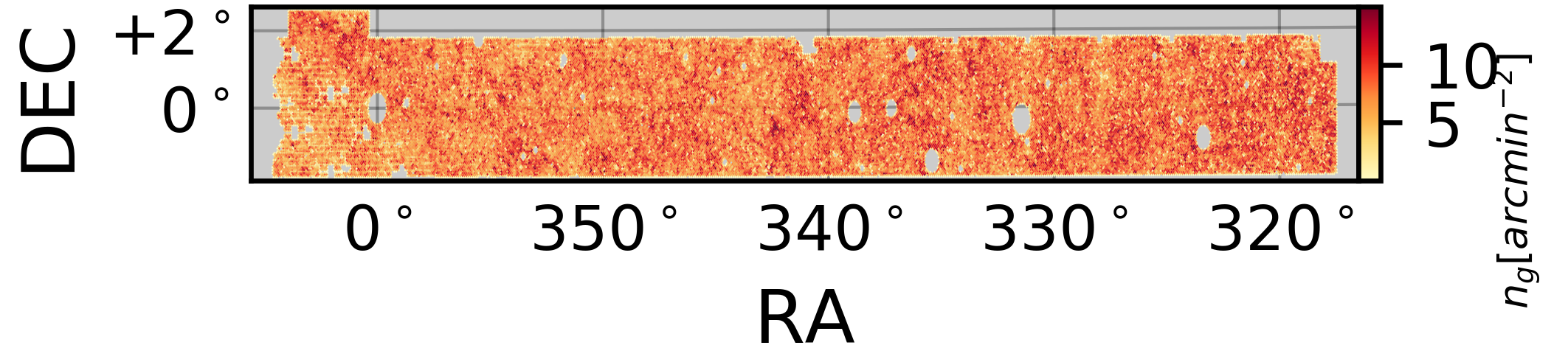}
	\includegraphics[scale=1]{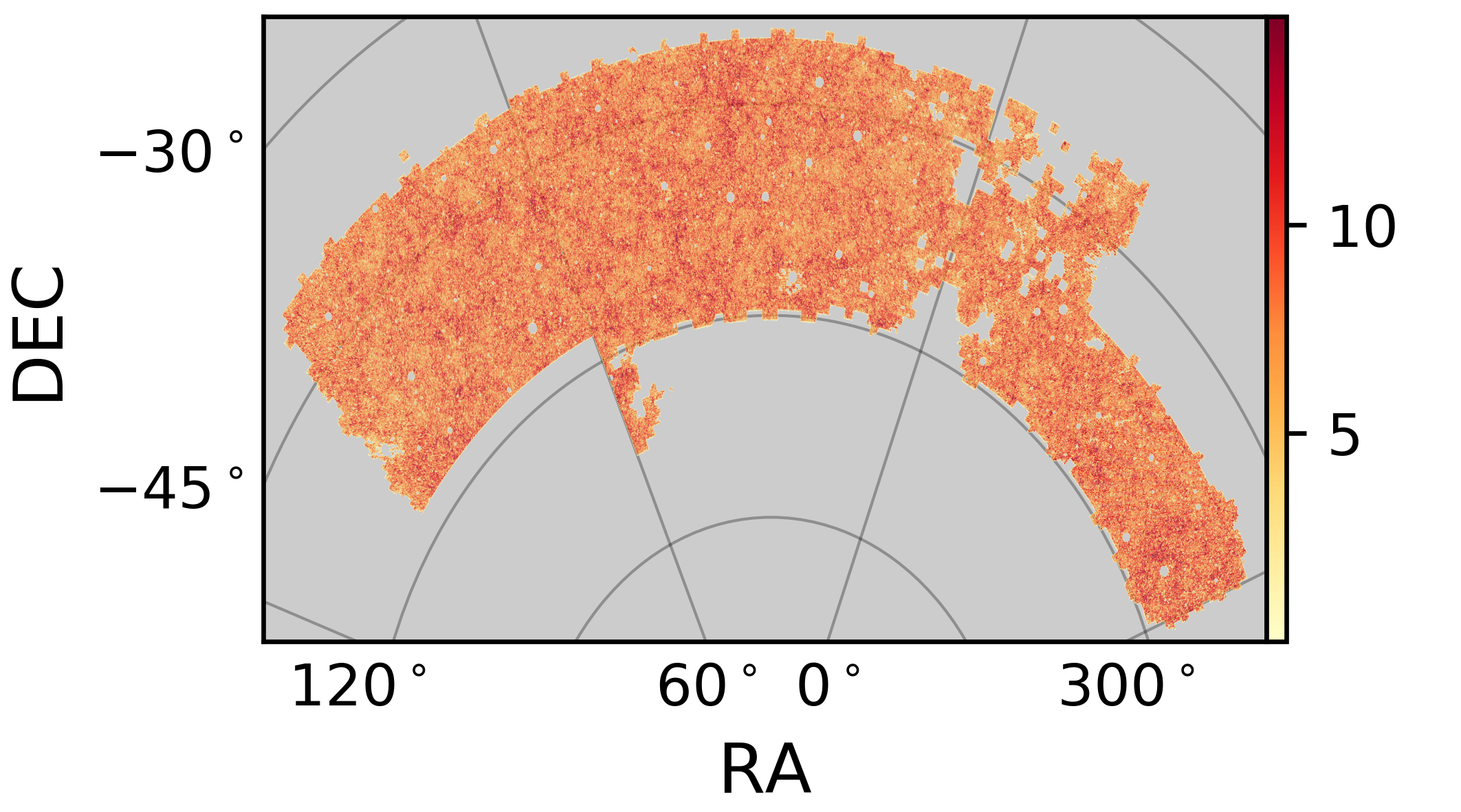}
	\caption{Spatial distribution of galaxies with $\magI\le 23$ for Y1-S82 (top) and Y1-SPT (bottom). Color bar displays density of galaxies per pixel at NSIDE=1024 ($\approx11$ arcmin$^2$).
	}
	\label{fig:spatdist}
\end{figure}

Figure~\ref{fig:spatdist} presents the projected galaxy distribution of Y1-S82 (top) and Y1-SPT (bottom)\footnote{These were produced with \href{https://github.com/pmelchior/skymapper}{skymapper} by Peter Melchior}. The distributions are fairly uniform and galaxy densities comparable. Holes caused by masking and dents on the footprint caused by unobserved regions can be seen in both regions.

\begin{figure}
	\includegraphics[scale=1]{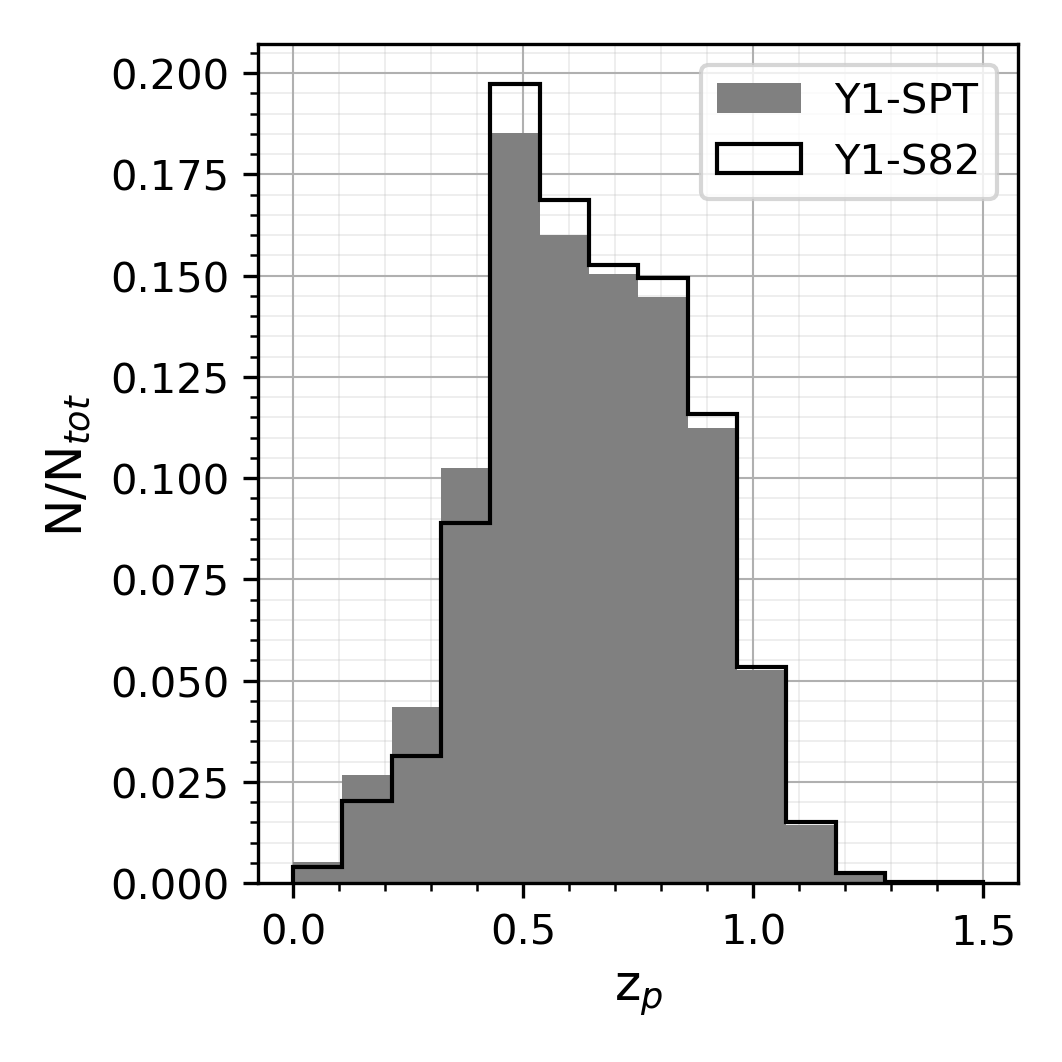}
	\caption{Normalized distribution of photometric redshifts for galaxies in Y1-S82 (dashed line) and Y1-SPT (solid line). }
	\label{fig:phz_dist}
\end{figure}

\begin{figure*}
	\includegraphics[scale=.98]{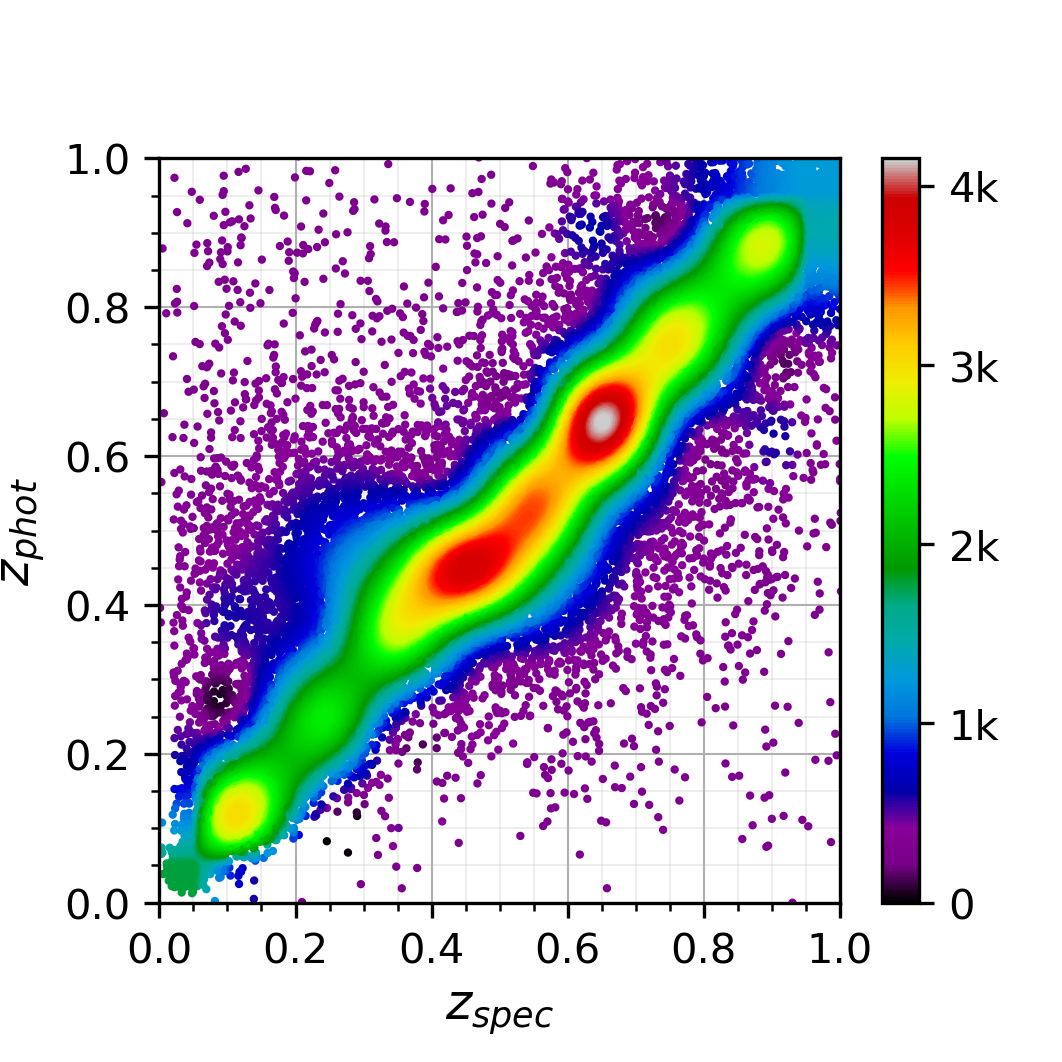}
	\includegraphics[scale=.98]{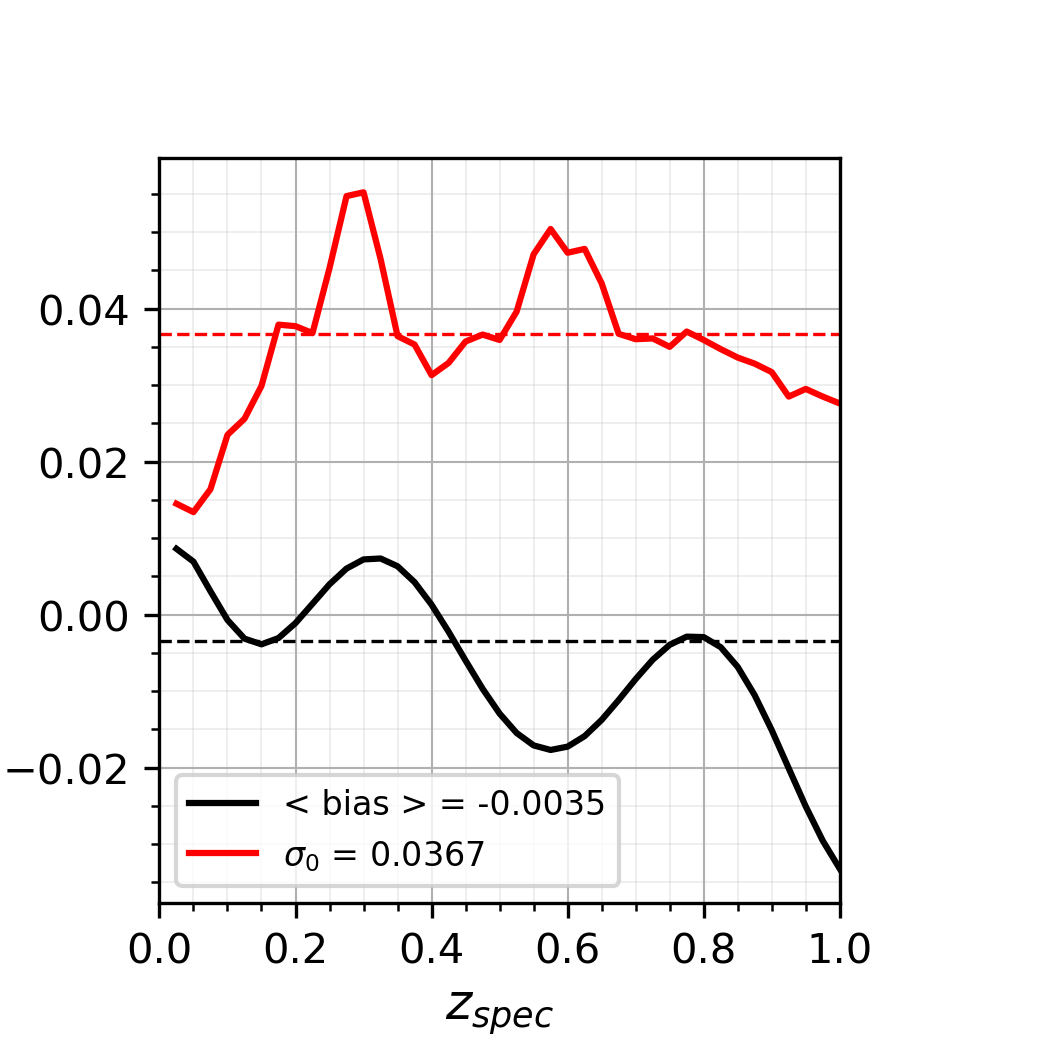}
	\caption{The DNF photometric redshifts compared to their associated spectroscopic redshifts for a validation sample of 50,476 galaxies within the DES-Y1 footprint. The right panel shows the bias and scatter of photometric redshifts relative to spectroscopic redshifts as a function of spectroscopic redshift.}
	\label{fig:zpstats}
\end{figure*}

Normalized photometric redshift distributions are shown in Figure~\ref{fig:phz_dist} for both regioggns. Our magnitude cut leads to a mean  $z_{phot} \sim0.63$
in both cases and very similar distributions. They both suffer from a counts drop around $z\sim$0.4 due to the lack of the $u$-band. This can also be seen in the left panel of Figure~\ref{fig:zpstats} where photometric redshifts are compared to spectroscopic redshifts for a validation sample of 50,476 galaxies built during DNF processing. There is an excellent correlation between spectroscopic and photometric redshifts; however, galaxies around $z_{spec}\sim$0.3 show a very large scatter, especially towards higher values of $z_{phot}$. 
This can also be seen in the right panel of Figure~\ref{fig:zpstats},
where we assessed the global quality of the photometric redshifts by characterizing the average bias and standard deviation of $(z_{ph}-z_{sp})/(1+z_{sp})$.
These points will be examined in detail in sections~\ref{sec:results} and \ref{sec:discussion}.

\begin{table}[h]
\centering
\begin{tabular}{lrccl}
\hline
Region & Galaxies & Area & Density & Mean  \\
  &  & (deg$^2$)    & (Gal./arcmin$^2$) & \phz \\
\hline\hline
Y1-S82 & \VacSetGals & \VacSetArea & \VacSetDens & \VacSetZmean \\
Y1-SPT & \VacSptGals & \VacSptArea & \VacSptDens & \VacSptZmean \\
\hline
\end{tabular}
\caption{VACs properties for both Y1-S82 and Y1-SPT.}
\label{tab:vacproperties}
\end{table}

	\section{The \wazp\ cluster finder algorithm}
    \label{sec:WAZP}

\begin{figure}
    \includegraphics[scale=0.60]{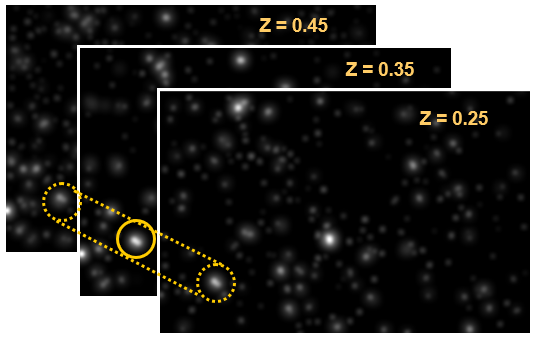}
    \caption{A schematic view of the \wazp\ algorithm. Here are shown the wavelet filtered galaxy density maps corresponding to three consecutive photometric redshift slices. It shows how a cluster can propagate along several slices.
    }
    \label{fig:wazp-scheme}
\end{figure}

The Wavelet Z-Photometric (\wazp) cluster finder is designed to detect galaxy clusters from multi-wavelength optical imaging galaxy surveys. It searches for projected galaxy overdensities in photometric redshift space without any assumption on the red sequence.
In a nutshell, \wazp\ first slices the galaxy catalog in photometric redshift space, and then generates smooth wavelet-based density maps for each slice where peaks are extracted (see Figure~\ref{fig:wazp-scheme}).
These overdensity peaks are then merged to create a unique list of clusters and associated galaxy members.
Hereafter, these various steps are described in detail.

\begin{enumerate}

\item {\it Slicing in photometric redshifts}.
By photometric redshift slices, we mean here the photometric redshift support over which individual galaxy redshift PDF's are integrated around a given redshift of interest. Therefore, at a given considered redshift, galaxies are weighted by that quantity.  These weights are used to build density maps at different redshifts or estimate richnesses as described in the next steps.

The adopted strategy to define photometric redshift slices is based on the statistical comparison of the "best-estimate" discrete photometric redshifts (taken here as the mean of the galaxy redshift PDF) and corresponding spectroscopic redshifts if available. 
Based on available spectroscopic samples, the mean bias and scatter of photometric redshifts ($z_{ph}$) relative to spectroscopic redshifts ($z_{sp}$) were derived.
Following the standard way to evaluate the performance of photometric redshifts \citep[e.g.,][]{Ilb06},
we computed statistics of $(z_{ph}-z_{sp})/(1+z_{sp})$ as a function of both $z_{sp}$ and $z_{ph}$. The location and width of a slice is then built in such a way that it includes 95\% of the galaxies of a given spectroscopic redshift. The separation between two slices corresponds to a fourth of their width assuring a sufficient overlap to avoid missing clusters being between two consecutive slices.

\item {\it Generation of galaxy number density maps.}
\wazp\ does not consider photometric redshifts as discrete values. Instead, it operates with redshift PDF's when provided, or generates them from the errors provided by the chosen photometric redshift algorithm. In each one of the slices defined above, galaxies are weighted by the integral of their redshift PDF over that slice. The resulting weighted RA-Dec distribution is then pixelized on a grid with a step of physical size 1/16$^{\rm th}$ of a Mpc. This image is finally filtered using the wavelet task {\texttt MR\_FILTER} from the multi-resolution
package MR/1 \citep{Sta98}. This task incorporates a statistically rigorous treatment of the Poisson noise, which allows us to keep significant structures in the desired scale range. Here we
select structures with scales in the range $0.5 - 2$~Mpc, typical of cluster scales, and apply a 3$\sigma$ iterative multi-resolution thresholding with a B-spline wavelet transform.

\item {\it Extraction of peaks.}
The smooth density maps obtained in the previous step are segmented, and in each object domain, one or more peaks are extracted. In the case of several peaks in one domain, depending on the distance between a peak and the closest saddle point, the peak can be merged or preserve its identity. Pixels of a domain are then distributed to peaks by proximity.

\item {\it Assessing peak significance}
The peak significance is chosen to be computed in a radius of $R_S=300$~kpc, a radius that encloses typical cluster cores \citep{adami98}. To perform background statistics, the survey is pixelized with pixel areas equal to $\pi R_S ^2$. Any pixel intersecting a bad region or an edge is removed.  Standard counts in cells are then applied to estimate the mean density ($N_{bkg-global}$) and standard deviation ($\sigma_{bkg}$). The significance, defined as $SNR = (N - N_{bkg-global})/\sigma_{bkg}$, where $N$ is the total density of galaxies in a cylinder centered at the peak position, with a length that is the width of the redshift slice and an angular radius $R_S$.

\item {\it Peak merging along the redshift direction.} As $z_{ph}$ slices overlap, one can expect clusters to be detected in several consecutive slices.  To build the final list of clusters, peaks of consecutive slices are associated, and only the slice in which the system has maximum significance is kept. Note that two clusters can be deblended along the line of sight if their distance in redshift is larger than 2$\times 3 \sigma_{dz/(1+z)}$ where $\sigma_{dz/(1+z)}$ denotes here the 68$^{\rm th}$ percentile of the $dz/(1+z)=(z_{ph}-z_{sp})/(1+z_{sp})$ distribution.

\item {\it Centering and cluster redshift.} The cluster center is defined as the location of the density map peak. However, if the brightest cluster member is found within the first neighbouring pixels, then this galaxy marks the center. This leads to a maximum shift of ~100~kpc from the peak location. Concerning the redshift, an initial value is derived as the mode of the sum of the galaxy redshift PDF's within a 0.5~Mpc radius around the cluster center. This value is refined iteratively based on the membership probabilities described below.

\item {\it Assignment of membership probabilities.} Membership probabilities ($P_{mem}$) are computed following the prescription given in \citet[][]{2016A&A...595A.111C}. In a nutshell, galaxies of the cluster field are piled up in a 3-dimensional grid (cluster-centric distance, magnitude, photometric redshift) where magnitudes and redshifts are included as probability distribution functions. The same is done for local background galaxies in (magnitude, photometric redshift) space. The local background galaxies are selected in a ring from 3 to 6~Mpc to the cluster center, whereas cluster field galaxies are selected within a 3~Mpc disk. The membership probability is the combination of the probability to be at the cluster redshift and the probability not to be a background galaxy. The final membership probability at a given cluster-centric distance, magnitude, and redshift is derived from the density ratio between the cluster field and the background field. Note that, as in \citet[][]{2016A&A...595A.111C}, no parametric modelling is used for the radial density, nor for the luminosity function.

\item {\it Richness and radius}
The cluster richness and radius are estimated jointly. The richness is the sum of the membership probabilities within a radius that corresponds to an overdensity of 200 times the mean galaxy background number density (similar to \citealt{hansen2005}). This is done considering galaxies, both in the field and in the cluster, down to a given fraction of $L^*$ luminosity.
Practically, galaxies brighter than ${\rm m}^{\star}(z_{cluster}) + \delta {\rm mag}$ are counted, where ${\rm m}^{\star}$ is the characteristic magnitude marking the knee of the luminosity function and $\delta {\rm mag}$ is a fixed quantity, chosen here to be 1.5. The adopted definition allows to produce "redshift independent richnesses", in the sense that the same cluster seen at two different redshifts would have the same richness. The evolution of the characteristic luminosity of the Luminosity Function can be described by a passively evolving population formed in a single burst \citep[e.g.,][]{2006ApJ...650L..99L}. In this study, we derive ${\rm m}^{\star}(z)$ from the passive evolution of a burst galaxy with a formation redshift $z_{form}=3$ taken from the PEGASE2 library \citep[\texttt{burst\_sc86\_zo.sed},][]{Fio97}. It is calibrated using the value of ${\rm K}^{\star}$($z=0.25$) derived by \citet{2006ApJ...650L..99L} from an observed cluster sample. The choice of $\delta {\rm mag}$ is critical as it sets the redshift limit (${ z}_{lim}$) of the final cluster sample through the relation ${\rm m}^{\star}({z}_{cluster}) + \delta {\rm mag} = {\rm mag}_{lim}$, where ${\rm mag}_{lim}$ is the survey apparent magnitude limit.

\end{enumerate}

	\section{Application to DES-Y1 survey}
    \label{sec:results}

\subsection{Running \wazp\ cluster finder}
\newcommand{\istar}{{\rm m_i}^{\star}}

As described in section~\ref{sec:DESdata}, the DES-Y1 survey is split into two regions ("SPT" and "S82"), for which two galaxy VACs are produced to feed the \wazp\ pipeline. These catalogs are built based on the $i$-band, chosen here as a reference band both for star-galaxy separation, and for defining apparent magnitude cuts.
 Given this selection, cluster detection was performed with the same setting independently from the position on the sky, assuming a sufficient homogeneity over the whole survey. This is an approximation as we have seen above that in some regions magnitude completeness limit can be lower by as much as 1~mag.

The redshift limit of the constructed \wazp\ sample is constrained by the depth of the survey reference band. It also depends on the adopted definition of the richness estimate and in particular, the adopted magnitude limit used to count galaxies entering the richness.
We assume here that the same cluster, seen at two different redshifts, would get the same richness by counting its galaxies down to an apparent magnitude $\istar(z)+\delta_{mag}$, where $\delta_{mag}$ is a fixed quantity and $\istar(z)$ is defined in section~\ref{sec:WAZP}. This can be achieved as long as this quantity remains lower than the $\magI$-band depth of the survey. In the present case, richnesses are chosen to be computed including galaxies down to $\istar+1.5$.

Based on the above considerations, given that some regions are not deeper than $\magI=22.27$, there is an upper redshift limit, $z=\WazpZmaxRich$, above which richnesses start to become incomplete depending on the survey location. At the limiting magnitude of the galaxy VAC, $\magI=23$, which corresponds to a redshift limit  $z=\WazpZmaxDetec$, richnesses are complete within only 2\% of the survey area. Based on the 10$\sigma$ $i$-band survey depth map (section \ref{sec:DESdata}), we have derived a map indicating our local cluster $z_{max}$ at each position of the survey, that is reported in the \wazp\ cluster catalog.
Detection is performed to slightly larger redshifts ($\sim 0.9 $), but for clusters that would be detected beyond their local $z_{max}$, their galaxy luminosity function is not sampled homogeneously across redshifts and therefore richnesses for these clusters would require some correcting factor. In this paper, we are not introducing such a correction, and therefore richnesses are consistent over the whole survey only up to $z=\WazpZmaxRich$.
As a lower redshift limit for cluster detection, we adopted in this paper the value $z_{min}=0.05$.

Besides the considerations on redshift limits above, we also need to assess the global quality of our photometric redshifts and determine the photometric redshift slicing strategy
on running \wazp. As can be seen in the right panel of  Figure~\ref{fig:zpstats}, the average bias remains relatively modest and the scatter roughly constant in the adopted redshift range. These properties should not prevent cluster detection in general. Note, however, that this point will be discussed in more details in section~\ref{sec:discussion}.

Operationally, the \wazp\ cluster finder runs on small sections of the sky. The LIneA Science Portal manages data tiling, launches the code on each tile on parallel cores and concatenates the final catalogs, both clusters and galaxy members. The data tiling consists in dividing the survey in overlapping rectangular tiles of typical area 20~deg$^2$. The overlaps are set to assure a tiling independent cluster detection for clusters with redshifts larger than 0.05 that would fall at the intersection of two tiles.

\subsection{The \wazp\ cluster catalog}

\begin{figure}
	\includegraphics[scale=1]{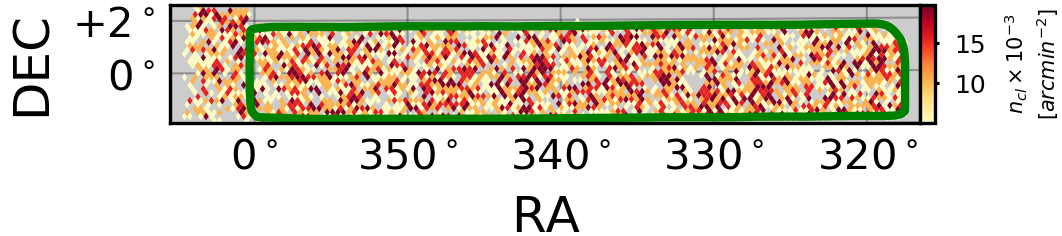}
	\includegraphics[scale=1]{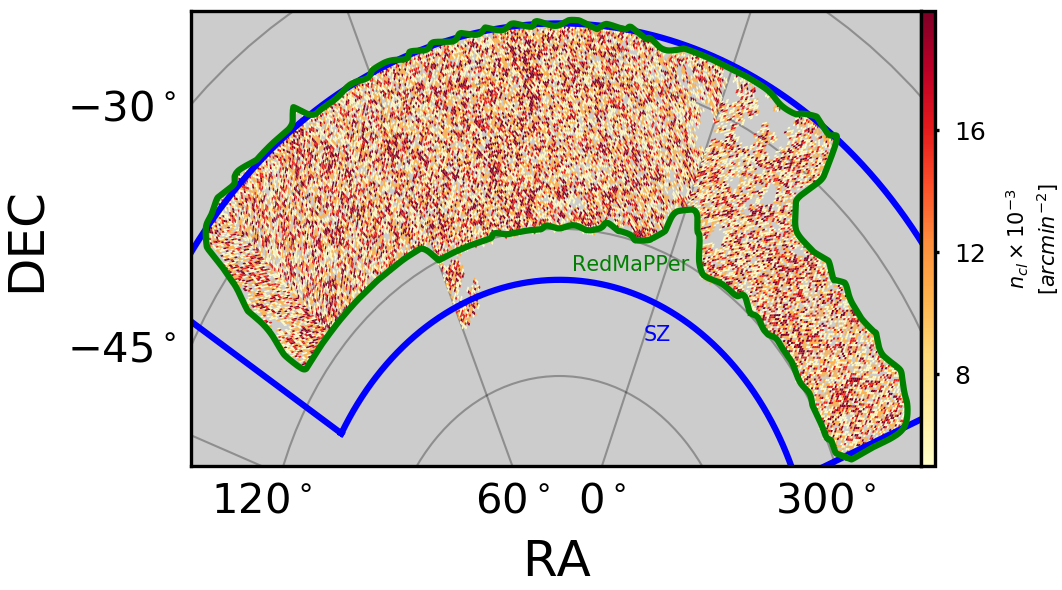}
	\caption{Projected density of \wazp\ clusters for both Y1-S82(top) and Y1-SPT(bottom) regions of DES first year data release.
	The green line is the contour of the \RM\ Y1 catalog and the blue line is the region of detection for the SPT survey (\citealt{Ble15}).
	}
	\label{fig:wazp_radec}
\end{figure}

The \wazp\ pipeline was run on the DES Y1-S82 and Y1-SPT regions defined
above.
This results in the detection of cluster candidates, not confirmed galaxy clusters.
However, these candidates will be referred to as "{\wazp} clusters" throughout the paper for simplicity.
For the combined sample, it led to the detection of \WazpClusters\
clusters in the redshift range $z=0.05$ to $z=0.91$ with richness
$N_{gals}\geq5$, corresponding to densities of \WazpSetDens\ and \WazpSptDens\ clusters/deg$^2$ for Y1-S82 and Y1-SPT respectively .
If we restrict to a sample with more reliable redshifts and complete richenesses i. e. $0.1 \le z\le \WazpZmaxRich$, 
we find \WazpClustersRMzlim\ clusters,
with a higher consistency in cluster densities with \WazpSetDensRMzlim\ and \WazpSptDensRMzlim\ clusters/deg$^2$ for the two DES regions. 
This result supports strongly the high homogeneity over the sky of the galaxy VAC construction, including photometric redshift computation, as well as the subsequent cluster detection. This can also be seen in Figure~\ref{fig:wazp_radec} where the projected distribution of detected clusters on the sky is shown. A description of the \wazp\ cluster catalog is provided on LIneA's website\wazpweblink.

\begin{figure*}
	\includegraphics[scale=1]{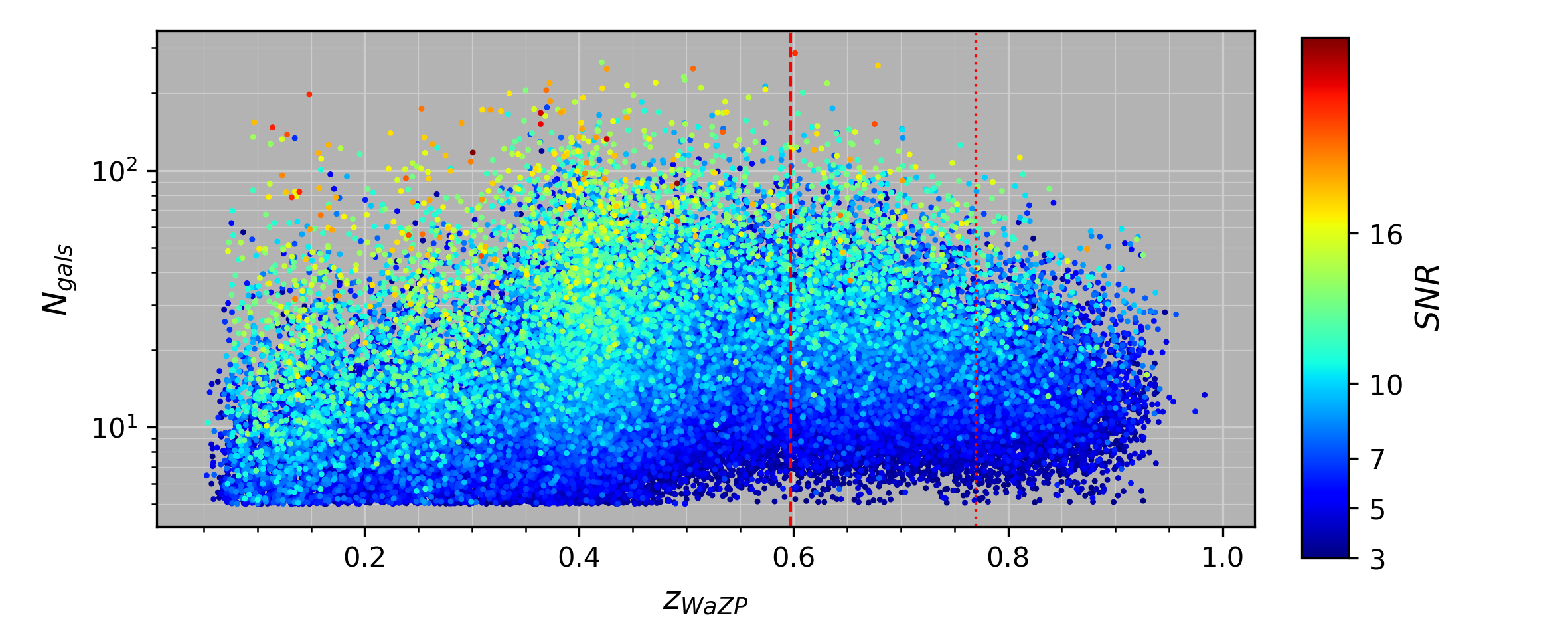}
	\caption{Richness as a function of redshift for \wazp\ clusters color coded by signal-to-noise ratio. The dashed line indicates the redshift (\WazpZmaxRich) above which cluster richnesses start to become incomplete and the dotted line marks the redshift (\WazpZmaxDetec) above which all clusters have incomplete richnesses. }
	\label{fig:wazp_zr}
\end{figure*}

The ranges of richnesses and redshifts covered by \wazp\ clusters are shown in Figure~\ref{fig:wazp_zr}. From the color coded SNR we can see that for a given richness, as expected, the SNR decreases with redshift. This is mainly due to the increasing scatter in the photometric redshifts leading to an increase of the mean background density of galaxies.
We see that above redshift $\WazpZmaxRich$, the number of rich clusters start to diminish rapidly,
and above $\WazpZmaxDetec$ there is only one cluster with richness greater than $100$.

\begin{figure}
	\includegraphics[scale=1]{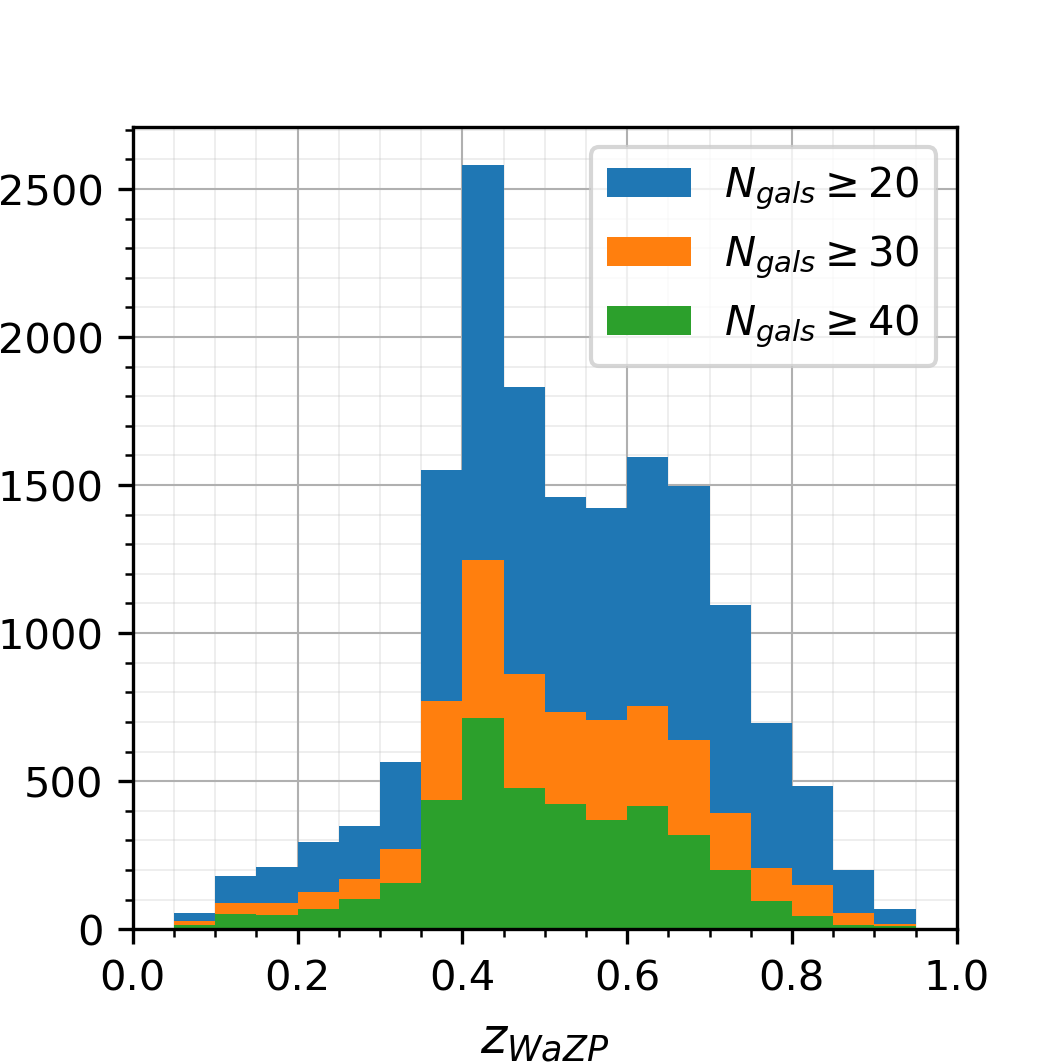}
	\caption{The counts distribution of \wazp\ clusters detected in the combined Y1-S82 and Y1-SPT DES regions as a function of cluster redshift for three different richness cuts.      }
	\label{fig:nzcl}
\end{figure}

The redshift distribution of \wazp\ clusters is shown in Figure~\ref{fig:nzcl}. The global bell shape of the counts looks as expected except for a sharp concentration of clusters at $z\sim 0.45$, similar to that observed on the galaxy photometric redshift distribution of Figure~\ref{fig:phz_dist}. This peak becomes more prominent for lower richness systems. In the next sections we investigate further the nature of this peak.

\begin{figure}
	\includegraphics[scale=1]{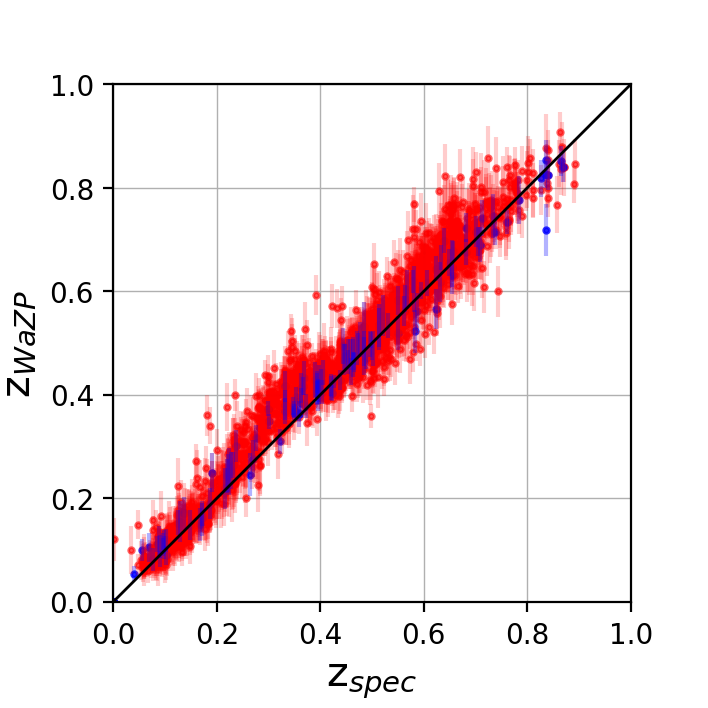}
	\caption{Comparison of cluster redshifts as derived by \wazp\ with estimated spectroscopic redshifts. \wazp\ cluster members are cross matched to all publicly available spectroscopic redshifts falling in the DES footprint. A cluster spectroscopic redshift is derived each time at least 5 concordant redshifts (within $\pm 2000$ km/s) are found within 0.5~Mpc to the cluster center (blue points, 131 clusters), or if the \wazp\ central galaxy has a spectroscopic redshift (red points, 1859 clusters).      }
	\label{fig:zcl}
\end{figure}

There are several ways to estimate the quality of the \wazp\ redshifts ($z_{\wazp}$). They can be compared to known cluster redshifts as shown in the next section, or, as done here, cluster members can be cross-matched with available spectroscopic galaxy samples.  The adopted procedure here is to search for all galaxies with spectroscopic redshifts within 0.5~Mpc around each detected cluster and likely to be cluster members. We considered that a cluster could be associated with a spectroscopic redshift if at least 5 galaxies were found within a range of $\pm 2000$ km/s. The selected velocity window is the one that maximizes the number of spectroscopic galaxies. The cluster spectroscopic redshift is then defined as the median of the redshifts in that window. We also associated a redshift in the case the central \wazp\ cluster galaxy has a spectroscopic redshift. This may lead to a few outliers but increases statistics by a factor of 10. Based on public spectroscopic surveys, 131 \wazp\ clusters covering the redshift range $z=0.05$ to $z=0.9$ could be associated a spectroscopic redshift with at least 5 concordant redshifts, and 1,859 clusters could be associated a spectroscopic redshift based on their central galaxy. In Figure~\ref{fig:zcl}, the comparison with \wazp\ redshifts is shown. Both spectroscopic redshift assignments led to the same statistical differences with \wazp\ redshifts: an average bias of $\sim $0.014 and scatter of $\sim $0.026.

\begin{figure}
	\includegraphics[scale=1]{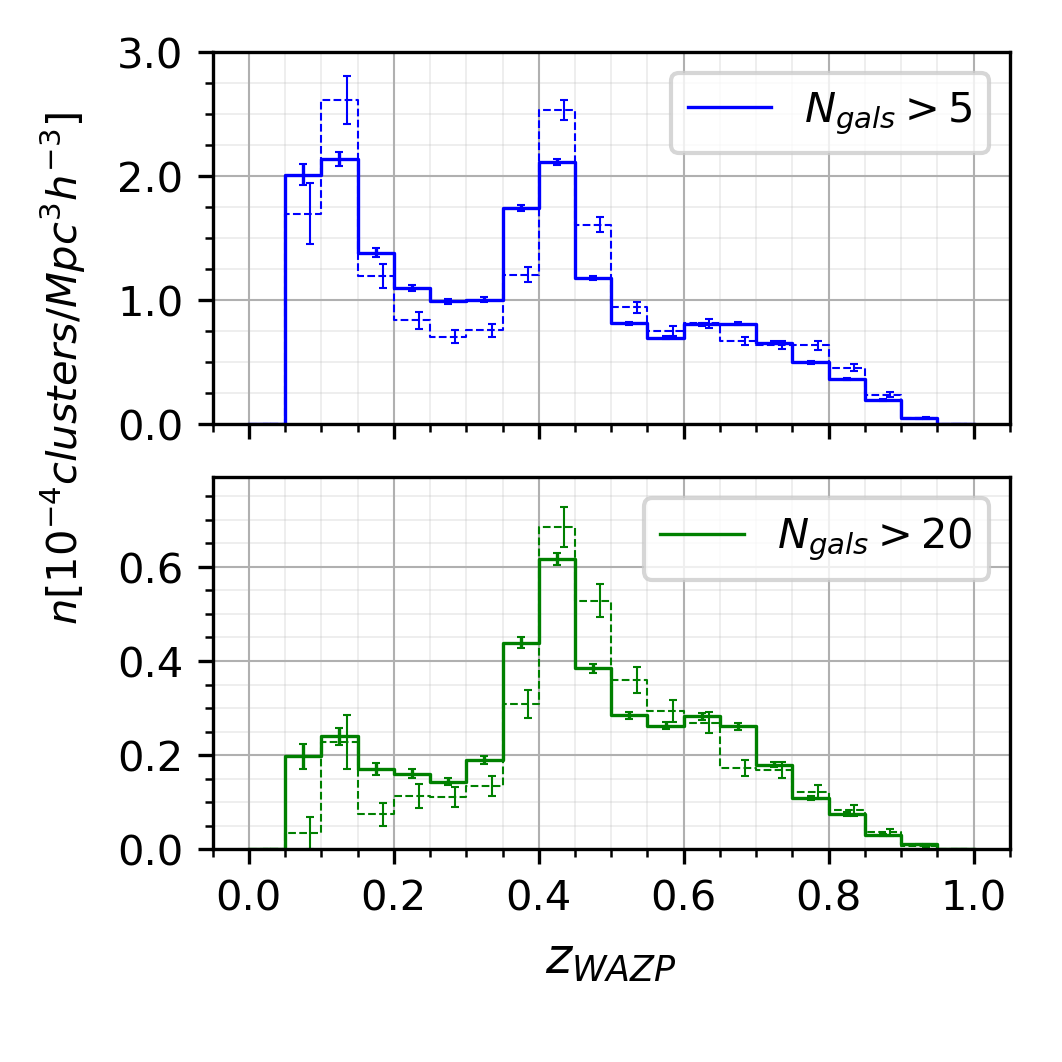}
	\caption{Volume density of \wazp\ clusters as a function of redshift in Y1-SPT (solid line) and Y1-S82 (dashed line) regions with different richness thresholds.
	The errorbars correspond to Poisson noise.}
	\label{fig:wazp_vol}
\end{figure}

Figure~\ref{fig:wazp_vol} shows the evolution of the volume density of clusters with redshift, with Poisson errorbars.
We can see that the density for both Y1 regions agree with each other for both richness cuts.

	\section{Comparison to other cluster catalogs}
	\label{sec:matches}

In this section, we compare the \wazp\ Y1A1 clusters identified in the previous section to those derived by other methods covering the same region. A first comparison is made with clusters detected by the South Pole Telescope (SPT) survey via the Sunyaev-Zel'dovich (SZ) effect \citep{Ble15}. A second comparison is done with clusters detected by the \RM\ optical cluster finder, based on the same photometric data but using an algorithm
searching for overdensities of red sequence galaxies \citep{Ryk14}.
These comparisons are based on matching clusters from two different catalogs.
This is a complex operation that has led to a variety of proposed algorithms \citep[e.g.,][]{Gerke2005, Knobel2009, Cucciati2010, Gerke2012, Adam2019}. We point out that adopting one algorithm or another or using different configurations of the same may result in different cluster associations. In particular, when a given cluster can potentially be connected to several counterparts. As detailed below, we have experimented with several of these approaches and finally adopted a hybrid iterative procedure to optimally solve multiple matches.
This appeared to be the only way to avoid having systems left unmatched due to a wrong matching of their obvious counterpart. Several such cases appeared in particular with interacting clusters for which richness rankings were reversed. The resulting pairing seems, from intensive visual inspection, optimal for addressing statistically the different properties of the commonly detected clusters (centering, redshift), and evaluating systems without any counterparts. In particular, our matching procedure allowed us to decrease the number of incorrect matches of rich clusters on each side significantly.

In this paper, we use a cylindrical matching where we require the angular distance of cluster centers to be smaller than some defined length (be it their respective radii or a fixed physical distance), and their redshift separation to be constrained by the typical redshift errors from both samples.
 In carrying out this comparison, some issues have to be considered. First, the cluster radius definition for each cluster sample may be different. Second, we must define the redshift window to be used. It should be large enough to take into consideration the errors in photometric redshift assigned to  clusters in both samples. However, a large window may lead to an increased number of multiple matches that need to be resolved as discussed in more detail below. Finally, it is crucial to ensure that the projected area of overlap of the samples is properly taken into account. To do that, footprints of the samples are used to flag clusters falling outside the overlapping regions or near their edges. This flag is useful when unmatched clusters are near the edges, in which case they can be removed from the matching statistics.

As mentioned above, we compare the combined Y1-S82 and Y1-SPT clusters identified by \wazp\ with those in the \SZ\ and \RM\ samples. In the first case, we take the \SZ\ sample as reference, treating the \SZ\ clusters as true representatives of the underlying mass distribution and test how well these systems are recovered.
In the \RM\ case, we investigate the unmatched cases in both directions to understand the specificities or the possible limitations of each algorithm.

	\subsection{\wazp\ versus SPT clusters}
	\label{sec:spt-clus}
\input{files/wazp_mt_sz_info.tex}

The \SZ\ sample \citep{Ble15} covers an area of 2,500~deg$^2$ (seen in bottom panel if Figure~\ref{fig:nzcl}),
within which 516 clusters (out of 677 candidates) were detected with signal-to-noise above 4.5.
In \citealt{Ble15}, it is stated that the catalog is highly complete for $M_{500^c}\ge7\times10^{14}M_\odot h^-1$ and $ z\ge0.25$. It is also mentioned that there were a number of optical followups to confirm these clusters.
Therefore, this catalog will be utilized to validate the detectability of \wazp\ regarding massive clusters.
We only use the
\SZinDES\ \SZ\
clusters that have information on mass and redshift, and are located within the overlap with DES (external envelope of the DES Y1-SPT region). We should stress that the redshifts assigned to \SZ\ clusters from \cite{BOCQUET} are both
spectroscopic (106) and photometric (225).

 We considered a one-way match,
 taking \SZ\ clusters as reference
 and we looked for \wazp\ clusters falling within SPT clusters radii.
 The adopted radius is $R_{200}$, the radius where the average cluster overdensity is 200 times the critical density, i.e., $\Delta = 3M_{200}/4\pi R_{200}^3\bar{\rho}_{crit}(z)=200$. It was computed from the available values of $M_{200}$ and converted to an angular radius $\theta(R_{200})$ using \cite{Ade15} cosmology. Here, $\theta_{200}=R_{200}/D_A(z)$, where $D_A(z)$ is the comoving angular diameter distance to the cluster redshift $z$.
For a match to happen, the redshift separation also has to fall within the interval defined by the combined redshift errors. This interval was defined as $3\sigma$ where $\sigma$ is the sum of redshift errors provided by the two catalogs. The resulting matching cylinder is quite large in order to account for possible centering offsets between the two wavelength domains and large redshift discrepancies. The statistical comparison, {\it a posteriori}, of the differences in centering and redshifts of the matched systems, allows us to evaluate the adopted matching criteria.

\begin{figure}
	\includegraphics[scale=1]{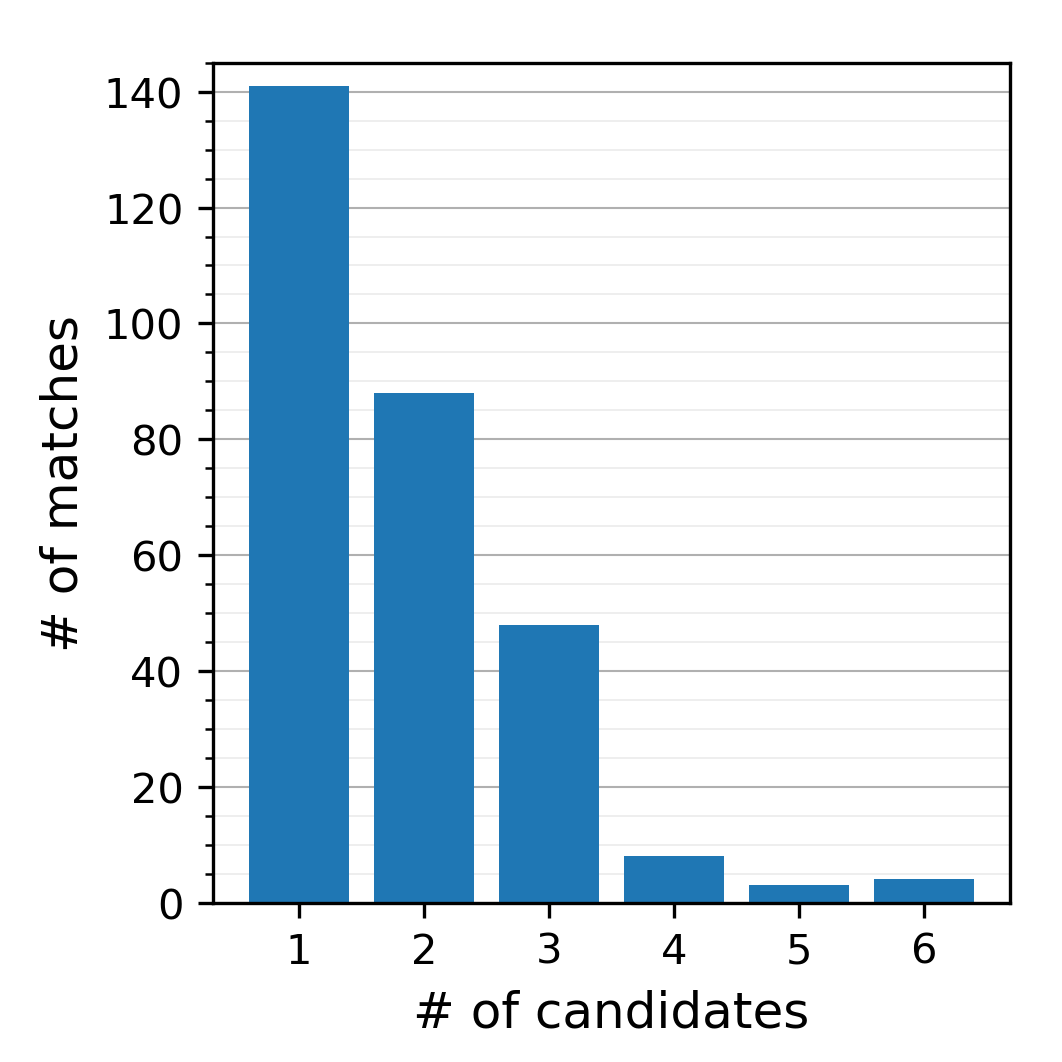}
	\caption{Number of \wazp\ clusters found to match individual \SZ\ clusters.}
	\label{fig:SZ_multi}
\end{figure}

Applying this method to the \SZ\ and \wazp\ samples, we find that \SZmtMMT\ \SZ\ clusters (out of \SZinDES) have at least one \wazp\ counterpart.
Among these, \SZmtOneCand\ have only one candidate for matching,
while the rest have a multiplicity function as shown in Figure~\ref{fig:SZ_multi}. The large fraction of multiple matches is not surprising considering the very different selection functions of the two samples, their relative densities and the adopted matching criteria.
The multiple \wazp\ matches were resolved by choosing the richest associated counterpart.

\newcommand{\unmatchedIII}{\texttt{SPT-CLJ2218-5532}}

Out of the \SZunmatched\ unmatched \SZ\ clusters,
\SZunmatchedEdge\ are located near the \wazp\ footprint edges
and \SZunmatchedZmax\ have redshifts beyond $z=\WazpZmaxDetec$, where \wazp\ cluster finder reaches its expected limit of completeness for DES-Y1 (\SZunmatchedZmaxII\ of those have $z>1.1$, and are completely beyond \wazp\ reach).
There was one unmatched \SZ\ cluster (\unmatchedIII) with $z=0.77$, just above the redshift limit for \wazp\ clusters with complete richnesses.
However, the local $z_{max}$ at this cluster position is $0.71$ and, upon visual inspection, we found no clear visible optical counterpart.

\begin{figure*}
	\includegraphics[scale=.98]{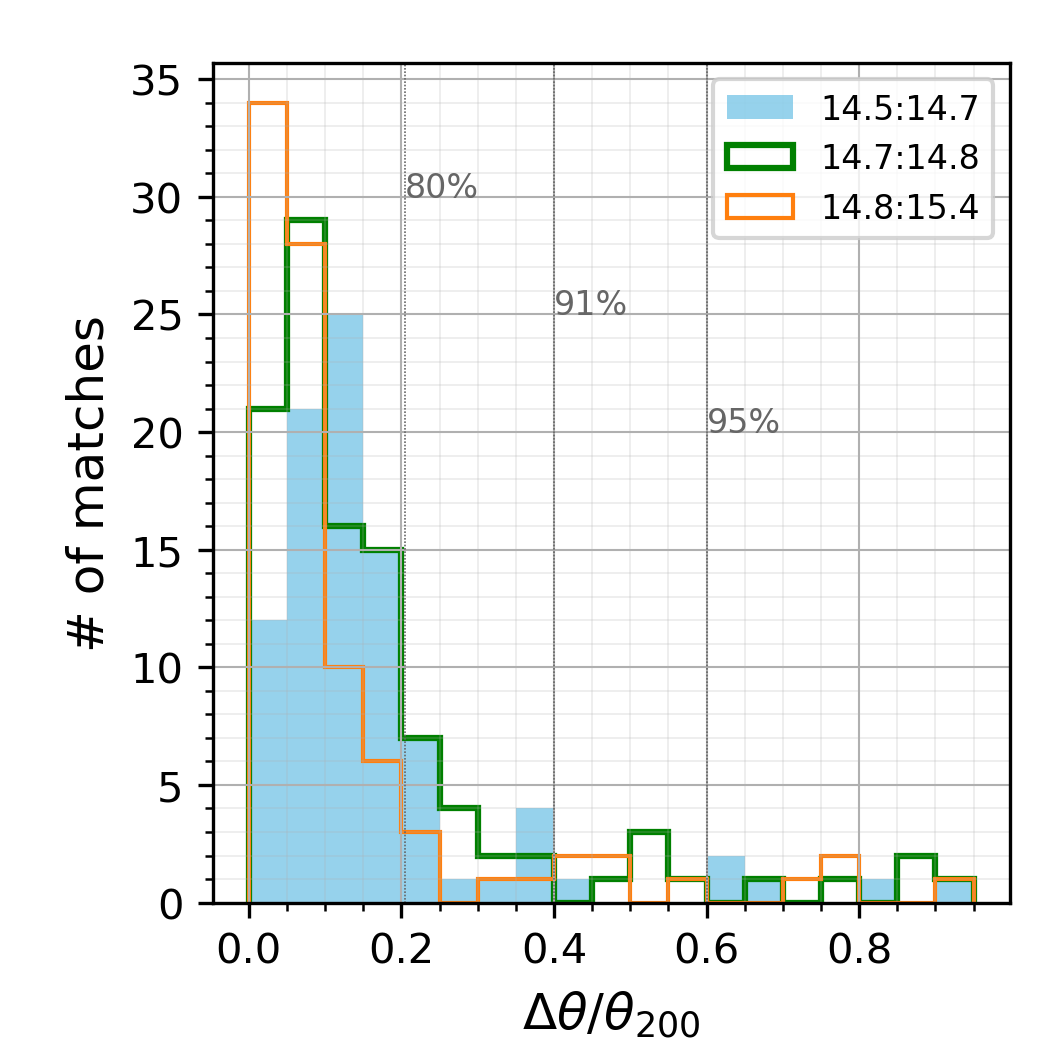}
	\includegraphics[scale=.98]{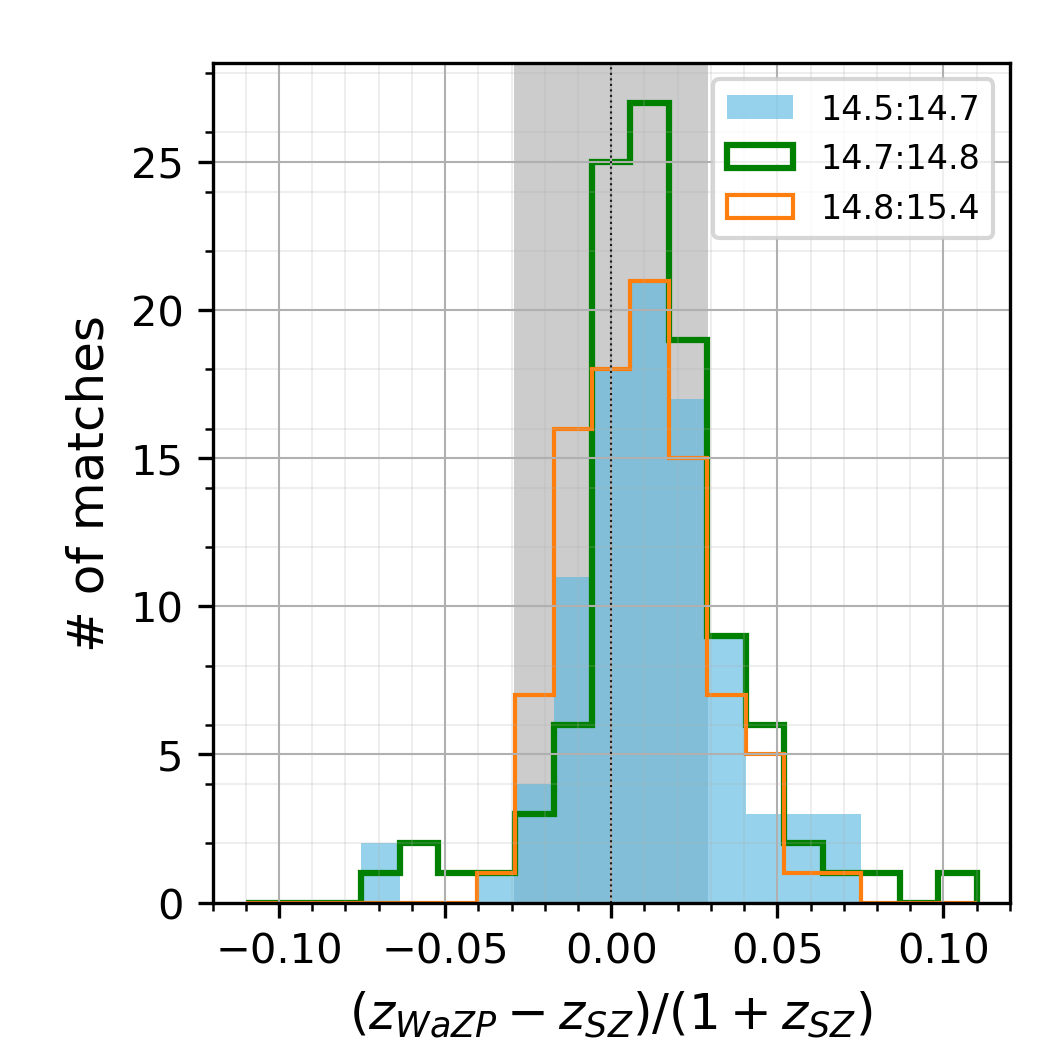}
	\caption{
    Distribution of angular distances (left panel) and redshift separations (right panel)
    of \wazp-\SZ\ matched clusters binned by $\log(M_{200}[\Munit])$. The gray shaded region on the right plot is the combined average photometric redshifts uncertainties of the matched clusters from both catalogs divided by $1+\zsz$.}
	\label{fig:SPTmatch_dAdz}
\end{figure*}

In Figure~\ref{fig:SPTmatch_dAdz}, we show the characteristics of the matching for three mass bins, both in terms of angular separation (left panel) and redshift separation (right panel). 
The average distance of \wazp-\SZ\ centers is $\langle\Delta\theta\rangle=0.16\,R_{200}$, with 80\% of clusters within $0.2\,R_{200}$ and 95\% within $0.8\,R_{200}$.
If we consider the different mass bins in the figure,
there is a small systematic improvement on $\langle\Delta\theta\rangle$ for higher masses ($0.177$, $0.178$ and $0.134$ respectively),
even though there is only $\approx 100$ clusters per mass bin.
We also see that there is a reasonable agreement in redshift,
with \SZmtPercWithinScat\% of matches within the average redshift uncertainties of the clusters $\langle z^{err}\rangle = \sqrt{ \langle z^{err}_{\wazp}\rangle^2 + \langle z^{err}_{SZ}\rangle^2 }$ (gray shaded region).
There is also a slight improvement on redshift scatter and bias was we look at higher mass bins.

\begin{figure*}
	\includegraphics[scale=.98]{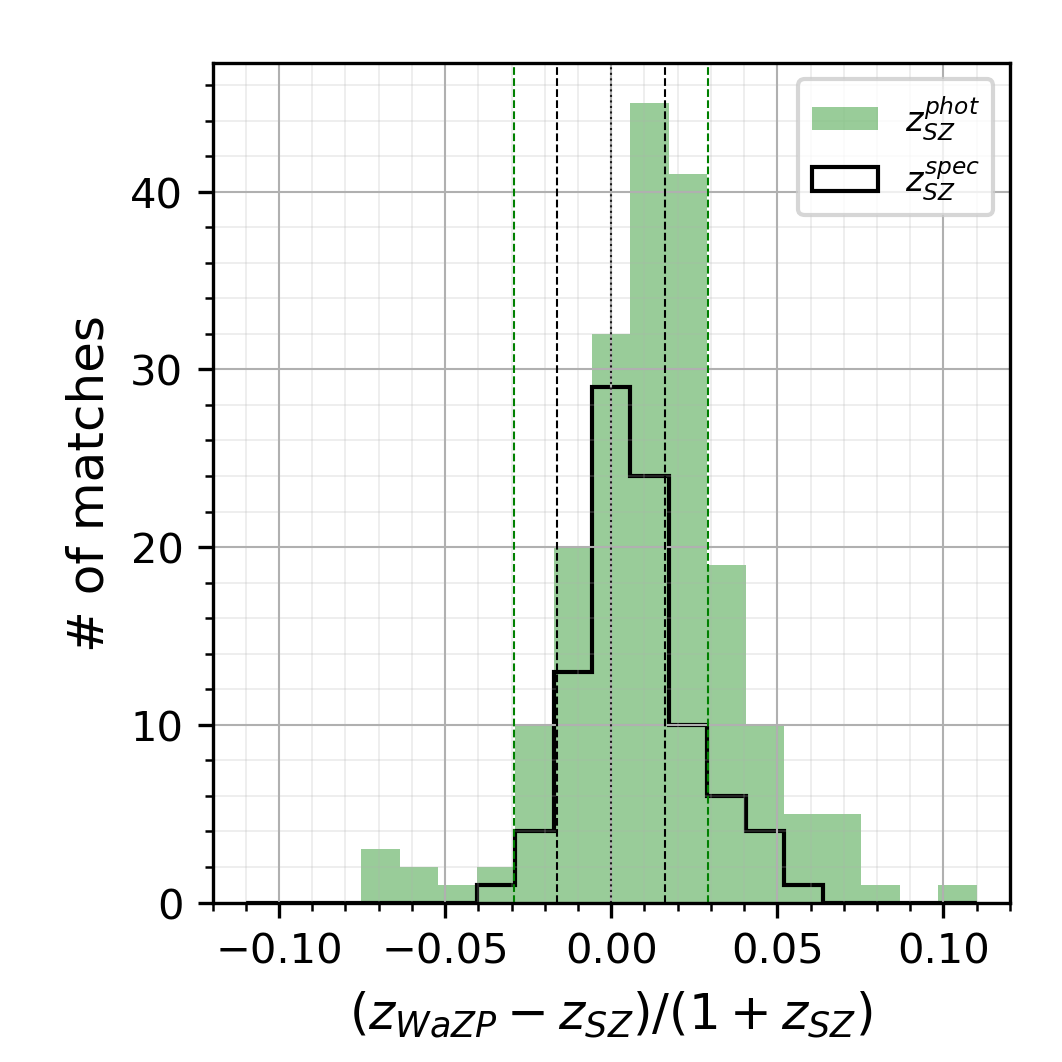}
	\includegraphics[scale=.98]{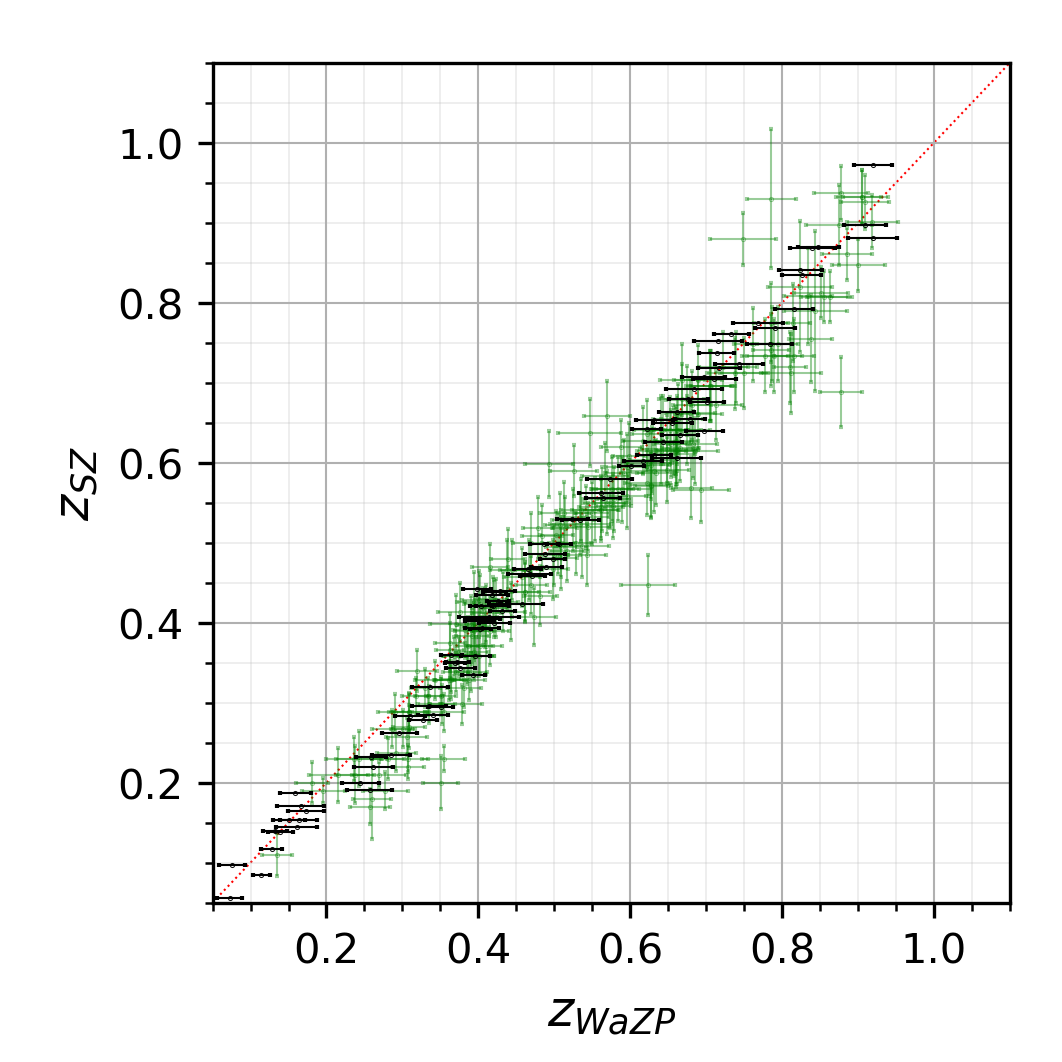}
	\caption{
    Distribution of redshift separations
    (left) and relation between the redshifts (right) of \wazp-\SZ\ matched clusters split into spectroscopic and  photometric redshift samples.}
	\label{fig:SPTmatch_dz_specphot}
\end{figure*}

We also compare the photometric estimates of \wazp\ redshifts with those assigned to \SZ\ clusters.
As 93 of the matched \SZ\ clusters have been assigned a spectroscopic redshift \citep{BOCQUET}, we can assess the accuracy of the estimated \wazp\ redshifts. Figure~\ref{fig:SPTmatch_dz_specphot} shows the distribution of redshift separations (left) and relation of redshifts (right), splitting the \SZ\ cluster redshifts into spectroscopic and photometric subsamples. As can be seen, \wazp\ redshifts show a good agreement with \SZ\ spectroscopic redshifts and some residual relative bias when compared with \SZ\ photometric redshifts. A quantitative description is provided in table~\ref{tab:szzprop}, which gives in column~(1) the type of sample; in column~(2) the number of clusters and in columns~(3) and (4) the bias and the scatter, defined as the mean and standard deviation of $(z_{\wazp}-z_{\SZ})/(1+z_{\SZ})$,
and column (5) is the combined redshift errors $\langle z^{err}\rangle/(1+\zsz)$.
When compared to photometric redshifts, we measure a relative bias of $\sim\zbiasszSZphot$ and a scatter $\sim \zscaterszSZphot$, a value similar to the combined redshift error \zerrrTotSZNORMphot. However, when comparing to \SZ\ clusters with spectroscopic redshifts, \wazp\ redshifts are almost unbiased (bias $= \zbiasszSZspec$), and show a significantly lower scatter ($\sigma = \zscaterszSZspec$, also compatible with the combined errors \zerrrTotSZNORMspec), corresponding to roughly half the average galaxy photometric redshift scatter. 
It is interesting to note that in all samples, the scatter was very close to the redshifts uncertainties,
even though the matching conditions only imposed a redshift difference of three times the errors.
From the right panel of Figure~\ref{fig:SPTmatch_dz_specphot}, it can also be seen that the moderate average bias is actually mainly due to low redshift clusters ($z\lesssim 0.4$).

\begin{table}[h]
\centering
\begin{tabular}{lrccc}
Sample
& N & bias & scatter & errors
\\
\hline\hline
All redshifts & \SZmtNUMtot  & \zbiasszSZ     & \zscaterszSZ     & \zerrrTotSZNORM     \\
Phot z only   & \SZmtNUMphot & \zbiasszSZphot & \zscaterszSZphot & \zerrrTotSZNORMphot \\
Spec z only   & \SZmtNUMspec & \zbiasszSZspec & \zscaterszSZspec & \zerrrTotSZNORMspec \\
\hline\hline
\\
\end{tabular}
\caption{Bias, scatter and combined redshift errors of \wazp-\SZ\ matched clusters for different subsamples as described in the text. }
\label{tab:szzprop}
\end{table}


\subsection{\wazp\ versus \redMaPPer}
	\label{sec:redmapper}

\input{files/wazp_mt_rm_info.tex}

In this paper we also match \wazp\ clusters with those detected by the \RM\ algorithm.
Here we use the \RM\ volume limited cluster catalog of DES-Y1 presented
in \citet{McC19} which consists of \RMClusters\ clusters with richness
$\lambda \ge $5 found in the Y1-SPT and Y1-S82 regions (seen in
Figure~\ref{fig:nzcl}) over the redshift range 0.1-0.95.
This volume limited catalog considers clusters for which the local
$z$-band depth assures a complete galaxy catalog down to the adopted
magnitude limit of the richness definition. The variable $z$-band depth
translates into a variable redshift limit ($z_{max}$) map that
characterizes the cluster sample.
When evaluating the recovery rates of clusters at high redshifts, we also
consider the {\it full} \RM\ cluster catalog over the same region,
defined by a constant $z_{max} = $ 0.95 and a richness threshold of 20.

\subsubsection{Differences in the detection algorithms}
\label{sec:redmapper_sub1}

\RM\ is an optical cluster finder based on the detection of spatial overdensities of red sequence galaxies \citep{Ryk16}. Although \wazp\ does not make any assumption relative to the cluster galaxy population when searching for galaxy overdensities, we do expect these two algorithms to yield similar samples up to redshifts $\sim$0.7, at least when considering the richest systems. However, a number of differences can be expected in the cluster characterization for several reasons.
First, as it was stressed above, galaxies used for searching for overdensities are not selected in the same way. \RM\ selects them based on colors whereas \wazp\ selects them based on redshifts. Second, the two algorithms differ in defining cluster centers. In the case of \RM\, centers are associated to a bright galaxy with some probability of being a central galaxy, whereas \wazp\ defines the center as a centroid. Note however, that, as described above, \wazp\ moves the center to the brightest cluster member position if its distance is less than 100~kpc, which happens here for 68\% of the \wazp\ clusters. Third, \RM\ redshifts are assigned based on an empirical modelling of red sequence colors, whereas \wazp\ assigns redshifts based on a concentration in photometric redshift space including all galaxy types at the cluster location.
Finally, we also expect differences on how each cluster finder performs in terms of deblending, or in terms of fragmentation and over-merging.

The above effects make the matching between the two samples non-trivial,
since the key elements to perform a proximity matching, like centering and redshift, can have distinct behavior, that may not lead to a unique solution.
Despite its complexity, it should be able
to provide us with a measure of the statistical consistency of the two catalogs.
It should also help us infer a lower limit for centering uncertainty,
as both cluster finders have optical centering estimations.
Finally, by carefully dealing with footprint coverage and edge effects, it should allow us to identify a number of missing systems and provide feedback on the respective selection functions, on possible ways to improve cluster detection algorithms and improve aspects of the construction of the underlying galaxy catalog.

\subsubsection{Matching procedure}
\label{sec:redmapper_sub2}

In contrast to what was done in the comparison between \wazp\ and SPT clusters, here, each cluster sample is considered as a reference to the other.
Therefore, not only we consider a one-way match using \RM\ as a reference (\RM-matched),
but we also analyse the case where \wazp\ (\wazp-matched) is the reference catalog. These one-way matches are used to investigate the fraction of missed detections. In addition to the one-way matches, in order to estimate differences in cluster properties (e.g., centering, redshift, richnesses), we also carry out a two-way (unique) match, for which it is required that both one-way matches point to the same cluster.

We recall that the cluster matching is performed within a specified redshift window. Following what is done in section~\ref{sec:spt-clus} when matching with \SZ\ clusters, this window was first defined considering the sum of the redshift errors provided for each cluster in the different samples.
However, while visually inspecting a sample of unmatched systems, it was noticed that for relatively low redshifts ($z\lesssim 0.4$), obvious pairs (i.e., sharing exactly the same center without any other overdensity on the line of sight) were not associated due to large redshift discrepancies. This fact is not surprising due to the systematic errors in photometric redshifts occurring at low redshifts.
In order to take this effect into account, we used an empirical approach to define the redshift window for matching. We first matched systems by angular center proximity only, without a redshift window, but imposing the center angular positions to be closer than 0.05~Mpc, computed at the largest redshift of the cluster pair.
The clusters were ranked by richness,
and when multiple candidates were found (about 10\% of the time for \wazp\ clusters and 24\% for \RM),
the richest candidate was selected.
As judged by eye inspection, with this criterion, most matches refer to the same system.

\begin{figure}
	\includegraphics[scale=1]{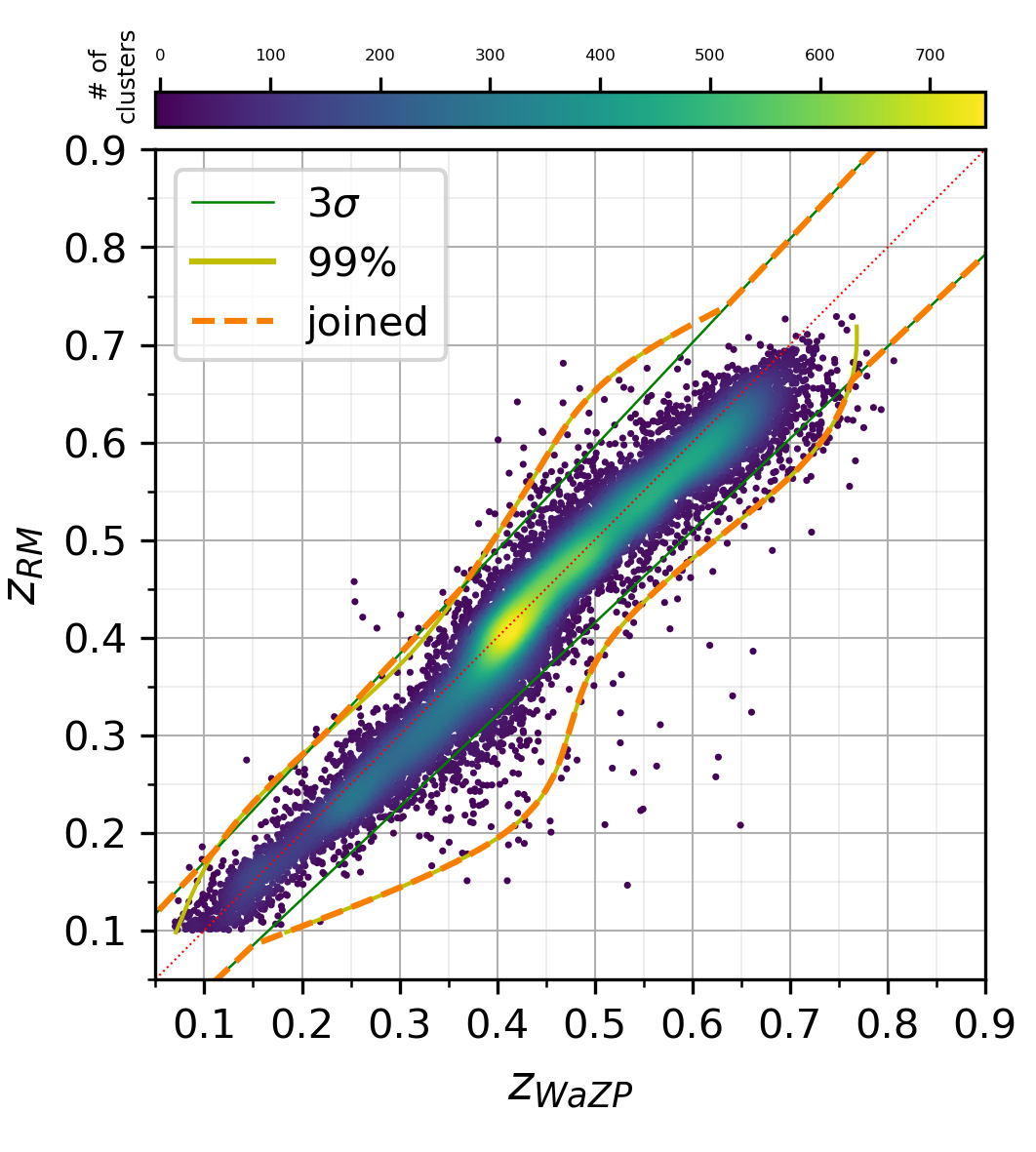}
	\caption{Redshift relation for \wazp-\RM\ matched clusters within
    $50$kpc. Yellow lines are the $99\%$ percentiles, the green lines are
    the $3\sigma$ levels, and the orange dashed lines are the union of
    the two, which will be used for the last step of matching.
    }
	\label{fig:RM_zXz_005mpc}
\end{figure}

Figure~\ref{fig:RM_zXz_005mpc} presents the resulting relation between redshifts for \wazp\ and \RM\ matched clusters.
As can be seen, this relation shows some deviations from a linear relation, in particular at \wazp\
redshift $\sim 0.4$, where \rm redshifts are distributed from 0.2 to 0.5, so a large scatter. This
issue is discussed in more details in the following section. Based on this plot, we defined a new
redshift window to carry out the matching, which corresponds to the union of the 99 percentile of the redshift differences using \RM\ as reference (as it is covering a smaller redshift baseline) and a $3\sigma(1+z)$ scatter.

Matching the catalogs by considering the resulting large redshift window combined to larger angular radii than in figure~\ref{fig:RM_zXz_005mpc} unavoidably leads to a large fraction of multiple associations. Resolving these multiples by selecting the richest available system on both sides resulted in many false matches. An emblematic case that appeared several times in our visual inspections, is the case of interacting clusters of similar richnesses. Both cluster finders would detect the two components but not necessarily with the same richness ranking. In that case the matching could lead to one mis-match and one unmatched cluster, or more mis-matches due to a cascade effect.
However, in the case of absence of an interacting system or if neighbouring systems are much poorer, the richness ranking is more adequate.

We found that for maximizing the number of correct associations, the best option is to go beyond a single matching rule.
Therefore, we decided to perform the matching following a several steps process where the most unambiguous pairs are matched first and then proceed to the rest of the list. The steps are detailed below.

It was determined empirically that a four-step process is optimal, where, at each step, the matching would only be performed on clusters not previously matched.
In the first step, we do not consider the clusters' redshift, and match all clusters that have the exact same centering (which happens when the two cluster finders are centered on the same galaxy).
By construction,
each one way match finds the same corresponding pairs,
therefore all cluster pairs found here will also be a match in the two way matching.
This led to a total of \mtRMstepI\ matched clusters.
In the second step, remaining clusters are matched within an angular distance of $300$~kpc (computed at the lowest redshift of the pair) from each other,
and a redshift difference less than $1\sigma_z$
(computed from Figure~\ref{fig:RM_zXz_005mpc}).
When more than one candidate is found, the richest one is considered to be the correct match.
This step also results in having the same number of matched clusters for both one way matches,
adding an extra \mtRMstepII\ matched clusters in each catalog.
The third step expands on the second one with a $3\sigma_z$ window,
leading to additional \mtRMstepIII\ matches in each catalog.
In the last step, we match the remaining clusters with the empirical redshift window
shown in Figure~\ref{fig:RM_zXz_005mpc} and use the radius provided by each cluster finder as a parameter for angular distance.
Here, the matching is not symmetric and results in another
\mtRMstepIVrm\ and \mtRMstepIVwz\ matched clusters for \RM\ and \wazp\ respectively.
Hence, we obtained a total of \mtRMmtRM\ one-way matches for \RM\ and and \mtRMmtWZ\ for \wazp.
We note that, in this four-step matching, if we do not remove matched clusters at each step and allow for multiple matches, we obtain \mtRMmmtRM\ \RM\ matched clusters and \mtRMmmtWZ\ \wazp\ matched clusters.

Finally, two-way matches are obtained when the two one-way matches point to each other.
This results in \mtRMcross\ \RM-\wazp\ cross-matched clusters.

\subsubsection{Comparison of the matched clusters}
\label{sec:redmapper_sub3}

\begin{figure*}
	\includegraphics[scale=.98]{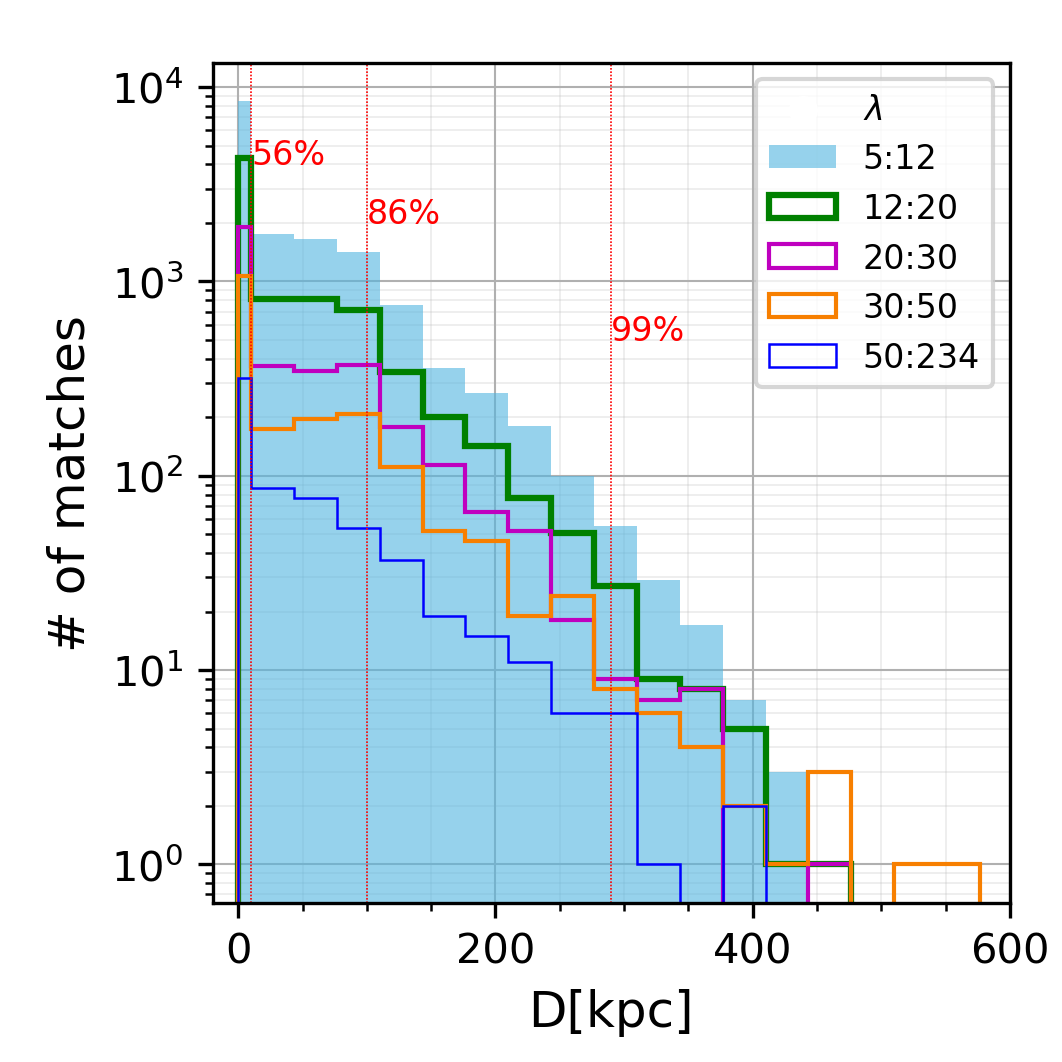}
	\includegraphics[scale=.98]{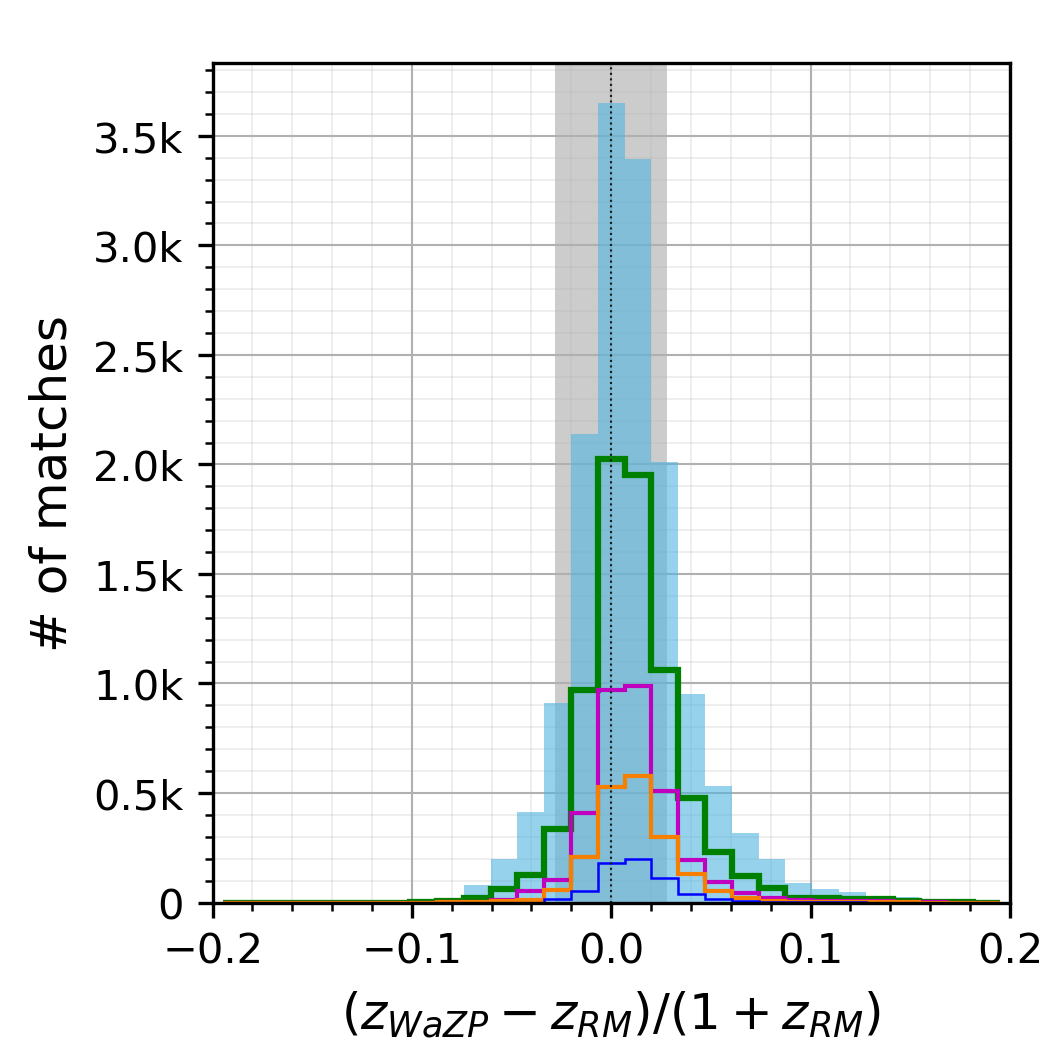}
	\caption{
    Distribution of distances for \wazp-\RM\ matched clusters binned by $\lambda$.
    On the left is the angular separation between cluster centers (converted into physical units at the mean redshift of the cluster pair).
    The first bin on the left plot represents all cluster pairs that have the exact same position.
    On the right is redshift differences and the gray shaded region is the average combined redshift uncertainties divided by $1+\zrm$.
    }
	\label{fig:RMmatch_dAdz}
\end{figure*}

We start by comparing the individual properties of the matched clusters (i. e. centering, redshift and richness).
Cross-matched clusters (\mtRMcross\ pairs) will be used in this evaluation,
as a reliable one-to-one correspondence between clusters is required.
Figure~\ref{fig:RMmatch_dAdz} shows the distribution of angular separation (left panel) and the redshift difference (right panel) of two-way matched clusters.
As it can be seen, all matched clusters are well within the mean radius of the clusters ($\sim 600$kpc), that was used in the last step of the matching. 
In 56\% of the cases, clusters have the exact same center.
Those occur when \wazp\ defines the same BCG as \redMaPPer\ to be its center.
The average distance of central position for the clusters that do not share the same center is of 86kpc,
with 86\% of matched clusters within 100kpc of each other and over 99\% within 300kpc.
The angular separation only shows a very weak tail beyond the typical cluster core radius. In addition, we note that the centering statistics do not seem to depend significantly on the richness.

\begin{figure*}
	\includegraphics[scale=.98]{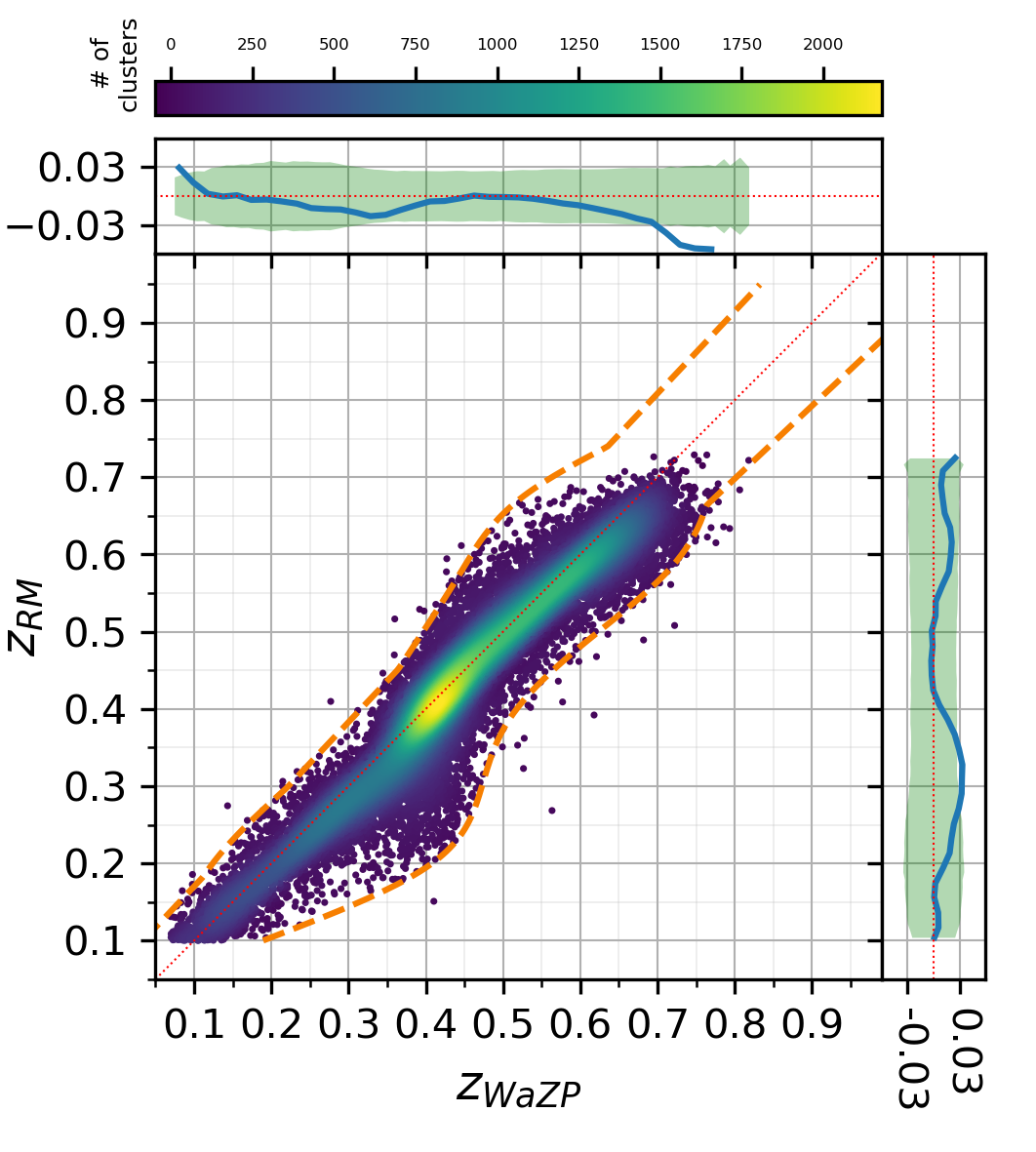}
	\includegraphics[scale=.98]{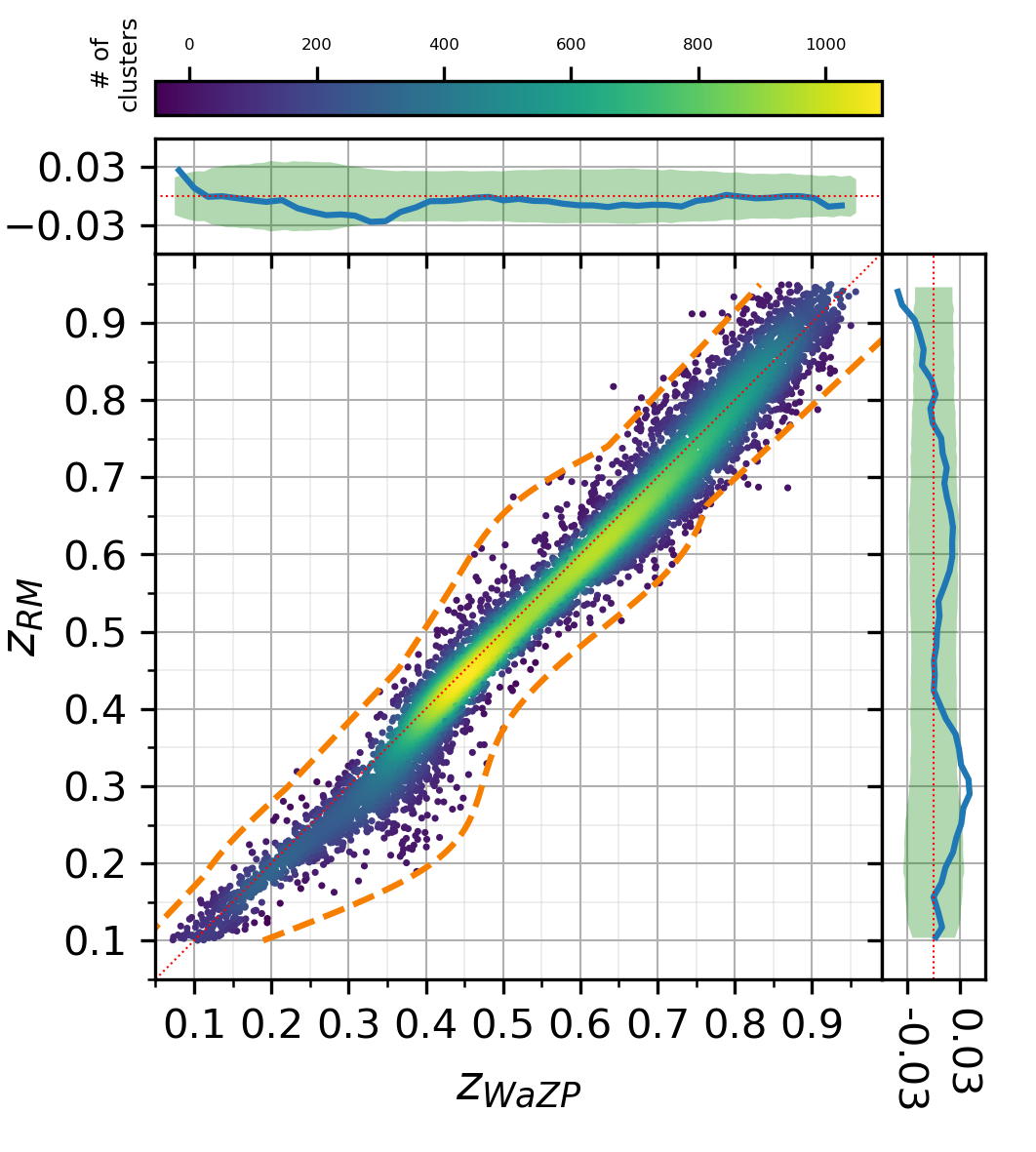}
	\caption{
    Redshift comparison of \wazp\  and \RM\ matched clusters by the four-step match, with the colors corresponding to a count map.
    The left panel is the matching to the \RM\ volume-limited catalog with $\lambda\ge 7$. The right panel is the matching to the \RM\ full catalog $\lambda\ge 20$, showing many additional matches at redshifts above $\sim$0.7.
    The matching of both catalogs was done as described in section~\ref{sec:redmapper_sub2}, with the yellow dashed lines being the windows defined in Figure~\ref{fig:RM_zXz_005mpc}.
    The top and right blue lines are the bias of the redshift relations defined by $(z_{RM}-z_{WaZP})/(1+z_{WaZP})$ and $(z_{WaZP}-z_{RM})/(1+z_{RM})$ respectively.
    The green shaded regions in those panels are the average combined uncertainties the clusters redshifts
    $\langle z^{err}\rangle = \sqrt{ \langle z^{err}_{WaZP}\rangle^2 + \langle z^{err}_{RM}\rangle^2 }$.
    Comparing both panels, we can see that most outliers present at $z_{WaZP}\sim 0.4$ are low richness clusters.
    }
	\label{fig:RMmatch_z}
\end{figure*}

Turning to the distribution of the redshift separations, we find that for the vast majority of pairs ($>99\%$), the redshift separation is well within one third of the redshift window used in the last step of the matching, which has an average size of $\sim \pm 0.15(1+z)$.
Additionally, we find over \zerrrTotRMNORMpairs\ of pairs with redshift separation within the combined uncertainty of the cluster redshifts $\langle z^{err}\rangle = \sqrt{\langle z^{err}_{WaZP}\rangle^2 + \langle z^{err}_{RM}\rangle^2 }=\zerrrTotRMNORM$ (gray shaded region).
We note, however, a small average redshift bias of $\zbiasrmRM$ (Table~\ref{tab:rmzprop})  exists between \wazp\ and \RM\ , with redshifts derived by \wazp\ being on average slightly larger than \RM\ redshifts. In Figure~\ref{fig:RMmatch_z}, \RM\ and \wazp\ redshifts are compared, showing that the bias is mainly due to a significant fraction of $z\sim 0.3$ clusters that were pushed to $z_{wazp} \sim 0.4 $, an effect that is discussed in the next section. 
Additionally, we see on the right panel that this effect occurs mainly on poor ($\lambda<5$) clusters.
Overall, these results show that  most matches are well within the ranges adopted in the matching procedure. It strongly supports the idea that we are detecting on average the same systems.
We also note that the scatter is very similar to the combined redshift uncertainties in all richness limited samples (Table~\ref{tab:rmzprop}),
and the biases are well within these values.

\begin{table}[h]
\centering
\begin{tabular}{lrccc}
$\lambda$ bin
& \# clusters
& $\frac{\zwazp-\zrm}{1+\zrm}$ & $\frac{\sigma_z}{1+\zrm}$
& $\frac{\sqrt{(\zrm^{err})^2+(\zwazp^{err})^2}}{1+\zrm}$
\\
\hline\hline
All           &  \rmMTcounts     & \zbiasrmRM    & \zscatterrmRM    & \zerrrTotRMNORM         \\
\rmRichBinI   &  \rmRICHcountI   & \zbiasrmRMI   & \zscatterrmRMI   & \zerrrTotRMNORMrmbinI   \\
\rmRichBinII  &  \rmRICHcountII  & \zbiasrmRMII  & \zscatterrmRMII  & \zerrrTotRMNORMrmbinII  \\
\rmRichBinIII &  \rmRICHcountIII & \zbiasrmRMIII & \zscatterrmRMIII & \zerrrTotRMNORMrmbinIII \\
\rmRichBinIV  &  \rmRICHcountIV  & \zbiasrmRMIV  & \zscatterrmRMIV  & \zerrrTotRMNORMrmbinIV  \\
\rmRichBinV  &  \rmRICHcountV  & \zbiasrmRMV  & \zscatterrmRMV  & \zerrrTotRMNORMrmbinV  \\
\hline\hline
\\
\end{tabular}
\caption{Bias, scatter and uncertainty of \wazp-\RM\ matched clusters.}
\label{tab:rmzprop}
\end{table}

\begin{figure}
	\includegraphics[scale=1]{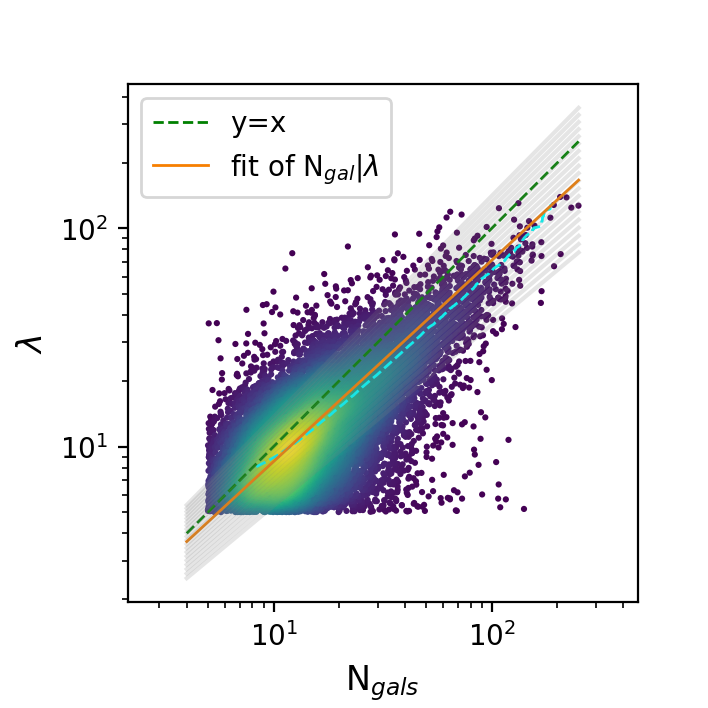}
	\caption{
    Comparison of richnesses computed by \wazp\ and \RM\ for 13,664 two-way matched clusters with
    redshifts in the range 0.1-0.6, with same centering and redshift offsets $\le 0.02(1+z)$. The
    orange line is a power-law fit of the richness relation. The dashed blue line represents the thresholds to be applied to the two richnesses to obtain the same cluster densities in the two cluster samples independently from any matching.
    }
	\label{fig:RMmatch_rich}
\end{figure}

We now compare the values of richnesses as derived from the two algorithms. We do not expect them to be equal in average as they are derived with different definitions. \RM\ richness ($\lambda$) considers red sequence galaxies down to 0.2L$^*$ in the $z$-band, whereas \wazp\ richness ($N_{gals}$) considers all galaxies down to 0.25$L^*$ in the $i$-band. Moreover these quantities are not necessarily computed in the same angular radius. Despite these differences, we expect some correlation between these two richness estimates. To perform this comparison we restrict to two-way matched clusters with same centers and redshift offsets $\le 0.02(1+z)$. We also considered clusters in the redshift range 0.1 - 0.6 in order to assure richnesses to be complete for both cluster finders. 
Figure~\ref{fig:RMmatch_rich} shows a strong correlation between the richnesses of the two cluster finders, with \wazp\ richnesses on average systematically larger than \RM\ ones.
To quantify the effect, we performed a linear fit in $log$ space. To do so, cluster richnesses were first sliced in $\lambda$ and for each slice the mode of the smoothed distribution of $N_{gals}$ computed.
This procedure minimizes the effect from Malmquist bias when constraining the relation between richnesses.
The fit of the resulting $(N_{gals},\lambda)$ pairs led to the relation:

\begin{equation}
\log\lambda=(0.92\pm 0.09) \log N_{gals} + (0.01 \pm 0.11).
\label{eq:lambda_ngals}
\end{equation}

\begin{figure}
	\includegraphics[scale=1]{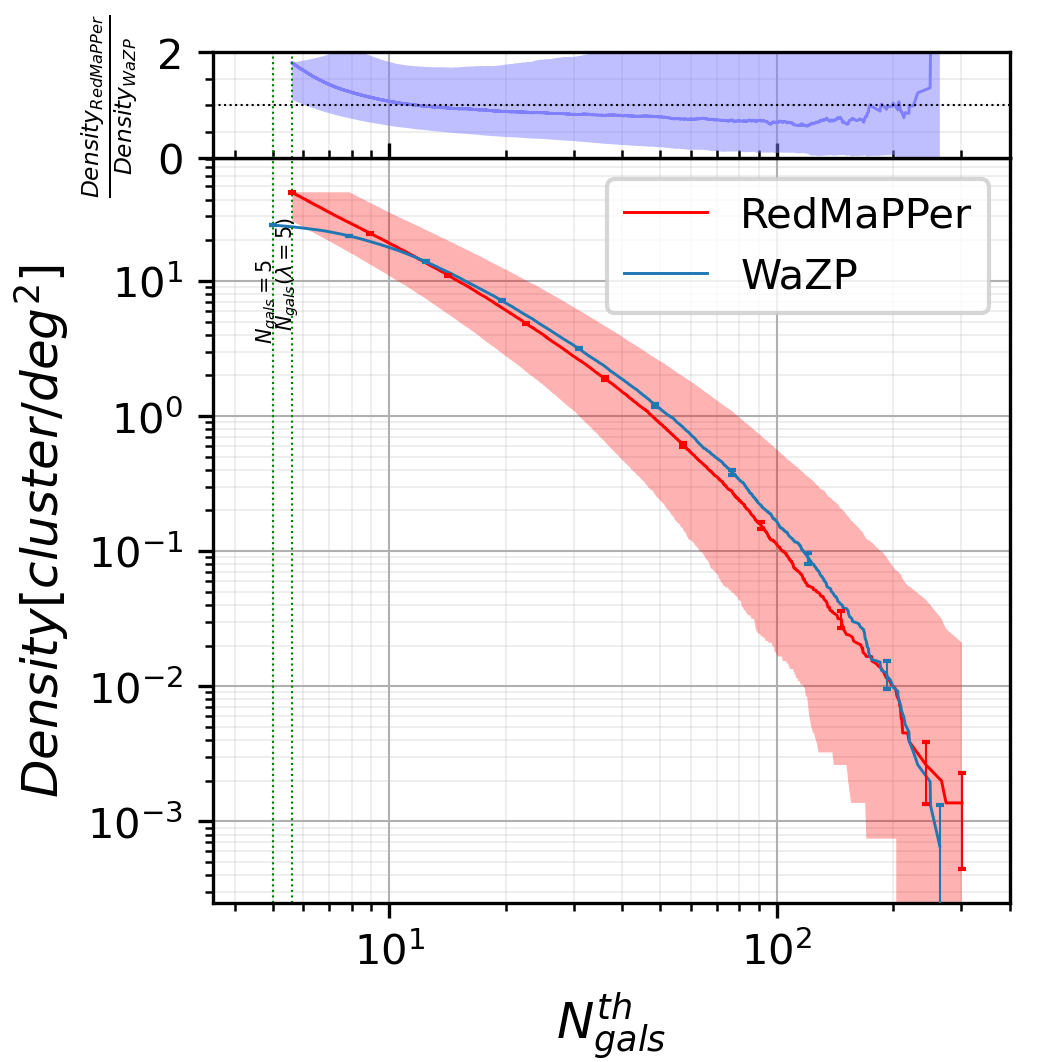}
	\caption{
    Density of \wazp\ and \RM\ catalogs as a function of $\rich$ threshold. The thresholds for
    \RM\ were computed by converting $\lambda$ to $\rich$ using Eq.~\ref{eq:lambda_ngals}, with the shaded regions being the uncertainties propagated and errobars from Poisson noise. The top panel shows the ratio between the densities.
    }
	\label{fig:RM_rich_dens}
\end{figure}

An independent comparison of the two estimated richnesses is based on the cluster density given a richness threshold. This is shown as the blue dashed line in Figure~\ref{fig:RMmatch_rich}. Each point of this line provides the threshold in $N_{gals}$ and in $\lambda$ richnesses to obtain the same density in the two cluster samples, and is independent of any matching. 
It is remarkable that this measurement is very close to the mean relation between the two richness estimators. 
Another way to look at this is to compare directly the densities of the two cluster samples. In Figure~\ref{fig:RM_rich_dens}, we compare  \wazp\ and \RM\ cluster densities considering clusters with redshifts in the range 0.1 - 0.6 and with richnesses above a given threshold, where the threshold in $\rich$ and $\lambda$  are related following Eq.~\ref{eq:lambda_ngals} (e.g. 
$\lambda\ge20$ is equivalent to $\rich\ge\wzRichIII$). Cluster densities are very similar over a very wide range of richnesses. This result supports the idea that, on average, the ranking of the two cluster samples by their richness is similar.

\begin{figure*}
	\includegraphics[scale=.98]{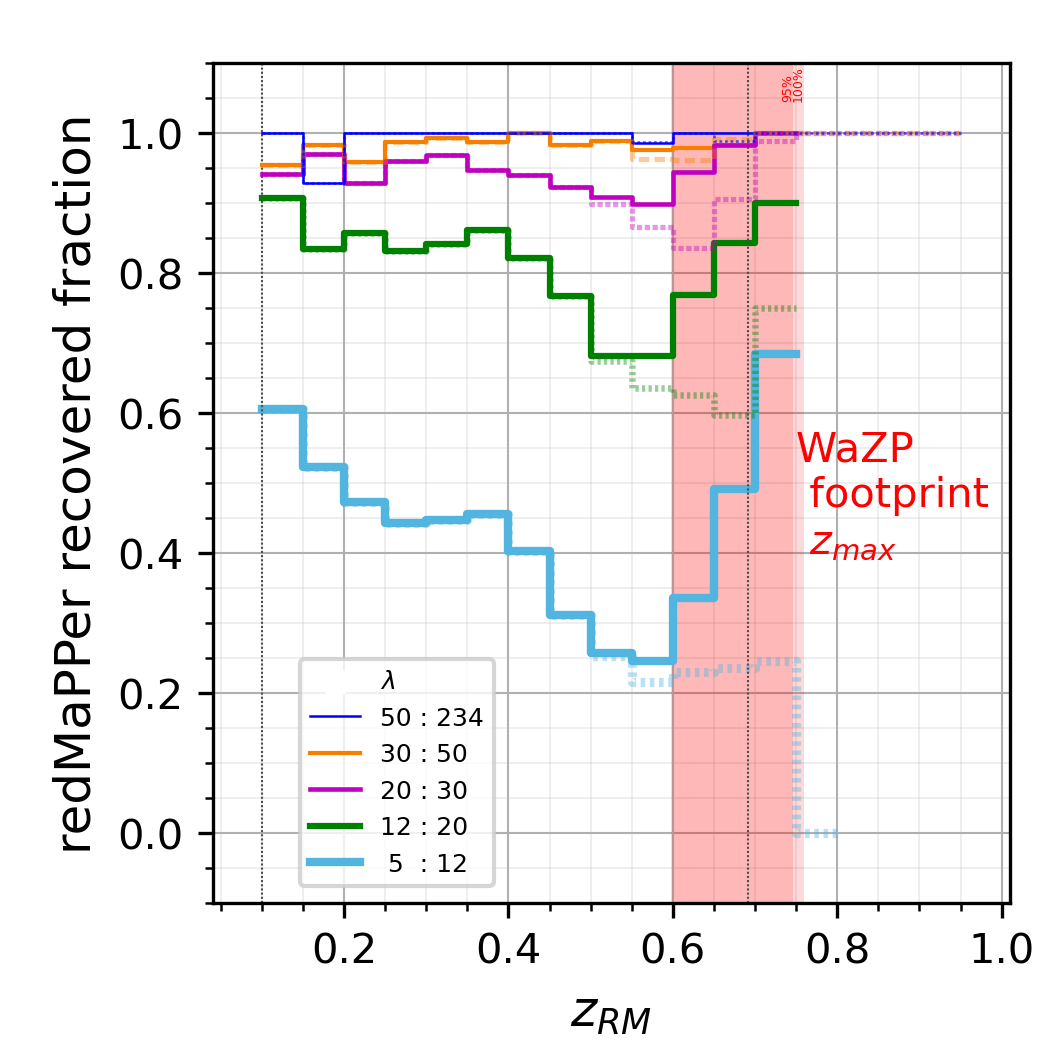}
	\includegraphics[scale=.98]{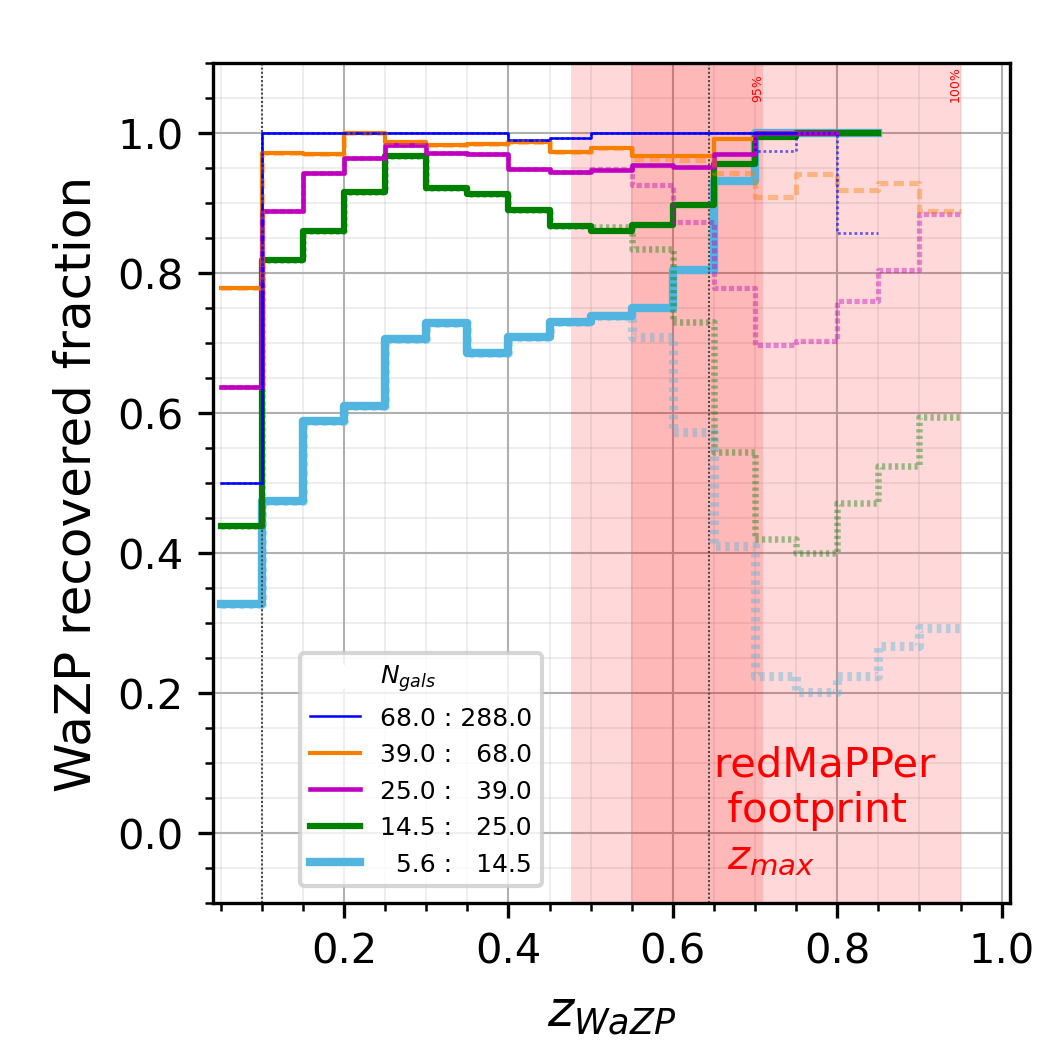}
	\caption{Fraction of clusters detected by both \RM\ and \wazp\
    binned by richness as a function of redshift.
    The dotted line is the minimum redshift of the \RM\ catalog, the gray shaded area is the region where \RM\ footprint decreases due its local $z_{max}$ and the dashed line is the mean of $z_{max}$.
    The solid lines are the recovery rates for the matching with \RM\ volume limited catalog,
    and the light shaded lines correspond to matching with \RM\ volume limited plus \RM\ full ($\lambda\ge 20$) catalog.
    On the left \RM\ is used as the reference catalog (binning by
    $\lambda$) and on the right
    \wazp\ is the reference catalog (binning by $N_{gals}$).
	}
	\label{fig:RMmatch_cp}
\end{figure*}

\subsubsection{Statistics of unmatched clusters}

We now evaluate the recovery rates between catalogs.
Our main goal in this section is to check if each cluster finder could have missed a detection,
therefore we take a very conservative approach to label clusters as unmatched.
In principle, these rates could be computed considering a two-way matching. However, in that case, if a cluster of the first sample appears to be fragmented in the second one, the extra cluster will be counted as not recovered. Here, we wish to separate the absence of a counterpart from fragmentation,
which should be treated separately. Therefore, we defined the recovery rate as the fraction of clusters having one or more counterparts in the matched cluster sample.
Obviously, the absence of a match does not exclude completely the existence of a counterpart. Some systems could suffer from a strong mis-centering, larger than tolerated by the matching criteria. Other systems could also suffer from edge effects that may occur at the periphery of the survey or close to a masked region within the survey. To minimize the latter, we do not consider unmatched clusters outside the intersection of the two cluster sample footprints (constructed as \textsc{nside} $=4096$ \texttt{Healpix} maps)
or unmatched cluster located in edge pixels.
In addition,
as \RM\ removes clusters with over 20\% of their area masked, an equivalent consideration had to be made when looking for \wazp\ counter-parts.
Hence,
unmatched clusters whose cover fraction was less than 80\% on the other catalog's detection fraction footprint were also discarded from the analysis.
These cover fractions were computed using the same weighted methodology as \citep{Ryk14} considering the other catalog footprint.
Although this is not a major contribution for the values of the recovery fraction of \wazp\ clusters, 
ignoring this effect leads to lower recovery close to the footprint edges and holes.
It is also important to note that, because both catalogs have a footprint with a variation on $z_{max}$ at different locations, the computation of whether the cluster
is inside the footprint or in a edge pixel and its cover fraction depends on the cluster position and redshift.

These cuts, based on the footprints and cover fraction, removed \mtRMexcludedFtEdCFrm\ and \mtRMexcludedFTEdCFwz\ \RM\ and \wazp\ clusters without counterpart, respectively, from our recovery rate analysis.
These clusters certainly contain information regarding each cluster finder selection function and limitations,
however the study of these objects require a different analysis on a object-by-object case and will be done in a future work.

With these considerations, for one-way matching including the possibility of multiple associations,
the recovery rate analysis is based on the total of \mtRMtotalCPrm\ \RM\ and \mtRMtotalCPwz\ \wazp\
clusters with \mtRMmmtRM\ and \mtRMmmtWZ\ matched respectively.

The recovery rate for each catalog as a function of redshift in different richness bins is shown in Figure~\ref{fig:RMmatch_cp}.
The left panel shows the fraction of \RM\ clusters recovered by \wazp\ and the right panel the other way around.
The gray shaded area is the redshift region where the \RM\ (left panel) and \wazp\ (right panel) footprints decreases in size, with the dashed line being the median value of this footprint redshift limitation (i. e. where the area drops to 50\% of the total footprint).
The different shades correspond to the 95\% and 100\% percentiles of the $z_{max}$ distribution.
We binned \RM\ clusters into 5 samples,
2 bins for lower richness ($\lambda<20$) clusters, that were not used for cosmological constraints \citep{Cos20}, and 3 sample with higher richness.
\wazp\ clusters were binned on the corresponding richness using our fit on Eq.~\ref{eq:lambda_ngals}. 
One can see that \RM\ clusters
with $\lambda\ge 20$ are mostly ($\ge 90\%$)
recovered up to the \wazp\ redshift limit. Similarly, at the same level of richness ($\rich\ge\wzRichIII$),
we find that \wazp\ clusters are also recovered at more than $90\%$ up to the \RM\ redshift limit of $\sim 0.7$.
We also note rapid increase in the recovery rate of clusters with richness in both cases.
Considering the $0.1<z<0.7$ range,
the overall recovery rate of \RM\ clusters is 93.3\%, 98.4\%, 99.7\% for $\lambda\ge20$ bins ($\rmRichBinIII$, $\rmRichBinIV$ and $\rmRichBinV$ respectively).
For \wazp\ clusters, similarly, we have 95.4\%, 97.9\% and 99.7\% for $\rich\ge\wzRichIII$ bins ($\wzRichBinIII$, $\wzRichBinIV$ and $\wzRichBinV$ respectively).
Major differences occur when considering clusters less rich than $\lambda\sim 20$ ($\rich\sim\wzRichIII$). It is remarkable that even in the $\lambda$ range 5 to 20 ($\rich$ \wzRichI\ to \wzRichIII), clusters are still recovered at rates between 50 and 60\%, depending on the redshift.
The dotted vertical line corresponds to the minimum redshift of \RM\ clusters ($z=0.1$),
hence the low recovery rate for \wazp\ clusters in the first redshift bin.

We note that the recovery rates reach 100\% at high redshifts in both panels of Figure~\ref{fig:RMmatch_cp}.
This is an effect of the footprints variable redshift limit leading to smaller effective area as redshift increases,
with only $3\%$ ($2\%$) of the \RM\ (\wazp) area remaining at $z=0.7$ ($0.76$).
Consequently,
at those redshifts,
all unmatched clusters are removed from the analysis,
resulting in an artificially perfect recovery.
To obtain a more relevant \wazp\ recovery rate at high redshifts, we combined \RM\ volume limited catalog ($\lambda \ge 5$) with the \RM\ full catalog (uniform $z_{max}$ of $0.97$ and $\lambda \ge 20$) and matched it to the \wazp\ catalog using the procedure described above.
These results are represented by the shaded lines in Figure~\ref{fig:RMmatch_cp}.
We see now that the \RM\ recovery rate for $\lambda\ge 20$ clusters extends to higher redshifts.
Comparing the  \wazp\ recovery fraction to the matching with \RM\ volume limited only,
we have a general decrease at high redshifts. The lower recovery rate of $N_{gals}\le\wzRichIII$ clusters is directly correlated with poorer clusters ($\lambda\le 20$) missing in the \RM\ full catalog.
Richer clusters ($\rich\ge\wzRichIII$) are affected by the scatter down of the richness relation between both cluster finders.

We conclude from this analysis that, statistically, rich systems are found by both cluster finders, independent of their redshifts, with very few individual differences that are investigated in the next section.

    \section{Discussion}
    \label{sec:discussion}

We have shown that all \SZ\ clusters with redshifts $\le$ \WazpZmaxDetec\ intersecting our footprint are recovered by \wazp\ cluster finder applied on DES-Y1 data. We have also shown that more than 90\% of clusters with richnesses above $\sim$20 detected by \RM\ (or 25 by \wazp ) are matched to those detected by \wazp\ (or by \RM ). In this section we compare some properties of the the two optical cluster samples. In particular, we explore differences such as redshift discrepancies, overmerging / fragmentation,  and the reasons for unmatched systems on both sides.

\subsection{Redshift discrepancies between \wazp\ and \RM}

Whereas the comparison of \wazp\ redshifts with spectroscopic redshifts assigned to \SZ\ clusters (see table~\ref{tab:szzprop}) showed moderate bias and small scatter, the comparison with \RM\ clusters revealed  stronger discrepancies.
This is related to the fact that \RM\ (and \wazp) clusters are on average much less massive than SPT clusters. This is confirmed if we restrict the cross match between \wazp\ and \RM\ to richer clusters. In that case, as shown in the right panel of Figure~\ref{fig:RMmatch_z}, a reduced redshift bias and scatter is observed between the two samples.

The fraction of redshift outliers (defined by a redshift difference $\ge 3\sigma_z(1+z)$) is less than 5\% of our clusters, with 78\% (82\%) of them having $\rich\le25$ ($\lambda\le20$). These redshift outliers are mainly produced at $z_{\wazp} \sim 0.4$.
This also reflects on \wazp\ cluster number counts (Figure~\ref{fig:nzcl}), that showed a peak at redshift $\sim$0.4 that becomes more prominent when considering the poorest clusters. From the same cluster number counts, a deficit of clusters at $z\sim0.3$ can also be noticed. These points support the idea that in the redshift range 0.15 - 0.35, \wazp\ detects on average the same clusters as \RM\ but shifts a fraction of these to $z \sim 0.4$.

\begin{figure}
	\includegraphics[scale=1]{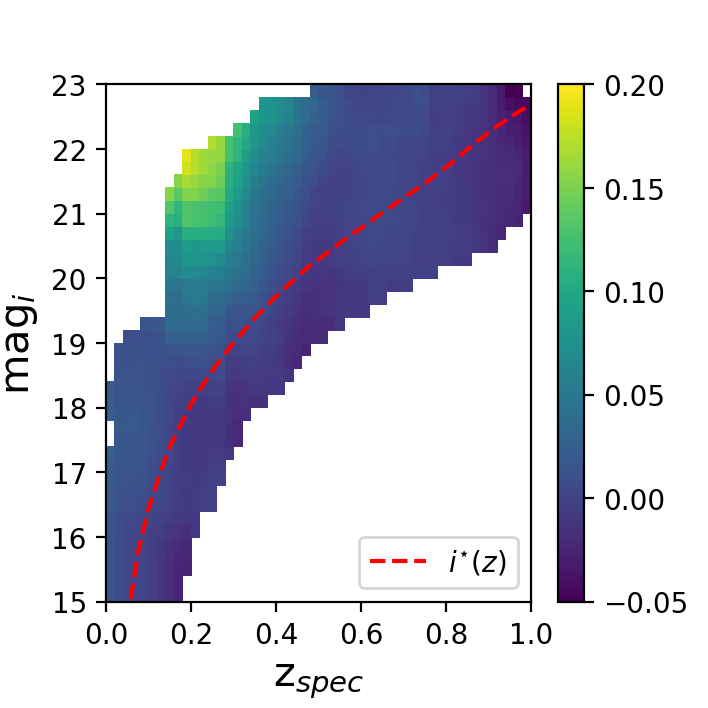}
	\caption{Bias of the galaxy DNF photometric redshifts relative to spectroscopic redshifts as a function of spectroscopic redshift and magnitude. Computation is based here on the SPT region. We require a minimum of 100 spectroscopic redshifts at each redshift - magnitude position. Pixels not satisfying this condition remain white on these maps.   }
	\label{fig:zpstats_mag}
\end{figure}

From global statistics of photometric redshifts, only moderate bias is measured (see right panel of Figure~\ref{fig:zpstats}). However, this global bias includes galaxies of all magnitudes down to $\magI^*+1.5$. In order to understand the shift in redshift of a fraction of \wazp\ clusters, one needs to investigate the photometric redshift bias at least as a function of both redshift and magnitude.
This is what is shown in Figure~\ref{fig:zpstats_mag}. We binned our spectroscopic sample in $i$-band magnitude and spectroscopic redshift and computed, for each bin with at least 100 galaxies, the median and standard deviation of $(z_{phot}-z_{spec})/(1+z_{spec})$.
The amplitude of the photometric redshift bias is shown as a color code. In most regions of this diagram, the bias is moderate, consistent with the global bias. However, for redshifts between 0.15 and 0.35 and $i$-band magnitudes fainter than $\sim20.$, we find a strong bias that reaches values of 0.1-0.2.

The origin of this strong bias is two-folded. It is first due to the lack of $u$-band and to the transition of the 4000\AA\ break between $g$ and $r$ band at redshift $z\sim 0.3-0.4$.
Second, it is due to the lack of faint ($\magI \geq 20$) red galaxies at redshifts below $\sim 0.35$ in our spectroscopic training sample. We stress that for these faint, low-redshift and red galaxies, the bias may be even larger as it cannot be estimated properly. The consequence is that their photometric redshift is overestimated, pushed to redshifts where the training set samples better the same (magnitude, colors) space. This effect has already been stressed in several other studies \citep[e.g., Figure 25 of][]{Ryk16}.

\begin{figure}
	\includegraphics[scale=1]{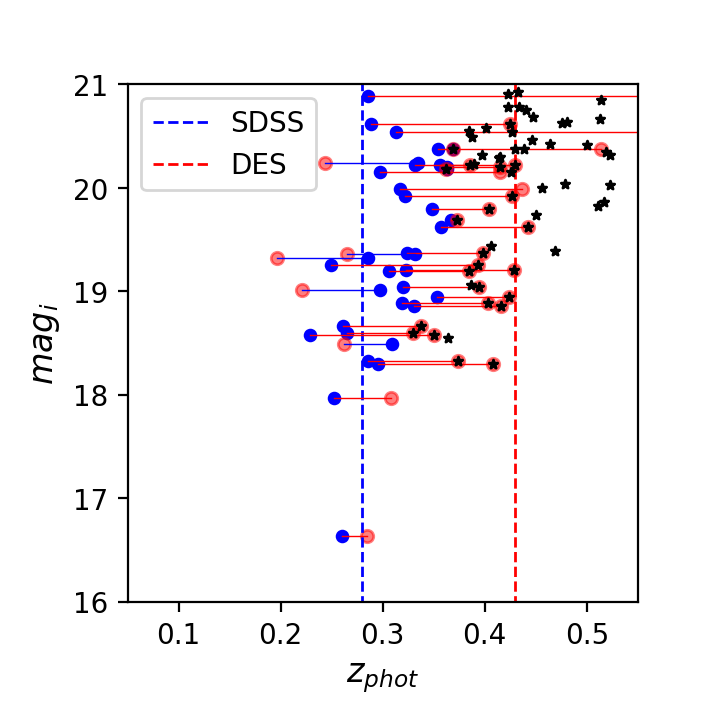}
	\caption{
	This figure illustrates how the bias in photometric redshifts shown statistically in right panel of Figure~\ref{fig:zpstats} may affect cluster members at $z\sim0.3$.
    It displays the members of a galaxy cluster detected in Y1-S82 by \wazp\ at redshift $z=0.43$ (black stars). This cluster is also found by \RM\ but with a redshift $z=0.28$.
    When using SDSS photometry \wazp\ recovers the cluster at the same redshift as \RM\ (blue dashed line), and associated members are shown in blue. These SDSS based members have a different photometric redshift when computed with DNF based on DES photometry (i.e., without the $u$-band). The latter are shown in red, and are linked to a blue point when it is the same galaxy.
    The common red and black symbols show that the two detections have many members in common. However, the BCG (at magnitude 16.5), for instance, was missed as its redshift is not biased. As the DNF+DES-Y1 bias is strongly magnitude dependent, it pulls members apart in two different redshift bins.
    }
	\label{fig:bias_ex}
\end{figure}

This statement can actually be tested by comparing how a DES redshift biased \wazp\ cluster is detected in the Sloan Digital Sky Survey, that is covered in $u,g,r,i,z$ bands in the overlapping Stripe 82 region. As an example, we selected from our DES-Y1-S82 run, one cluster of richness above 60 with a redshift 0.43, whereas the same cluster is detected by \RM\ at a redshift of 0.28.  \wazp\ was run on a small section of SDSS-S82 around that cluster with the same settings but based on SDSS DR-12 photometric redshifts from \cite{Beck2016}. Based on these redshifts, \wazp\ recovers a much lower redshift ($z=0.28$) for the cluster, which is consistent with \RM\ and with the available BCG spectroscopic redshift. In Figure~\ref{fig:bias_ex} we show the redshifts and magnitudes of galaxies classified as cluster members for the two detections, based on DES and on SDSS. One can first notice a large overlap between the members. Then, one can clearly see that these common members are systematically shifted to larger redshifts within the DES. As this effect is stronger for fainter objects, one can also notice for instance that the BCG, at a magnitude of 16.5, has an unbiased redshift. The consequence is that the BCG was not considered as a member in the DES based membership.
To conclude, \wazp\ based on DNF-DES photometric redshifts seems able to recover clusters at $z \sim0.3$, but redshift, membership and therefore richness may be severely affected.

New approaches are currently being investigated to correct for the impact on cluster detection and characterization of the photometric redshift bias effect.

\subsection{Relative fragmentation}

In the previous section, the relative completeness of the two optical cluster finders presented in figure~\ref{fig:RMmatch_cp} is meant to highlight the fraction of clusters without any counterpart. For those clusters tagged as matched, this matching does not assure a one-to-one correspondence for the matched clusters, but only that a cluster from one sample has at least one counterpart in the matched sample. 
Here, we examine clusters from one sample that are matched to more than one cluster in the opposite sample. In the case of a two-way match, the richest counterparts are selected letting the additional possibilities unmatched.  
The extra component(s) involved in the one-way match only could be interpreted as a cluster sub-structures in the other sample, or as a missed cluster, depending on the adopted definition of each cluster finder.

In terms of one-way matching, from figure~\ref{fig:RMmatch_cp}, we found that 
 96\% of \wazp\ clusters in the redshift range 0.1 - 0.6 with $N_{gals} \geq 25$ have a \RM\ counterpart, and conversely, 
 94\% of \RM\ clusters with $\lambda\geq 20$ have a \wazp\ counterpart in the same redshift range. 
 If we now consider two-way matches, only 87\% of \wazp\ clusters have a \RM\ counterpart, whereas 91\% of \RM\ clusters are two-way matched, about the same fraction as for the one-way matching.  
 The larger decrease of matches for \wazp\ clusters 
 when going from one to two-way matching suggests that they are in average relatively more fragmented (or \RM\ clusters relatively more merged). This is what we investigate below.
 
 The apparent larger fragmentation of \wazp\ clusters could be due to the presence of very low richness clusters in the periphery of richer ones. To test this, we evaluated the relative fragmentation rate of the two cluster finders considering different richness cuts in the associated systems. The relative fragmentation rate is estimated as the fraction of matches that have more than one counterpart. If we start from \wazp\ clusters with  
 $N_{gals} \geq 25$, the relative fragmentation rate is 22\%, 6\% and 2\% when considering counterparts with, respectively, $\lambda \ge 5, 10, 20$. Conversely, starting from \RM\ clusters with $\lambda \geq 20$, the relative fragmentation rate is 30\%, 24\% and 14\% when considering counterparts with, respectively, $N_{gals} \ge 6, 12, 25$. The ratio (\wazp\ to \RM) of the fragmentation rates   
 increases strongly when considering richer multiple counterparts. We can conclude from this that \wazp\ tends to find pairs of relatively rich clusters more frequently than \RM. 
 
\begin{figure*}
	\includegraphics[scale=0.43]{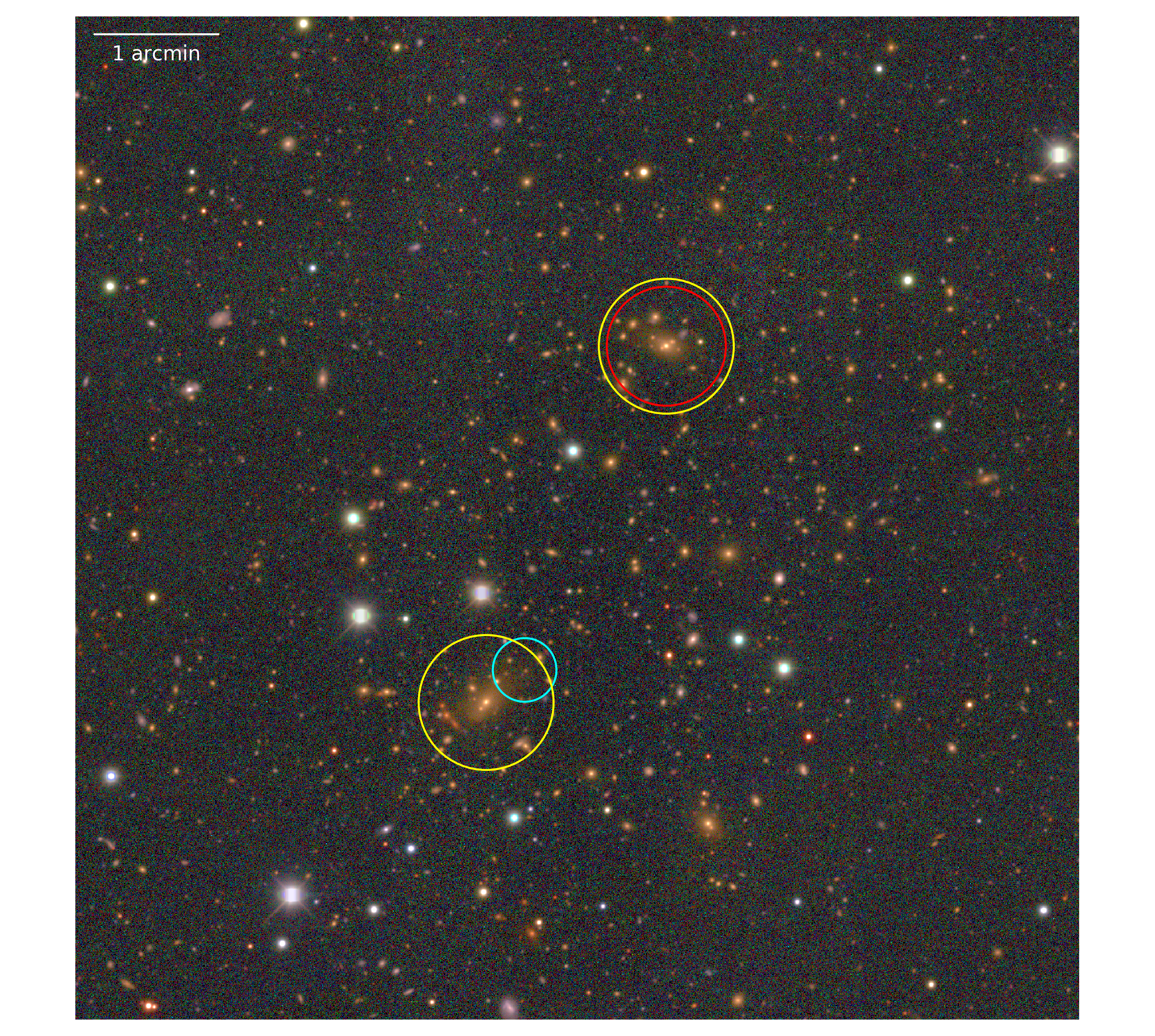}
	\includegraphics[scale=0.43]{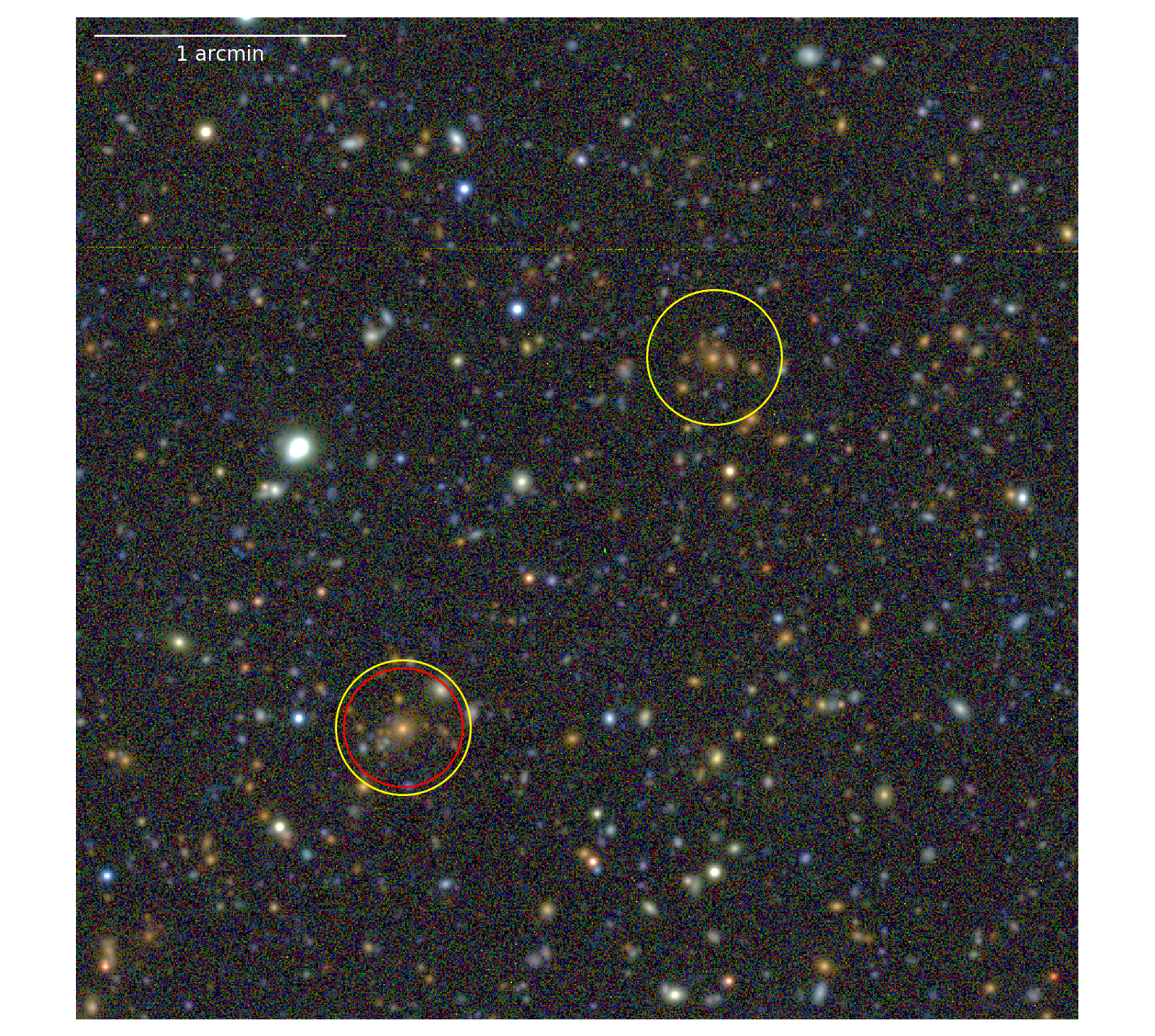}
	\caption{Examples of two clusters that are one-way matched but not two-way matched. In these examples, \wazp\ detects two clusters (yellow circles), whereas \RM\ detects one (red circle), with a cluster radius reaching the second component.   
	{\it In the left panel} (RA=34.312, Dec=-52.760), \wazp\ detected two clusters at $z=0.39 / 0.38$ separated by 1~Mpc. The top right detection is a \RM\ cluster with $z=0.34$ ($\lambda=119$ and $N_{gals}=118$). The bottom left detection ($N_{gals}=81$) also matches a SPT detection (cyan circle). 
	{\it In the right panel} (RA=28.125, Dec=-40.944), \wazp\ detected two clusters at $z=0.67 / 0.69$ separated by 1~Mpc. The bottom left detection is a \RM\ cluster with $z=0.68$ ($\lambda=32$ and $N_{gals}=47$). The top right detection has a richness $N_{gals}=66$. }
	\label{fig:unmatched1}
\end{figure*}

 Should multiple systems be seen as one or several clusters is a matter of cluster definition for each cluster finder. They may also suggest a wrong tuning of the detection algorithm leading to undesirable fragmentation within a clearly unique cluster. To address this point, we visually inspected the 50 richest \wazp\ cluster pairs and found that the vast majority do correspond to clear separate groups. In very few cases only \wazp\ detected two peaks clearly within the same cluster. In figure~\ref{fig:unmatched1} we show two relatively rich cases at redshifts 0.38 and 0.68. In both cases, the \RM\ cluster radii are only slightly larger than the distance between the two \wazp\ clusters, which assured the one way matching of both \wazp\ systems. It is likely here that the galaxies from the extra \wazp\ clusters were percolated to the most likely \RM\ cluster reducing their weight as members of a secondary cluster and eventually leading to only one detection (\citet[][]{Ryk14}, section 9). However, let us stress that the detection algorithms may be tuned to find different overdensities, leading to different samples with their own selection function.

\subsection{Unmatched systems}
Let us now turn to the \wazp\ or \RM\ clusters for which no counterpart was found using our one-way matching procedure. Our goal here is to provide some insight on the reasons why some systems, or types of systems would not be detected by one algorithm or the other. We should first stress that our matching procedure is designed in such a way that we have strongly limited the number of unmatched clusters that could be due e.g. to edge effects, variable depths of the used reference bands  or differences in estimated redshifts.

Treating edge effects properly appeared to be a critical issue as it concerns a significant number of detections due to the complex geometry of the masked regions. Moreover each cluster sample was not built using exactly the same footprint, in particular due to the different reference band used. We considered regions covered by the two footprints,  and also followed \RM's prescription and discarded clusters that would intersect empty regions of the galaxy catalogue by more than 20\% within a 1~Mpc radius. 
Note that this area fraction is actually weighted by a projected NFW profile as described in \citet{Rykoff_2012}.
Concerning the adopted tolerance in redshift difference, as shown above, we have carried out an empirical approach, precisely to avoid unmatched systems that would be detected on both sides but with a too large redshift discrepancy. This case may still happen in our matched catalog, but with a lower occurrence.

In order to qualify the unmatched clusters, we have carried out a visual inspection of the 60 richest ones (for each cluster finder) in the redshift range 0.1 to 0.65. 
These systems have richnesses $N_{gals} \ge 30$ and $\lambda \ge 25$. 

\begin{figure*}
	\includegraphics[scale=0.43]{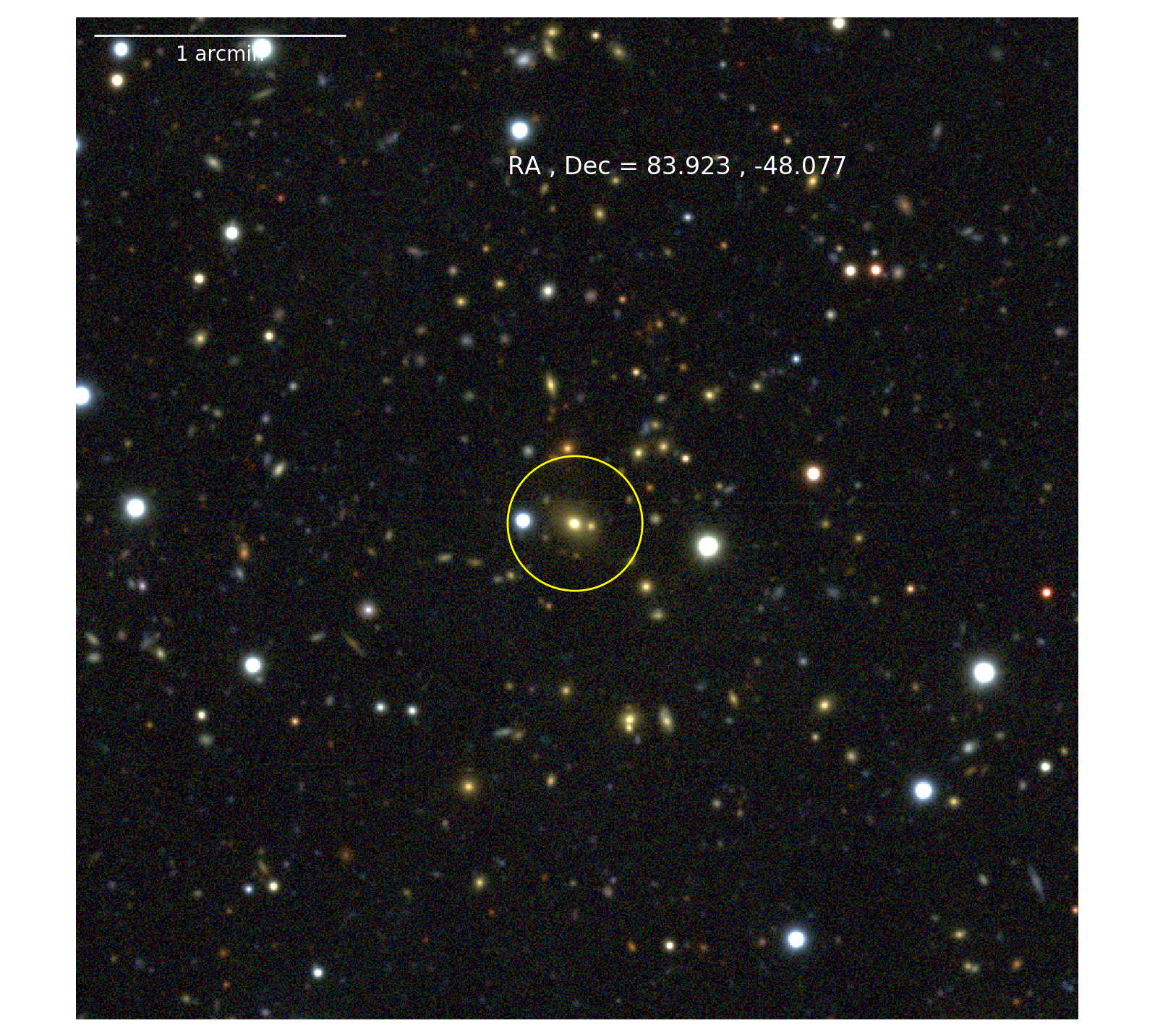}
	\includegraphics[scale=0.43]{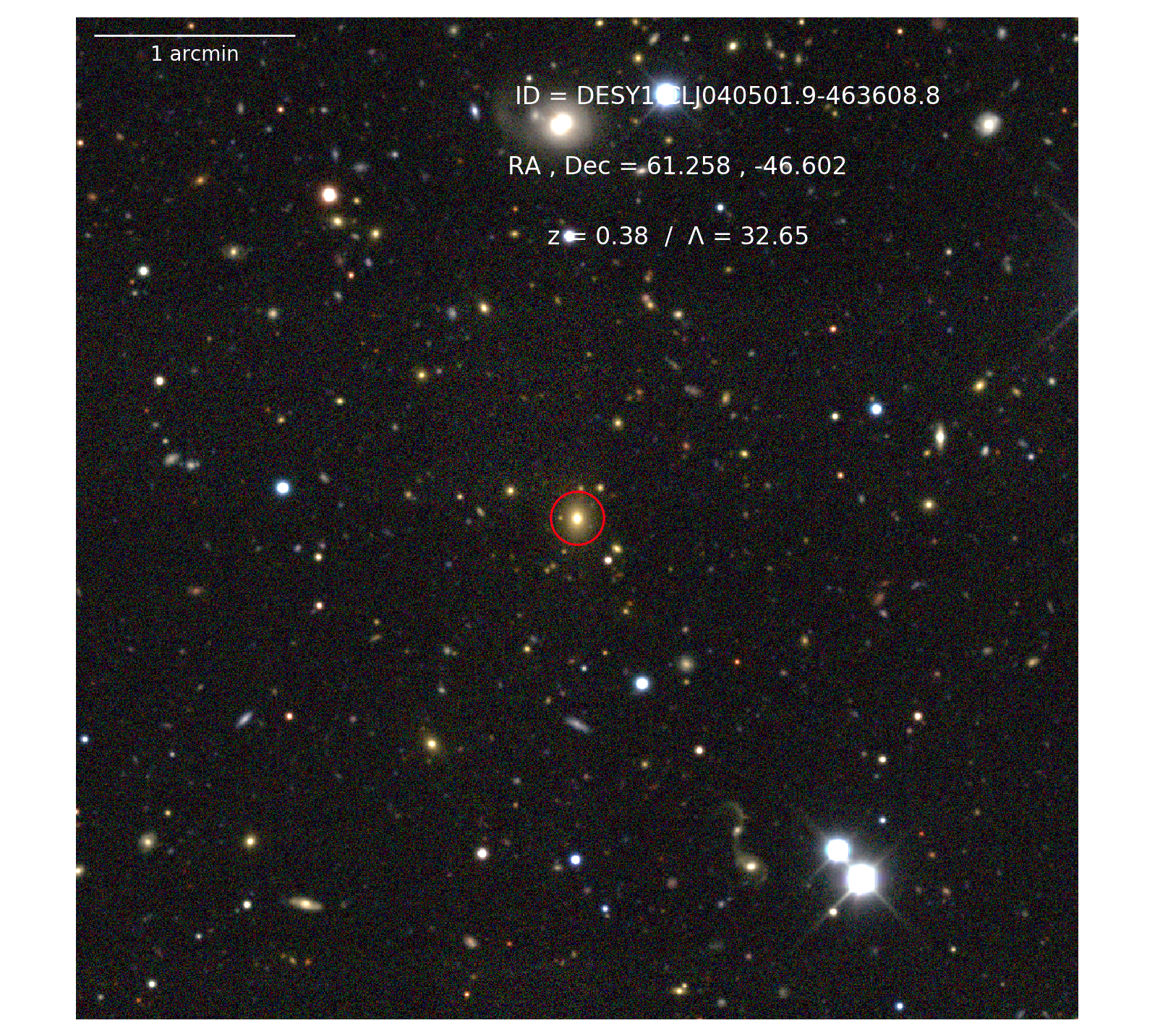}
	\caption{Examples of a \wazp\ cluster (left panel, yellow circle) and a \RM\ cluster (right panel, red circle) both tagged as unmatched. The redshifts and richnesses of these two clusters are $z=0.44$ / $N_{gals}=56.$ (left panel) and  $z=0.38$ / $\lambda=32.$ (right panel). 
}
	\label{fig:unmatched3}
\end{figure*}

Without trying to derive precise statistics from this inspection, unmatched systems clearly enter two categories common to the two cluster finders. The first one, corresponding to one third of the inspected systems, is made of clear concentrated overdensities of red galaxies (two examples are shown in Figure~\ref{fig:unmatched3}, one detected by \wazp\ and the second by \RM). For these systems, possible edge or depth effects were checked and discarded. Those not found by \RM\ have redshifts ranging uniformly from 0.3 to 0.6, whereas those not found by  
\wazp\ are concentrated in two redshift bins, around 0.25-0.35 (possibly due to the photometric redshift bias described above) and the second around 0.5-0.6. 

\begin{figure*}
	\includegraphics[scale=0.43]{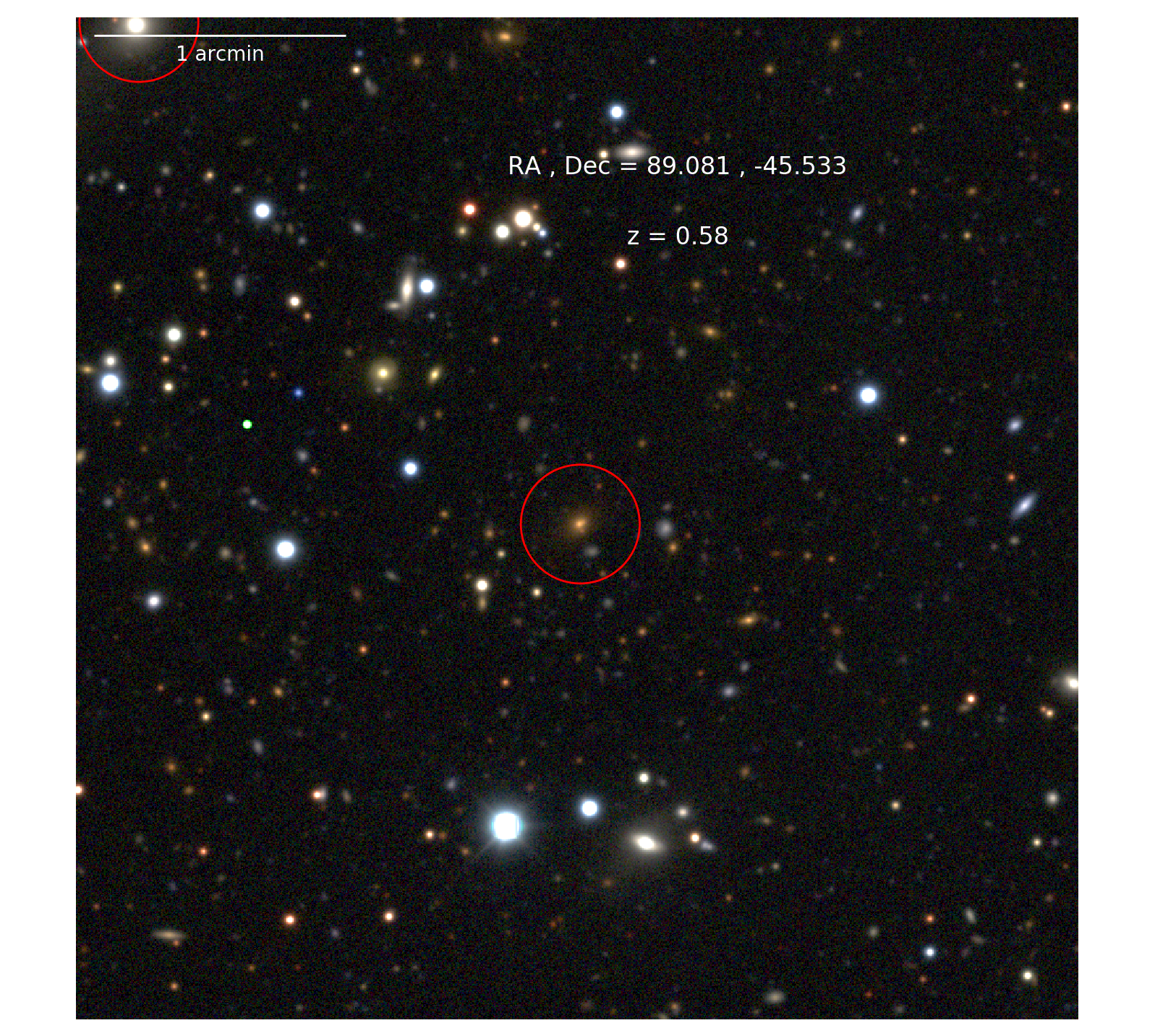}
	\includegraphics[scale=0.43]{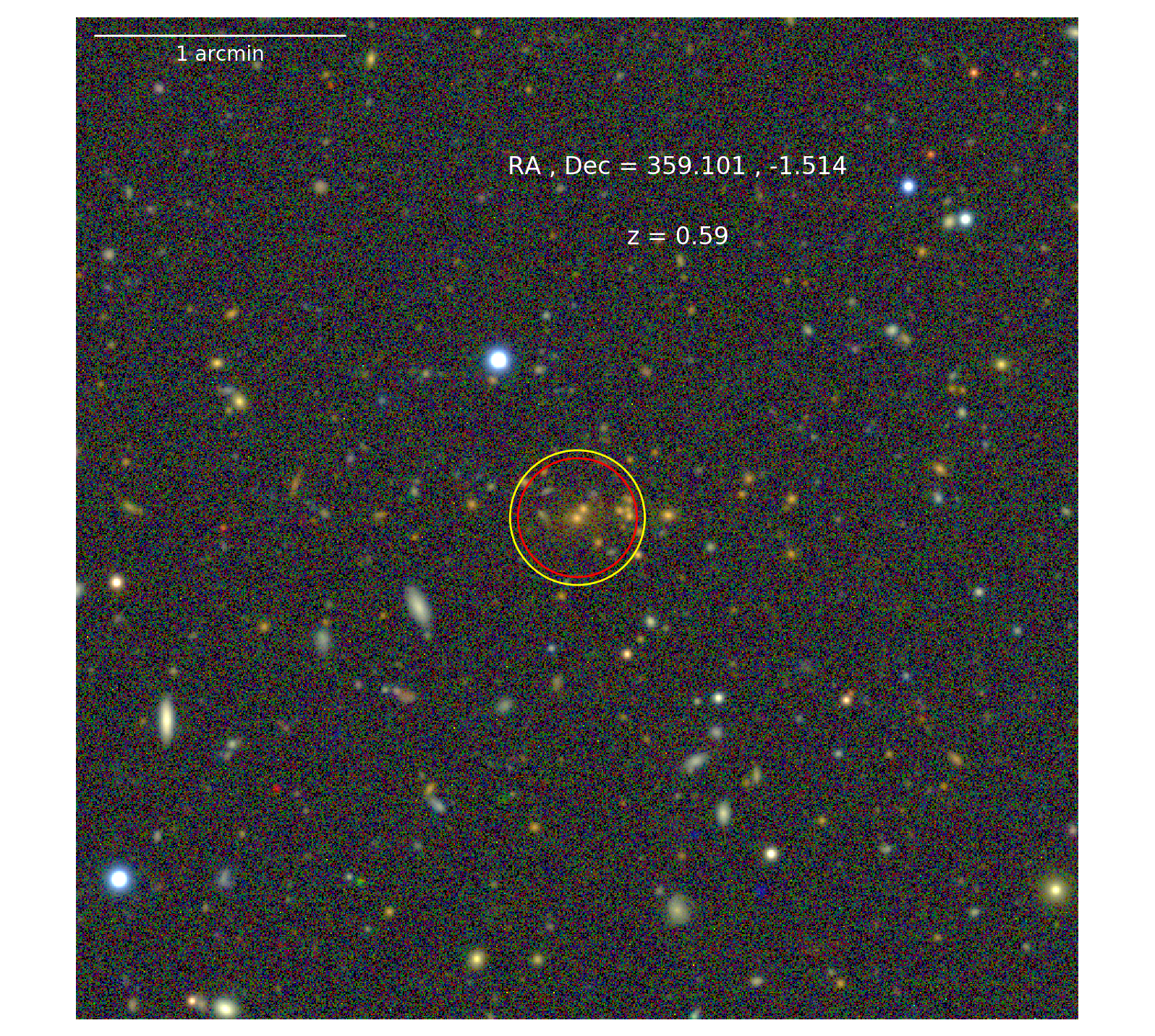}
	\caption{Example of a \RM\ cluster (red circle) not found by \wazp\ (left panel) compared to a \RM\ cluster with same richness and redshift that is matched to \wazp\ (right panel, yellow circle). The two \RM\ clusters have a richness $\lambda = 30$, and a redshift $z=0.58 - 0.59$. The matched \wazp\ cluster was found at $z=0.6$ with a richness $N_{gals}=37$.   
}
	\label{fig:unmatched4}
\end{figure*}

A second category covering more than half of the inspected clusters is composed of much looser systems, without any obvious central concentration, sometimes possibly fragments of larger scale filamentary structures, and in some few cases no apparent cluster at all. These loose systems may appear as poorer clusters even though they are selected among the richest unmatched, typically $\lambda$ (or $N_{gals}$) $\sim 30-35$.   
One typical example is shown in Figure~\ref{fig:unmatched4} where we compare the case of two \RM\ clusters at the same redshift ($z\sim 0.6$) and with similar richnesses ($\lambda \sim 30$). The concentrated system is well recovered by \wazp\ at the same redshift and with similar richness, whereas no counterpart was found for the looser one. Similar opposite situations occur when considering \wazp\ clusters as a reference. 

\begin{figure*}
	\includegraphics[scale=.98]{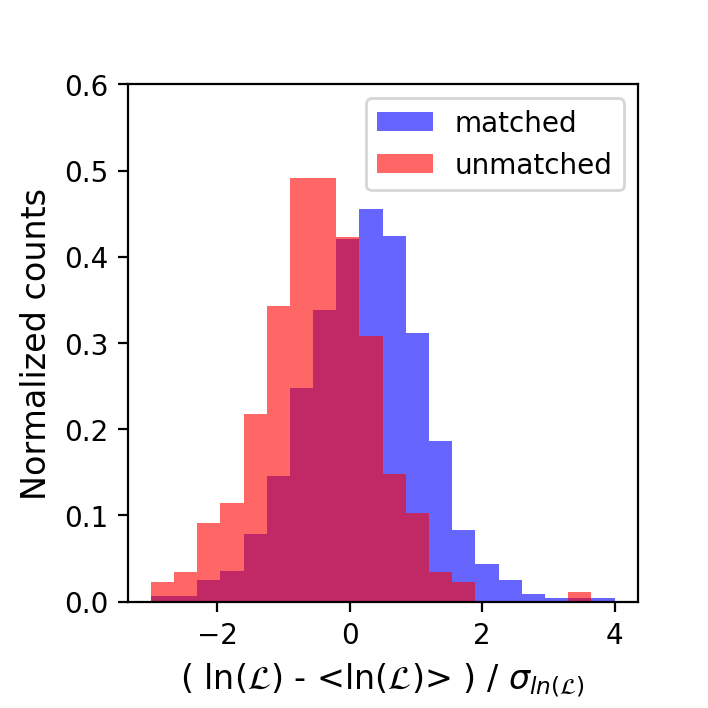}
	\includegraphics[scale=.98]{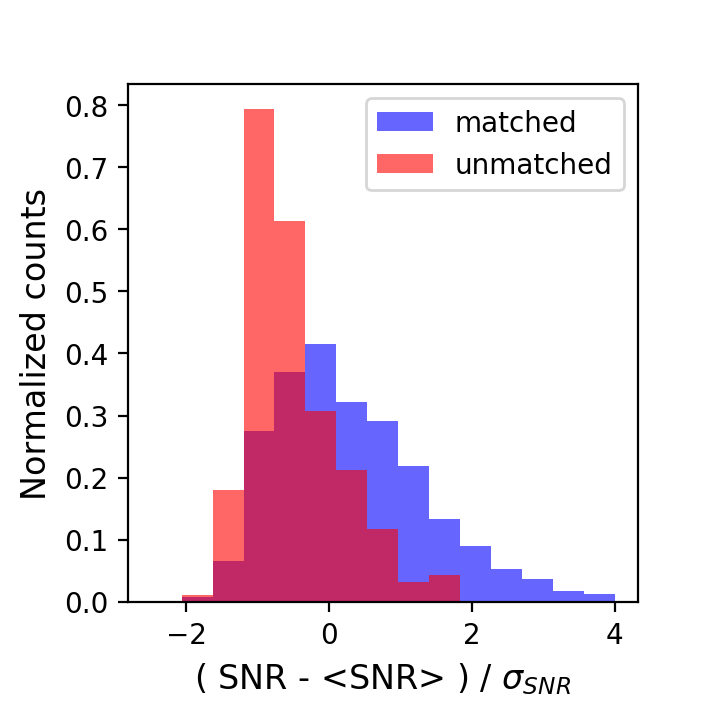}
	\caption{Comparison of the distributions of \RM\  cluster likelihoods relative to the median likelihood at the clusters redshift and richness (left panel) and \wazp\ cluster SNRs relative to the median SNR at the clusters redshift and richness (right panel) for matched  (blue) and unmatched (red) clusters. For both samples we have considered here clusters with richnesses $\lambda \ge 20$ and $N_{gals} \ge 25$ and redshifts in the range 0.1 - 0.6. Note that the distributions are normalized; in both cases, the unmatched clusters are $\sim 20$ times less numerous than the matched ones. 
}
	\label{fig:ppties_unmatched}
\end{figure*}

What seems to be common to these loose unmatched clusters is that they are often characterized, at a given richness, by a lower SNR (in the case of \wazp) and a lower likelihood (in the case of \RM). In order to verify this observation statistically, we have compared the \wazp\ SNR and \RM\ likelihood of the matched and unmatched clusters. To do this, as these two quantities depend in average on both redshift and richness, we have computed the median and 68 percentile of the SNR and likelihood in bins of redshift and richness. We can then compare how each matched and unmatched cluster deviates relative to its local (in redshift-richness space) median SNR or likelihood. The result, considering all clusters in the redshift range 0.1-0.6 and with $N_{gals}\ge 25$ and $\lambda \ge 20$, is shown in Figure~\ref{fig:ppties_unmatched}. Clearly, unmatched clusters have in average lower SNR or lower likelihood than the average, suggesting that these quantities should be considered in the cluster selection function.  

	\section{Conclusions}
    \label{sec:conc}

In this paper we present \wazp\, a new (2+1)D cluster finder based on photometric redshifts.
It is applied to DES-Y1A1 data and resulting samples are compared to those derived from the SPT survey based on the \SZ\ effect, and from the \RM\ cluster finder applied to the same photometric data.

Our conclusions can be listed as follows.

\begin{itemize}

    \item A galaxy Value Added Catalog was derived from the DES Y1A1 survey. It is controlled by the $i$-band that is used for star-galaxy separation and for determining the local depth. Depending on the location, complete galaxy samples can be built down to $\magI=22.25 - 23$, with a median (corresponding to half of the survey area) completeness magnitude limit of $\magI=22.7$.

    \item The \wazp\ cluster finder was applied to DES Y1A1 survey led to
    the detection of \WazpClusters\ clusters over \VacArea\ deg$^2$, with redshifts
    ranging from 0.05 to 0.9 and richness greater than $5$. Due to the $i$-band limiting magnitude
    of the survey and the adopted limiting magnitude for estimating richnesses, complete richnesses
    are derived for clusters in the redshift range 0.05-\WazpZmaxRich. Clusters detected at larger redshifts
    get increasingly incomplete richnesses depending on the local depth. 

    \item  Considering the SPT cluster sample intersecting the DES Y1A1 footprint in the redshift range 0.05 - \WazpZmaxDetec,
    \wazp\ was shown to recover all 293 \SZ\ clusters.
    Comparing redshifts of both cluster finders,
    we found a bias of $\zbiasszSZ$ and a scatter of the same order of redshift uncertainties ($\sim \zerrrTotSZNORM$). 
    When we restrict to \SZ\ clusters with an assigned spectroscopic redshift, all these quantities are lowered by 40\%.

    \item Cross-matching \wazp\ and \RM\ catalog led to the development of an iterative matching algorithm to minimize incorrect associations. Special care was taken to deal with edge effects and depth variations. It resulted in matching  \mtRMcross\ clusters in the two-way criteria with richnesses $\rich$ (and $\lambda$) down to 5. Considering one-way matching for clusters richer than $\rich=25$ ($\lambda=20$) with $0.1<z<0.6$, we showed that \wazp\ recovered 96\% \RM\ clusters, and, symmetrically, \RM\ recovered 94\% \wazp\ clusters.

    \item The centering offset between \wazp\ and \RM\ is less than 200~kpc in most cases (97\%), which is much less than the matching criteria used.

    \item Comparison of the estimated redshifts from \RM\ and \wazp\ shows an overall good agreement. However, a fraction of \wazp\ clusters suffer from a redshift bias, reflecting the underlying galaxy photometric redshift bias. Note that this effect does not seem to prevent detection in general.

    \item Despite different definitions, the comparison of \wazp\ and \RM\ richnesses shows a strong correlation. The scatter of this relation will be analyzed in details on a separate paper in a future work. We also computed a richness relation based on thresholds that provided the same densities of clusters. This relation is remarkably close to the fit from matched clusters, supporting the idea that, on average, the ranking of the two cluster samples by their richness is similar.

    \item The study of the relative fragmentation of \wazp\ and \RM\ clusters showed that \wazp\ tends to find pairs of relatively rich clusters more frequently than \RM.
    
    \item The visual inspection of 60 unmatched clusters richer than $N_{gals} \ge 25$ (for \wazp) and $\lambda \ge 20$ (for \RM), showed that (for both samples) ~1/3 of these are clear concentrations of red galaxies, which, in itself could help to improve the completeness of both samples. A second category representing more than half of the inspected unmatched systems is composed of much looser systems or more filamentary structures. This dominant category of unmatched clusters has been shown statistically to be  characterized by lower SNRs (for \wazp) and lower likelihoods (for \RM) than for the matched clusters. 
\end{itemize}

One of the main aspects of this work is to address the relative completenesses of two optical cluster finders based on different methods and applied to the same survey. As shown in this paper, the comparison of the resulting samples in itself is not a trivial task free of  assumptions. However, it is a useful guide to detect features of cluster finding algorithms when applied to large surveys with all their complexity (missing data, mis-classified sources, depth variation, etc.). In that sense, such comparisons are complementary to those performed with mock catalogs, that can be used, for instance, to address the sample purity in addition to completeness. 

A limitation of the present study is the quality of the photometric redshifts partly hampered by the poor representation of faint red galaxies in our spectroscopic training set, in particular at low redshifts ($z\leq 0.35$). In a companion paper we are exploring the impact of using a much larger spectroscopic training set, and the use of several photometric redshift codes.
We will also study the impact of the star-galaxy classification algorithm.  Finally, we will compare \wazp\ cluster samples based on $i$-band and $z$-band reference magnitudes with a special focus on high redshift cluster detection.

	\section*{Acknowledgments}


M.A. was supported by FAPESP (2013/26612-2) and CNPq (380101/2017-3, 165049/2017-0, 381135/2018-7).
C.B. acknowledges the support obtained from the Bonus-Qualité-Recherche 2013 \& 2017 Programs of the J.-L. Lagrange Laboratory (UMR 7293).

C.B. thanks J.-L. Starck for his help in using and configuring the MR/1 multi-resolution to optimally build galaxy density maps. C.B. also thanks S. Maurogordato, A. Cappi and E. Slezak for many stimulating discussions during the development of the WaZP cluster finder algorithm. 

Funding for the DES Projects has been provided by the U.S. Department of Energy, the U.S. National Science Foundation, the Ministry of Science and Education of Spain, 
the Science and Technology Facilities Council of the United Kingdom, the Higher Education Funding Council for England, the National Center for Supercomputing 
Applications at the University of Illinois at Urbana-Champaign, the Kavli Institute of Cosmological Physics at the University of Chicago, 
the Center for Cosmology and Astro-Particle Physics at the Ohio State University,
the Mitchell Institute for Fundamental Physics and Astronomy at Texas A\&M University, Financiadora de Estudos e Projetos, 
Funda{\c c}{\~a}o Carlos Chagas Filho de Amparo {\`a} Pesquisa do Estado do Rio de Janeiro, Conselho Nacional de Desenvolvimento Cient{\'i}fico e Tecnol{\'o}gico and 
the Minist{\'e}rio da Ci{\^e}ncia, Tecnologia e Inova{\c c}{\~a}o, the Deutsche Forschungsgemeinschaft and the Collaborating Institutions in the Dark Energy Survey. 

The Collaborating Institutions are Argonne National Laboratory, the University of California at Santa Cruz, the University of Cambridge, Centro de Investigaciones Energ{\'e}ticas, 
Medioambientales y Tecnol{\'o}gicas-Madrid, the University of Chicago, University College London, the DES-Brazil Consortium, the University of Edinburgh, 
the Eidgen{\"o}ssische Technische Hochschule (ETH) Z{\"u}rich, 
Fermi National Accelerator Laboratory, the University of Illinois at Urbana-Champaign, the Institut de Ci{\`e}ncies de l'Espai (IEEC/CSIC), 
the Institut de F{\'i}sica d'Altes Energies, Lawrence Berkeley National Laboratory, the Ludwig-Maximilians Universit{\"a}t M{\"u}nchen and the associated Excellence Cluster Universe, 
the University of Michigan, NFS's NOIRLab, the University of Nottingham, The Ohio State University, the University of Pennsylvania, the University of Portsmouth, 
SLAC National Accelerator Laboratory, Stanford University, the University of Sussex, Texas A\&M University, and the OzDES Membership Consortium.

Based in part on observations at Cerro Tololo Inter-American Observatory at NSF’s NOIRLab (NOIRLab Prop. ID 2012B-0001; PI: J. Frieman), which is managed by the Association of Universities for Research in Astronomy (AURA) under a cooperative agreement with the National Science Foundation.

The DES data management system is supported by the National Science Foundation under Grant Numbers AST-1138766 and AST-1536171.
The DES participants from Spanish institutions are partially supported by MICINN under grants ESP2017-89838, PGC2018-094773, PGC2018-102021, SEV-2016-0588, SEV-2016-0597, and MDM-2015-0509, some of which include ERDF funds from the European Union. IFAE is partially funded by the CERCA program of the Generalitat de Catalunya.
Research leading to these results has received funding from the European Research
Council under the European Union's Seventh Framework Program (FP7/2007-2013) including ERC grant agreements 240672, 291329, and 306478.
We  acknowledge support from the Brazilian Instituto Nacional de Ci\^encia
e Tecnologia (INCT) do e-Universo (CNPq grant 465376/2014-2).

This manuscript has been authored by Fermi Research Alliance, LLC under Contract No. DE-AC02-07CH11359 with the U.S. Department of Energy, Office of Science, Office of High Energy Physics.

	\section*{Data Availability}
The \wazp\ DES-Y1A1 cluster products with a full description of the catalogs and columns can be found are available in its online supplementary material \wazpweblink.

\bibliography{refs}

	\section*{Affiliations}
    1. Laborat\'orio Interinstitucional de e-Astronomia - LIneA, Rua Gal. Jos\'e Cristino 77, Rio de Janeiro, RJ - 20921-400, Brazil
\\  2. Laboratoire d'Annecy De Physique Des Particules (LAPP), 9 Chemin de Bellevue, 74940 Annecy, France
\\  3. Université C\^{o}te d'Azur, OCA, CNRS, Lagrange, UMR 7293, CS 34229, 06304, Nice Cedex 4, France
\\  4. Observat\'orio Nacional, Rua Gal. Jos\'e Cristino 77, Rio de Janeiro, RJ - 20921-400, Brazil
\\  5. Departamento de F\'isica Matem\'atica, Instituto de F\'isica, Universidade de S\~ao Paulo, CP 66318, S\~ao Paulo, SP, 05314-970, Brazil
\\  6. Fermi National Accelerator Laboratory, P. O. Box 500, Batavia, IL 60510, USA
\\  7. Instituto de Fisica Teorica UAM/CSIC, Universidad Autonoma de Madrid, 28049 Madrid, Spain
\\  8. Institute of Cosmology and Gravitation, University of Portsmouth, Portsmouth, PO1 3FX, UK
\\  9. CNRS, UMR 7095, Institut d'Astrophysique de Paris, F-75014, Paris, France
\\  0. Sorbonne Universit\'es, UPMC Univ Paris 06, UMR 7095, Institut d'Astrophysique de Paris, F-75014, Paris, France
\\  11. Department of Physics and Astronomy, Pevensey Building, University of Sussex, Brighton, BN1 9QH, UK
\\  12. Department of Physics \& Astronomy, University College London, Gower Street, London, WC1E 6BT, UK
\\  13. Instituto de Astrofisica de Canarias, E-38205 La Laguna, Tenerife, Spain
\\  14. Universidad de La Laguna, Dpto. Astrofísica, E-38206 La Laguna, Tenerife, Spain
\\  15. Department of Astronomy, University of Illinois at Urbana-Champaign, 1002 W. Green Street, Urbana, IL 61801, USA
\\  16. National Center for Supercomputing Applications, 1205 West Clark St., Urbana, IL 61801, USA
\\  17. Institut de F\'{\i}sica d'Altes Energies (IFAE), The Barcelona Institute of Science and Technology, Campus UAB, 08193 Bellaterra (Barcelona) Spain
\\  18. INAF-Osservatorio Astronomico di Trieste, via G. B. Tiepolo 11, I-34143 Trieste, Italy
\\  19. Institute for Fundamental Physics of the Universe, Via Beirut 2, 34014 Trieste, Italy
\\  20. Centro de Investigaciones Energ\'eticas, Medioambientales y Tecnol\'ogicas (CIEMAT), Madrid, Spain
\\  21. Department of Physics, IIT Hyderabad, Kandi, Telangana 502285, India
\\  22. Santa Cruz Institute for Particle Physics, Santa Cruz, CA 95064, USA
\\  23. Department of Astronomy, University of Michigan, Ann Arbor, MI 48109, USA
\\  24. Department of Physics, University of Michigan, Ann Arbor, MI 48109, USA
\\  25. Institute of Theoretical Astrophysics, University of Oslo. P.O. Box 1029 Blindern, NO-0315 Oslo, Norway
\\  26. Jet Propulsion Laboratory, California Institute of Technology, 4800 Oak Grove Dr., Pasadena, CA 91109, USA
\\  27. Institut d'Estudis Espacials de Catalunya (IEEC), 08034 Barcelona, Spain
\\  28. Institute of Space Sciences (ICE, CSIC),  Campus UAB, Carrer de Can Magrans, s/n,  08193 Barcelona, Spain
\\  29. Kavli Institute for Cosmological Physics, University of Chicago, Chicago, IL 60637, USA
\\  30. School of Mathematics and Physics, University of Queensland,  Brisbane, QLD 4072, Australia
\\  31. Center for Cosmology and Astro-Particle Physics, The Ohio State University, Columbus, OH 43210, USA
\\  32. Department of Physics, The Ohio State University, Columbus, OH 43210, USA
\\  33. Center for Astrophysics $\vert$ Harvard \& Smithsonian, 60 Garden Street, Cambridge, MA 02138, USA
\\  34. Australian Astronomical Optics, Macquarie University, North Ryde, NSW 2113, Australia
\\  35. Lowell Observatory, 1400 Mars Hill Rd, Flagstaff, AZ 86001, USA
\\  36. Department of Astrophysical Sciences, Princeton University, Peyton Hall, Princeton, NJ 08544, USA
\\  37. Instituci\'o Catalana de Recerca i Estudis Avan\c{c}ats, E-08010 Barcelona, Spain
\\  38. Physics Department, 2320 Chamberlin Hall, University of Wisconsin-Madison, 1150 University Avenue Madison, WI  53706-1390
\\  39. Institute of Astronomy, University of Cambridge, Madingley Road, Cambridge CB3 0HA, UK
\\  40. Instituto de F\'\i sica, UFRGS, Caixa Postal 15051, Porto Alegre, RS - 91501-970, Brazil
\\  41. School of Physics and Astronomy, University of Southampton,  Southampton, SO17 1BJ, UK
\\  42. Computer Science and Mathematics Division, Oak Ridge National Laboratory, Oak Ridge, TN 37831
\\  43. Department of Physics, Stanford University, 382 Via Pueblo Mall, Stanford, CA 94305, USA

\label{lastpage}
\end{document}

%% file: files/info.tex
\newcommand{\VacSetGals}{4,721,380}
\newcommand{\VacSetArea}{143.66}
\newcommand{\VacSetDens}{9.13}
\newcommand{\VacSetZmean}{0.65}
\newcommand{\VacSptGals}{45,206,403}
\newcommand{\VacSptArea}{1,387.47}
\newcommand{\VacSptDens}{9.05}
\newcommand{\VacSptZmean}{0.63}
\newcommand{\VacArea}{1,511.13}
\newcommand{\VacGals}{49,927,783}

%% file: files/wazp_info.tex
\newcommand{\WazpClusters}{60,547}
\newcommand{\WazpClustersRMzlim}{39,439}
\newcommand{\WazpClustersZmaxDetec}{53,406}
\newcommand{\WazpSptDens}{39.45}
\newcommand{\WazpSetDens}{40.47}
\newcommand{\WazpZmaxDetec}{0.76}
\newcommand{\WazpZmaxRich}{0.60}
\newcommand{\WazpSptDensZmaxDetec}{35.28}
\newcommand{\WazpSetDensZmaxDetec}{35.28}
\newcommand{\WazpDensDetecRMzlim}{25.76}
\newcommand{\WazpDensRMzlim}{25.76}
\newcommand{\WazpSptDensRMzlim}{25.71}
\newcommand{\WazpSetDensRMzlim}{26.25}

%% file: files/wazp_mt_sz_info.tex
\newcommand{\SZmtNUMtot}{292}
\newcommand{\SZmtNUMphot}{200}
\newcommand{\SZmtNUMspec}{92}

\newcommand{\zbiassz}{0.017}
\newcommand{\zscatersz}{0.038}
\newcommand{\zbiasszWZ}{0.012}
\newcommand{\zscaterszWZ}{0.025}
\newcommand{\zbiasszSZ}{0.012}
\newcommand{\zscaterszSZ}{0.026}

\newcommand{\zbiasszphot}{0.021}
\newcommand{\zscaterszphot}{0.043}
\newcommand{\zbiasszWZphot}{0.014}
\newcommand{\zscaterszWZphot}{0.028}
\newcommand{\zbiasszSZphot}{0.015}
\newcommand{\zscaterszSZphot}{0.029}

\newcommand{\zbiasszspec}{0.010}
\newcommand{\zscaterszspec}{0.024}
\newcommand{\zbiasszWZspec}{0.007}
\newcommand{\zscaterszWZspec}{0.017}
\newcommand{\zbiasszSZspec}{0.007}
\newcommand{\zscaterszSZspec}{0.017}

\newcommand{\zerrWZszMT}{0.016}
\newcommand{\zerrSZwzMT}{0.017}
\newcommand{\zerrrTotSZNORM}{0.025}

\newcommand{\zerrWZszMTspec}{0.015}
\newcommand{\zerrSZwzMTspec}{0.000}
\newcommand{\zerrrTotSZNORMspec}{0.015}

\newcommand{\zerrWZszMTphot}{0.017}
\newcommand{\zerrSZwzMTphot}{0.024}
\newcommand{\zerrrTotSZNORMphot}{0.030}

\newcommand{\SZinDES}{331}
\newcommand{\SZmtMMT}{292}
\newcommand{\SZmtOneCand}{141}
\newcommand{\SZmtPoorXX}{14}
\newcommand{\SZmtOneCandPercRichXX}{62}
\newcommand{\SZunmatched}{39}
\newcommand{\SZunmatchedEdge}{12}
\newcommand{\SZunmatchedZmax}{27}
\newcommand{\SZunmatchedZmaxII}{9}
\newcommand{\SZmtPercWithinScat}{79}

%% file: files/wazp_mt_rm_info.tex
\newcommand{\rmMTcounts}{28,621} 

\newcommand{\rmRichI}{5} 
\newcommand{\rmRichII}{12} 
\newcommand{\rmRichIII}{20} 
\newcommand{\rmRichIV}{30} 
\newcommand{\rmRichV}{50} 
\newcommand{\rmRichVI}{234} 

\newcommand{\wzRichI}{5.6} 
\newcommand{\wzRichII}{14.5} 
\newcommand{\wzRichIII}{25} 
\newcommand{\wzRichIV}{39} 
\newcommand{\wzRichV}{68} 
\newcommand{\wzRichVI}{288} 

\newcommand{\rmRichBinI}{5-12} 
\newcommand{\rmRichBinII}{12-20} 
\newcommand{\rmRichBinIII}{20-30} 
\newcommand{\rmRichBinIV}{30-50} 
\newcommand{\rmRichBinV}{50-234} 

\newcommand{\wzRichBinI}{5.6-14.5} 
\newcommand{\wzRichBinII}{14.5-25} 
\newcommand{\wzRichBinIII}{25-39} 
\newcommand{\wzRichBinIV}{39-68} 
\newcommand{\wzRichBinV}{68-288} 

\newcommand{\rmRICHcountI}{15,099} 
\newcommand{\rmRICHcountII}{7,531} 
\newcommand{\rmRICHcountIII}{3,440} 
\newcommand{\rmRICHcountIV}{1,918} 
\newcommand{\rmRICHcountV}{632} 

\newcommand{\zbiasrm}{0.014} 
\newcommand{\zscatterrm}{0.037} 
\newcommand{\zbiasrmWZ}{0.009} 
\newcommand{\zscatterrmWZ}{0.026} 
\newcommand{\zbiasrmRM}{0.010} 
\newcommand{\zscatterrmRM}{0.027} 
\newcommand{\zbiasrmI}{0.013} 
\newcommand{\zscatterrmI}{0.041} 
\newcommand{\zbiasrmWZI}{0.009} 
\newcommand{\zscatterrmWZI}{0.028} 
\newcommand{\zbiasrmRMI}{0.010} 
\newcommand{\zscatterrmRMI}{0.029} 
\newcommand{\zbiasrmII}{0.014} 
\newcommand{\zscatterrmII}{0.035} 
\newcommand{\zbiasrmWZII}{0.010} 
\newcommand{\zscatterrmWZII}{0.024} 
\newcommand{\zbiasrmRMII}{0.010} 
\newcommand{\zscatterrmRMII}{0.025} 
\newcommand{\zbiasrmIII}{0.015} 
\newcommand{\zscatterrmIII}{0.031} 
\newcommand{\zbiasrmWZIII}{0.010} 
\newcommand{\zscatterrmWZIII}{0.022} 
\newcommand{\zbiasrmRMIII}{0.011} 
\newcommand{\zscatterrmRMIII}{0.023} 
\newcommand{\zbiasrmIV}{0.016} 
\newcommand{\zscatterrmIV}{0.029} 
\newcommand{\zbiasrmWZIV}{0.011} 
\newcommand{\zscatterrmWZIV}{0.020} 
\newcommand{\zbiasrmRMIV}{0.011} 
\newcommand{\zscatterrmRMIV}{0.021} 
\newcommand{\zbiasrmV}{0.019} 
\newcommand{\zscatterrmV}{0.026} 
\newcommand{\zbiasrmWZV}{0.013} 
\newcommand{\zscatterrmWZV}{0.018} 
\newcommand{\zbiasrmRMV}{0.014} 
\newcommand{\zscatterrmRMV}{0.019} 

\newcommand{\zerrWZrmMT}{0.026} 
\newcommand{\zerrRMwzMT}{0.010} 
\newcommand{\zerrrTotRMNORM}{0.028} 
\newcommand{\zerrrTotRMNORMpairs}{75\%} 

\newcommand{\zerrWZrmbinI}{0.028} 
\newcommand{\zerrRMrmbinI}{0.011} 
\newcommand{\zerrrTotRMNORMrmbinI}{0.031} 
\newcommand{\zerrWZrmbinII}{0.025} 
\newcommand{\zerrRMrmbinII}{0.009} 
\newcommand{\zerrrTotRMNORMrmbinII}{0.027} 
\newcommand{\zerrWZrmbinIII}{0.022} 
\newcommand{\zerrRMrmbinIII}{0.008} 
\newcommand{\zerrrTotRMNORMrmbinIII}{0.024} 
\newcommand{\zerrWZrmbinIV}{0.019} 
\newcommand{\zerrRMrmbinIV}{0.007} 
\newcommand{\zerrrTotRMNORMrmbinIV}{0.021} 
\newcommand{\zerrWZrmbinV}{0.016} 
\newcommand{\zerrRMrmbinV}{0.007} 
\newcommand{\zerrrTotRMNORMrmbinV}{0.018} 

\newcommand{\maxUMrmRICH}{103.78} 
\newcommand{\maxUMwzRICH}{225.29} 

\newcommand{\maxUMmmtrmRICH}{55.78} 
\newcommand{\maxUMmmtwzRICH}{154.29} 

\newcommand{\mtRMmtRM}{30,663} 
\newcommand{\mtRMmmtRM}{32,467} 
\newcommand{\mtRMmtWZ}{28,775} 
\newcommand{\mtRMmmtWZ}{33,498} 
\newcommand{\mtRMcross}{28,621} 

\newcommand{\RMClusters}{83,238} 
\newcommand{\RMClustersZmin}{0.10} 
\newcommand{\RMClustersZmax}{0.95} 
\newcommand{\RMClustersLambdaVII}{83,238} 
\newcommand{\mtRMexcludedFtEdrm}{9,330} 
\newcommand{\mtRMexcludedFtEdwz}{18,480} 
\newcommand{\mtRMexcludedCFrm}{6,953} 
\newcommand{\mtRMexcludedCFwz}{257} 
\newcommand{\mtRMexcludedFtEdCFrm}{16,283} 
\newcommand{\mtRMexcludedFTEdCFwz}{18,737} 
\newcommand{\mtRMremainUNMATCHEDrm}{34,488} 
\newcommand{\mtRMremainUNMATCHEDwz}{10,482} 
\newcommand{\mtRMtotalCPrm}{66,955} 
\newcommand{\mtRMtotalCPwz}{43,980} 
\newcommand{\mtRMstepI}{15,534} 
\newcommand{\mtRMstepII}{6,431} 
\newcommand{\mtRMstepIII}{3,783} 
\newcommand{\mtRMstepIVrm}{4,915} 
\newcommand{\mtRMstepIVwz}{3,027} 